\documentclass[a4paper,11pt]{article}

\usepackage{amssymb} 
\usepackage{amsmath} 
\usepackage{textcomp}
\usepackage[pdftex]{graphicx} 
\usepackage{subfigure} 
\usepackage{caption} 
\usepackage{lscape} 
\usepackage{vmargin} 
\usepackage{fancyhdr} 
\usepackage{longtable}
\usepackage[figuresright]{rotating} 
\usepackage{color}
\usepackage{hyperref}

\setmarginsrb{27mm}
	     {25mm}
	     {25mm}
	     {15mm}
	     {7mm}
	     {14mm}
	     {7mm}
	     {14mm}

\title{Solar Neutrino Spectroscopy}
\author{Michael Wurm\footnote{mail: {\it michael.wurm@uni-mainz.de}}\medskip\\
{\it PRISMA Cluster of Excellence and Institute of Physics,}\\
{\it Johannes Gutenberg University, 55099 Mainz, Germany}}

\date{\today}

\begin{document}

\maketitle

%
%

\begin{abstract}
\noindent More than forty years after the first detection of neutrinos from the Sun, the spectroscopy of solar neutrinos has proven to be an on-going success story. The long-standing puzzle about the observed solar neutrino deficit has been resolved by the discovery of neutrino flavor oscillations. Today's experiments have been able to solidify the standard MSW-LMA oscillation scenario by performing precise measurements over the whole energy range of the solar neutrino spectrum.

This article reviews the enabling experimental technologies: On the one hand mutli-kiloton-scale water Cherenkov detectors performing measurements in the high-energy regime of the spectrum, on the other end ultrapure liquid-scintillator detectors that allow for a low-threshold analysis. The current experimental results on the fluxes, spectra and time variation of the different components of the solar neutrino spectrum will be presented, setting them in the context of both neutrino oscillation physics and the hydrogen fusion processes embedded in the Standard Solar Model.

Finally, the physics potential of state-of-the-art detectors and a next-generation of experiments based on novel techniques will be assessed in the context of the most interesting open questions in solar neutrino physics: a precise measurement of the vacuum-matter transition curve of electron-neutrino oscillation probability that offers a definitive test of the basic MSW-LMA scenario or the appearance of new physics; and a first detection of neutrinos from the CNO cycle that will provide new information on solar metallicity and stellar physics.
\end{abstract}

\newpage

\tableofcontents

\newpage

%
%

\section{Introduction}

The detection of solar neutrinos is one of the great on-going themes of experimental neutrino physics. Almost 50 years have past since the first detection of neutrinos from solar fusion in the Homestake experiment \cite{Davis:1968cp}, and more than a decade since the SNO experiment provided the final confirmation of their flavor-changing oscillations \cite{Ahmad:2002jz,RevModPhys.88.030502}. Both prize-winning measurements mark the beginning and the end of a long period in which solar neutrino physics was defined by the experimental and theoretical efforts to resolve the puzzle of the apparent solar neutrino deficit \cite{Bahcall:1990cv,Bahcall:2002ng}. Over the decades, the field has evolved from the understanding of the dominant features of the signal towards the study of more intricate sub-leading effects. Today, the production rates and oscillation probabilities of the {$^7$Be} and {$^8$B} components in the solar neutrino spectrum are determined to the level of a few per cent \cite{Abe:2016nxk,Bellini:2011rx}, a precision inconceivable in the early days of the field \cite{Bahcall:1995gw}.

This dramatic improvement in observational accuracy was heralded by the arrival of huge underground detectors based on water and organic scintillator, able to perform time and energy-resolved measurements of the solar neutrino flux \cite{Fukuda:2002uc,Boger:1999bb,Alimonti:2008gc,Markoff:2003tg}. The motivation for their continuing observation program is twofold: With the basic MSW-LMA oscillations \cite{Bahcall:1999ed} established (sect.~\ref{sec:pee}), particle physicist are now scrutinizing the data for deviations from this standard three-flavor scenario, searching for instance for signs of non-standard neutrino interactions \cite{Friedland:2004pp,Minakata:2010be} or mixing of active with ultra-light sterile neutrinos \cite{deHolanda:2010am}. On the other hand, solar physics will profit from more precise flux data on all spectral components \cite{Serenelli:2012zw}, especially regarding the neutrinos from the subdominant CNO fusion cycle for which currently only upper limits exist \cite{Collaboration:2011nga}.

This paper reviews the current status of the field from an observational point of view, centered on the spectroscopic detectors that today form the forefront of solar neutrino experiments. Starting from short introductions on the relevant solar and neutrino oscillation physics (sects.~\ref{sec:solnuspec}+\ref{sec:solnuosc}), section \ref{sec:detectors} describes the main features of the four most proponent experiments forming the forefront of real-time solar neutrino observatories: the large-volume water Cherenkov detectors SNO \cite{Boger:1999bb} and Super-Kamiokande \cite{Fukuda:2002uc} and the liquid-scintillator detectors Borexino \cite{Alimonti:2008gc} and KamLAND \cite{Markoff:2003tg}. In section \ref{sec:solnumeas}, the essence of their observational results is reviewed, ordered by the components of the solar neutrino spectrum. Section \ref{sec:status} presents our current understanding of solar neutrino oscillation and stellar physics, as well as the open questions that remain and that could be answered by the present or up-coming generation of large-volume experiments. Finally, section \ref{sec:newexp} provides an overview of new ideas for solar neutrino detectors currently discussed within the community, highlighting their strengths and limitations with regard to specific topics.

%
%

\section{The solar neutrino spectrum}
\label{sec:solnuspec}

The prediction of the neutrino spectrum originating from solar fusion processes arises from two basic ingredients: On the one hand, the Standard Solar Model (SSM) describes the basic environment of the solar interior \cite{Bahcall:2002ng}, e.g. the density and temperature profile of the solar neutrinosphere. On the other hand, nuclear physics provides the network of fusion reactions featuring neutrinos in the final state, the cross-sections and the spectral shapes that in combination make up the solar neutrino spectrum \cite{Raffelt:1996wa}. In the following, a short overview of these two fundamental building plots is given, highlighting the parameters that enact the largest influence on the neutrino spectrum.

\subsection{The Standard Solar Model}
\label{sec:ssm}

\begin{figure}[ht]
\centering
\includegraphics[width=0.6\textwidth]{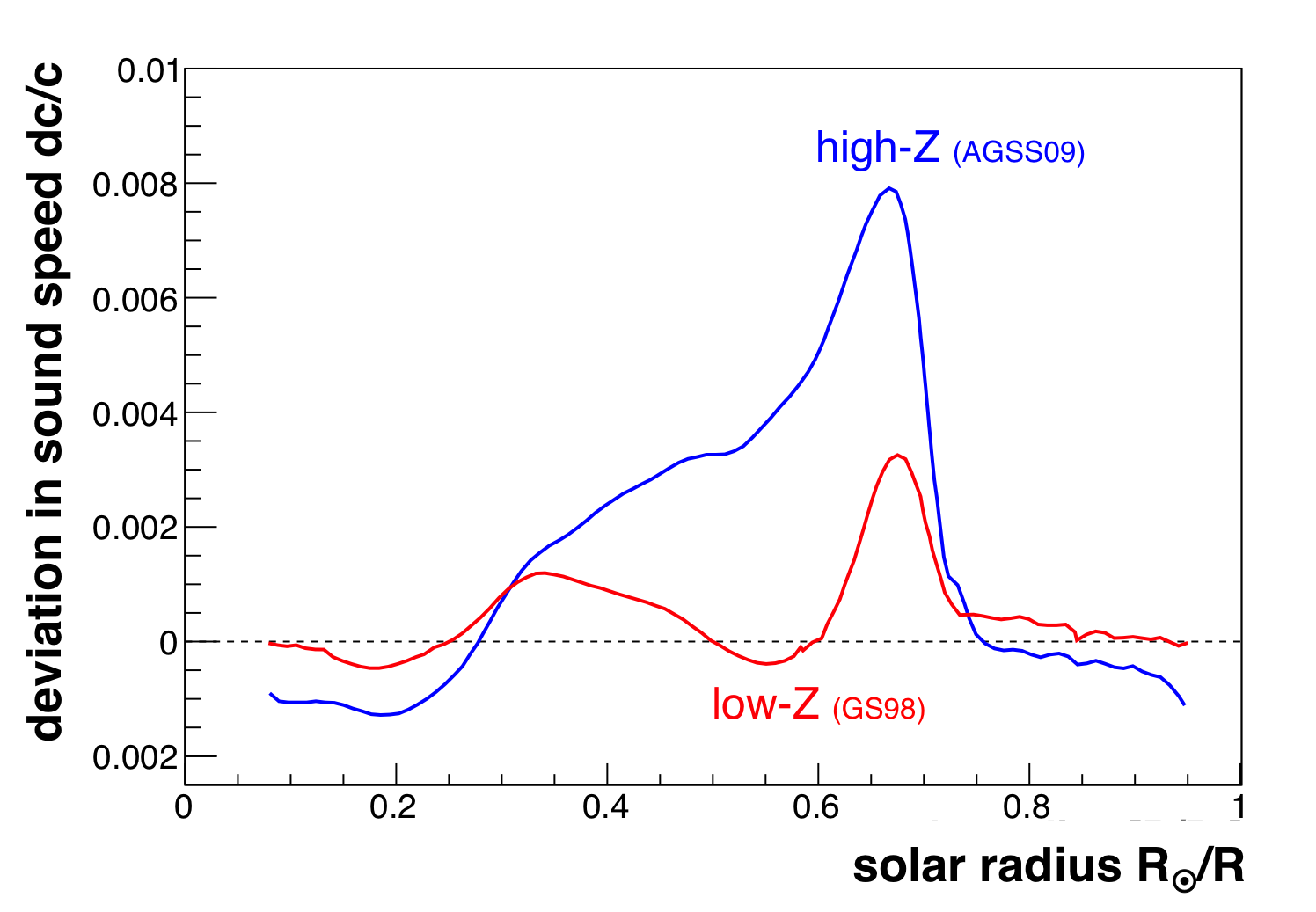}
\caption{Relative deviation of the radially dependent sound-speed profiles as measured by helioseismology and as calculated from the standard solar model. The curves displayed reflect different assumptions on solar metallicity \cite{Asplund:2009fu}: Noteworthy, the former {\it high-Z} model (GS98) \cite{Grevesse:1998bj} shows substantially better agreement with helioseismic data then the improved {\it low-Z} (AGSS09) model \cite{Asplund:2009fu}}
\label{fig:solarcsound}
\end{figure}

The SSM uses basic stellar theory and inputs from the available optical data of the Sun to construct a model of the solar interior. Based on this, the SSM makes predictions for a number of observables that are accessible for helioseismology and in particular for neutrino experiments. As its theoretical foundation, the model relies on a number of rather undisputed assumptions \cite{Raffelt:1996wa, Robertson:2012ib, Stix:2004}:
\medskip\\
{\bf Hydrogen fusion.} The Sun is a main sequence star aged 4.5 billon years. Its energy output is maintained by the release of nuclear binding energy. Thermonuclear burning of hydrogen to helium happens via the net reaction
\begin{eqnarray}
\label{eq:pp}
4p \to {^{4}{\rm He}} + 2e^+ + 2\nu_e.
\end{eqnarray}
A total of $Q=26.73$\,MeV is released per $^{4}{\rm He}$ nucleus produced\footnote{It should be noted that combined with the solar constant $S_\odot$ measuring electromagnetic luminosity, the energy output per net fusion alone can be used for a rather accurate measurement of the number of neutrinos emitted by the Sun, $\Phi_\nu = 2S/Q = 6.6 \cdot 10^{10} {\rm cm}^{-2}{\rm s}^{-1}$}. Conditions to undergo fusion processes are met only in the very center of the sun ($R\lesssim 0.2\,R_{\odot}$) where temperatures and densities exceed $7\cdot 10^6$\,K and 20\,g/cm$^3$, respectively. Hydrogen burning happens mainly via the proton-proton $(pp)$ chain \cite{Bethe:1939bt, Salpeter:1952ffc}, with a minor contribution by the catalyst Bethe-Weizs\"acker cycle \cite{Bethe:1939bt, Weizsaecker:1938}. 
\medskip\\
{\bf Local hydrostatic equilibrium} is maintained throughout the Sun. Despite the fact that the Sun constantly looses energy, the thermal pressure caused by fusion processes locally counterbalances the gravitational inward force for all radii up to the photosphere. On short time scales, the energy released in fusion is assumed to exactly compensate for the energy losses by electromagnetic (and neutrino) radiation. An equation of state is required for the description of the pressure, density and temperature of the solar plasma. Mainly consisting of fully ionized hydrogen and helium, the plasma can be described in good approximation as an ideal gas. Corrections that account for electron degeneracy, radiation pressure and plasma screening effects in the center as well as partially ionization of metals close to the solar surface are only on the per-cent level.
\medskip\\
{\bf Energy transport} inside the Sun is realized via two processes: In the inner layers of the Sun, the transport is by electromagnetic radiation. The radiative opacity of solar matter is determined by the interactions of the photons with the solar plasma: free-free Thomson scattering off electrons, bound-free scattering off the electronic hulls of partially ionized metals as well as inverse bremsstrahlung on hydrogen and helium nuclei. The resulting free mean path of photons is very low: Quanta originally created in the center of the Sun take several ten thousand years to arrive to the solar surface. The decreasing temperature towards the outer layers leads to a rise in the fraction of partially ionized atoms and thus to a further increase in opacity. From a radius of $\sim0.7\,R_{\odot}$, convection thus replaces radiation as the most effective means of transporting energy outward. This gives rise to a multilayer structure of convective cells of varying depths and dimensions, resulting in the distinctive pattern of rising hot cells and retreating cooler plasma that is observable in the granules on the solar surface.\medskip\\
{\bf Stellar evolution codes.} To arrive at the current state of the Sun, solar modelling starts out at the very beginning of solar nuclear burning when the Sun entered the main sequence, and tracks the evolution of the star over time (e.g. the accumulation of helium in the solar core) to the present. Its current age of roughly 4.5 billion years is determined from the relative abundances of long-lived radioactive isotopes in solar-system meteors, most famously the strontium/rubidium clock that relies on the decay of the rubidium isotope {$^{87}$Rb} ($T_{1/2}=4.88\cdot 10^{10}$\,yrs) \cite{Hahn:1943}.\\
A variety of stellar evolution codes is available for this formidable task \cite{Weiss:2008, Lebreton:2008iu}: The current state of the art relies on one-dimensional models with $\sim 10^3$ radial mesh points and time steps of no longer than 10\,Myrs. Radial resolution is chosen higher in critical regions (e.g. the bottom of the convection zone), and also the duration of time steps is adjusted to the stellar evolution rate. For a given time step, the equations of stellar structure are solved for each radial mesh point, taking into account the equation of state, the local energy release by fusion, and radiative opacities. In addition, a set of simple boundary conditions has to be met that are given by mass continuity and the relations between surface temperature, area and luminosity prescribed by the Stefan-Boltzmann-law. Using an informed guess on the start parameters, a radially continuous solution for density, temperature and pressure is found by iterative methods that takes into account the various interdependences of the determining quantities.\\
The resulting configuration is only a brief snap-shot of the stellar state variables. Following the slow burn-up of hydrogen in the core region, the central temperature of the Sun increases over the ages, which has to be taken into account by a sequence of subsequent sets of calculations.\medskip\\
{\bf Input parameters.} Solar modelling is a case of reverse engineering: Arriving at the present age of the Sun, the resulting stellar model has to fulfill a set of macroscopic parameters that we currently observe: Mostly mass, surface radius and surface luminosity. While these requirements are set {\it a-posteriori}, they can be regarded as input parameters as they are undisputed and strongly constrain the parameter space.\\
Additional input is required on the microscopic level: Solar structure critically depends on the initial elemental abundances, especially of elements heavier than helium, a.k.a. metals, that have to be derived from observation. While quantities for refractory elements can be gained from the analysis of meteorites, the volatile metals (C, N, O, Ne, Ar) are determined from abundances measured by absorption lines in the photosphere \cite{Grevesse:1998bj}. Note that the surface composition today is supposed to correspond in good approximation to the initial quantities of hydrogen $X_{\rm ini}$, helium $Y_{\rm ini}$ and metals $Z_{\rm ini}$. Changes are mainly induced in the core when elements undergo fusion. However, some minor diffusion of heavier elements towards the core is expected, and can be included in the evolution calculations \cite{Thoul:1993kt}.\\
The local energy output is calculated based on the framework of reactions underlying the thermonuclear fusion of protons to helium \cite{Raffelt:1996wa}. The dependence of the involved reaction cross-sections on ambient temperature is one of the critical parameters when correlating the solar energy outputs to the emitted neutrino fluxes \cite{Bahcall:1996vj} (sect.~\ref{sec:solnurates}).\\
Further important parameters are the Rosseland mean opacities to radiative transport that depend on local density, temperature, elemental composition and ionization levels \cite{Serenelli:2012zw, Seaton:2004vv}. The spectrally averaged values are usually derived form tables such as OPAL \cite{Seaton:2004vv, Iglesias:1996bh}, the contents in turn deriving from extensive calculations.
\medskip\\
{\bf Benchmarking the result.}  In order to validate the resulting configuration, predictions are made for the observables that have not been used during construction of the solar model. In the light of the current experimental efforts, this regards mainly two experimentally approaches: The observation of helioseismic waves by the surface oscillations they induce (see below) and the neutrino fluxes originating from different branches of the $pp$ chain and CNO cycle (sect.~\ref{sec:solnurates}). Uncertainties on these predictions are obtained by the computation of multiple model stars based on variations of the input parameters within their uncertainty ranges. The resulting ensemble of up to 10,000 model Suns reflects the expected mean values and spreads of the observables which are then summarized by the SSM \cite{Bahcall:2005va}.
\medskip\\
{\bf Solar neutrino rates.} Based on the radial profiles obtained for temperature, density, pressure and elemental abundances, the SSM returns the rates at which nuclear fusion processes are proceeding. Here, our main interest are the weak charged-current reactions involved in generating neutrinos. As will be described in more detail in section \ref{sec:ssm}, the SSM predicts both the total rates as well as the radial production profiles of these contributions \cite{Bahcall:2005va}. Moreover, it provides the radius-dependent densities of electrons and neutrons that are mandatory for the correct description of neutrino oscillations in matter (sect.~\ref{sec:oscmat}).
\medskip\\
{\bf Helioseismic predictions.} The SSM predicts as well a number of quantities that can be tested by helioseismology, especially by the observations of sub-surface pressure ($p$) waves (e.g.\cite{ChristensenDalsgaard:2002ur}): Since the early 1980s \cite{TurckChieze:2010gc}, the occurrence of these waves has been established and investigated by a variety of ground- and satellite-based instruments observing the induced motions of the solar surface. Evidence of $p$-waves is found both by periodic variations of the total solar irradiance on the time scale of minutes and by oscillatory patterns in the Doppler-shifts of absorption lines observed from the solar photosphere. Within the solar body, the radial propagation of waves depends critically on the local parameters of the plasma. In particular, the depth to which $p$-waves can penetrate into the deeper layers of the Sun is a function of wave frequency, which leads in turn to the formation of standing waves and the occurrence of preferred ridges in wave power spectra that relate preferred oscillation frequencies to the observed horizontal wave numbers \cite{ChristensenDalsgaard:2002ur}.\\
Information on the solar interior is mostly inferred by a direct modelling approach in which the eigenfrequencies and wave numbers are determined based on the SSM \cite{ChristensenDalsgaard:2002ur}. By varying the assumptions on the underlying parameters, the SSM output can be adjusted to match the observed helioseismic data. By this process, helioseismic observation can access quantities like the depth-dependent sound speed, solar differential rotation and the depth of the convective zone: At its bottom, the disappearance of radial motion causes a discontinuity for the propagation of $p$-waves that is clearly discernible in helioseismic measurements. Until recent years, the per-mill level agreement between the SSM predictions and results from helioseismology has been striking \cite{TurckChieze:2010gc,Bahcall:1996qw}, considerately strengthening the trust in the correctness of the SSM despite the observed solar neutrino deficit.
 \medskip\\
{\bf Puzzle on solar abundances.} Over the last decade, a noteworthy discrepancy arose in measurements of the volatile-element abundances in the solar photosphere: A re-evaluation of the Fraunhofer absorption line spectrum resulted in lower abundance values for these elements \cite{Asplund:2004eu}. Compared to earlier analyses \cite{Grevesse:1998bj}, the recent effort relies on a fully three-dimensional modelling of radiation hydrodynamics and transport, thereby refining the description of the Doppler and collision broadening of the absorption lines \cite{Asplund:2009fu}. As a consequence, model profiles provide now a substantially better fit to the observed line shapes. In many cases, this results in a lower line strength required compared to earlier analyses \cite{Asplund:2009fu}.\\
However, the reduced elemental abundances result in a substantial degradation of the formerly astonishing level of agreement between SSM predictions and the observables of helioseismology. Most striking is probably the substantially inflated level of (dis)agreement reached in the radial-dependent sound speed profile that is illustrated by figure \ref{fig:solarcsound}. This tension can be somewhat relieved if the lower abundances are counterbalanced by increased elemental opacities \cite{Asplund:2009fu, Serenelli:2009ww}.\\
While this dispute remains unresolved, the pragmatic solution currently adopted within the solar neutrino community is the use of two different sets of volatile element abundances (metallicities) when calculating the expected neutrino fluxes: The former standard solar com- position dubbed as the high-metallicity model ({\it high-Z} or GS98) described in \cite{Grevesse:1998bj}, and the novel, low-metallicity model ({\it low-Z} or AGSS09) derived from the new evaluation by \cite{Asplund:2009fu}.\\
Resolving this new puzzle is a great task that may lie within the reach of the next generation of solar neutrino experiments \cite{Robertson:2012ib} (sects.~\ref{sec:status}+~\ref{sec:newexp}).

\subsection{Solar neutrino rates}
\label{sec:solnurates}

Solar neutrinos are emitted in a number of reactions that are part of two competing frameworks: the $pp$-chain and the CNO cycle \cite{Raffelt:1996wa}. The total solar neutrino spectrum is built from a superposition of the fluxes emitted in this partial reactions. In this context, the SSM provides the environmental parameters of the solar core to determine the relative rates at which these reactions occur \cite{Bahcall:2002ng}.
\medskip\\
{\bf Hydrogen burning.} In the middle of its life time on the main sequence, the Sun feeds its electromagnetic radiation by thermonuclear burning of hydrogen to helium-4. In the process, neutrinos are created when protons are converted to neutrons in weak charged-current interactions. The net reaction displayed in equation~(\ref{eq:pp}) features two positrons and correspondingly to electron neutrinos ($\nu_e$) in the final state. Note that the flavor of these neutrinos is inevitably fixed to $\nu_e$ by (a) the low $Q$-values along the reaction chain that forbids the creation of a muon in the final state and (b) the conservation of lepton flavor in weak interactions.

\begin{figure}[ht]
\centering
\includegraphics[width=\textwidth]{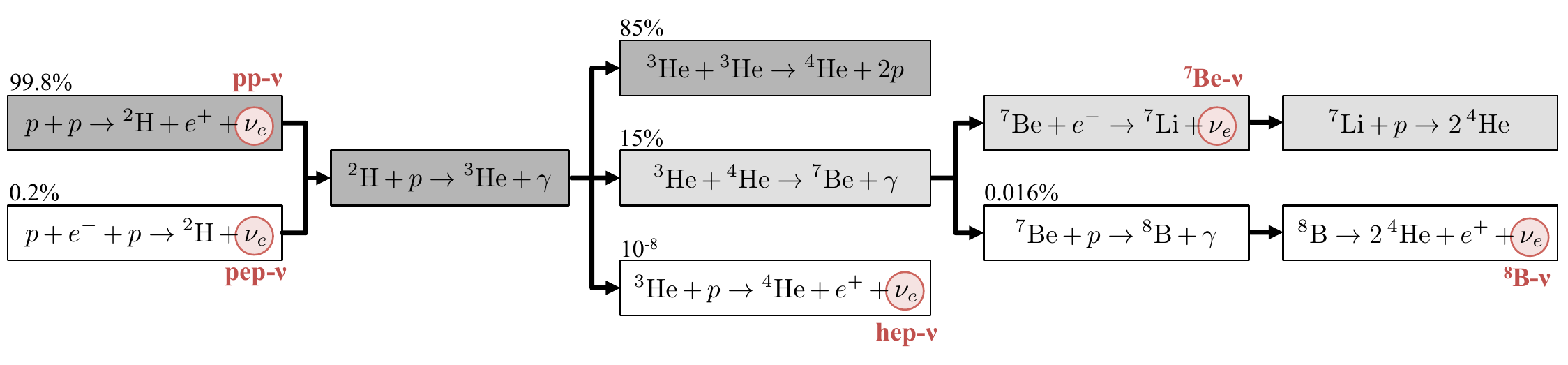}
\caption{The solar proton-proton fusion chain that accounts for 99\,\% of the solar energy release \cite{Raffelt:1996wa, Bethe:1939bt}. The neutrinos emitted in several of the reactions are marked in red.}
\label{fig:ppchain}
\end{figure}

\begin{figure}[ht]
\centering
\includegraphics[width=0.75\textwidth]{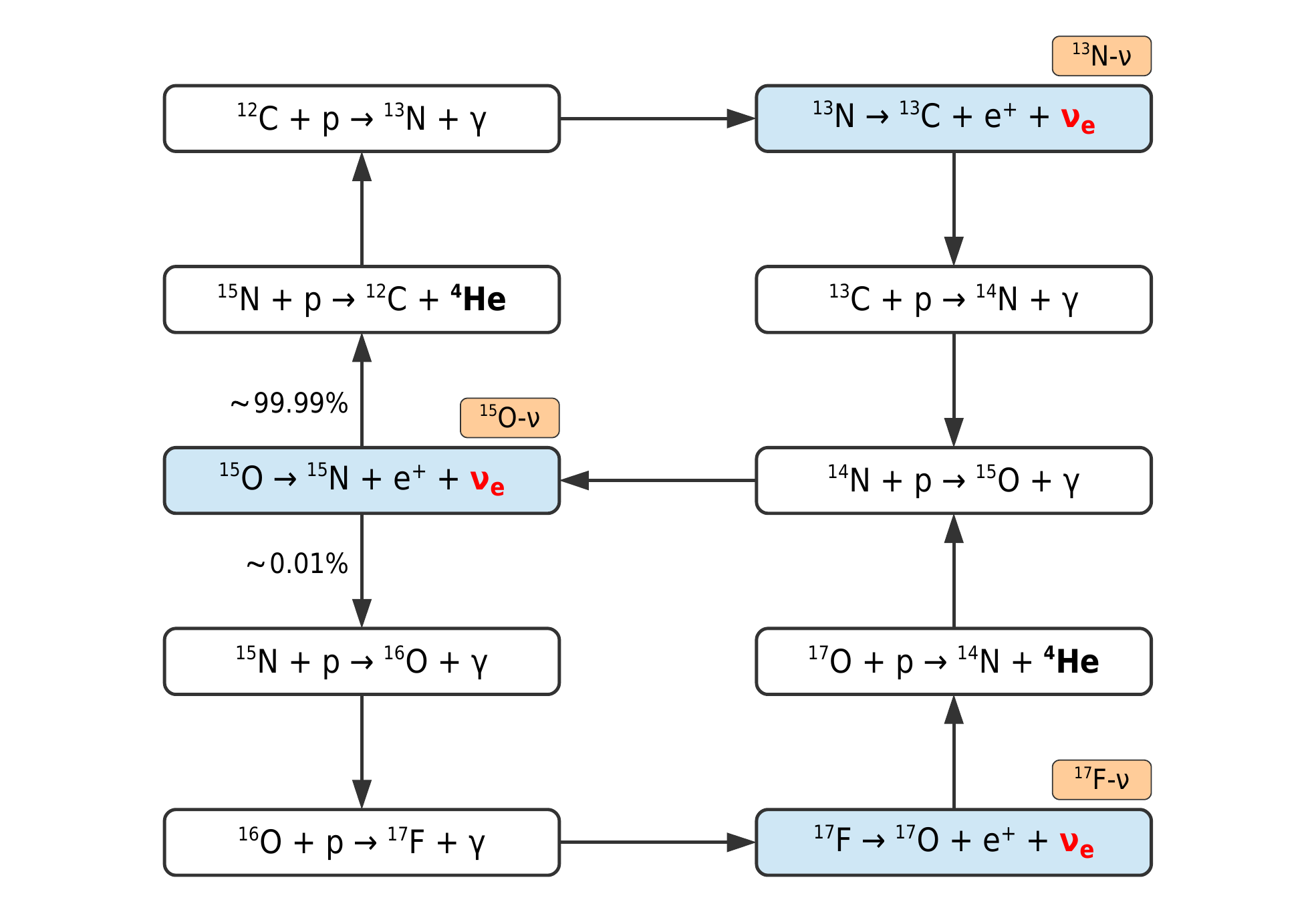}
\caption{The catalyst CNO cycle provides only for a small fraction of solar energy release \cite{Raffelt:1996wa, Weizsaecker:1938}. Reactions emitting neutrinos are shown shaded. {\it (Figure courtesy of S. Lorenz)}}
\label{fig:cnocycle}
\end{figure}

\noindent While equation (\ref{eq:pp}) describes the net fusion reaction, there are several different pathways realized in the Sun that all  lead eventually to the formation of {$^{4}$He}:
 \medskip\\
{\bf pp-chain.} The dominant hydrogen fusion process in the Sun is known as the proton-proton (or $pp$-) chain and was first  proposed by Hans Bethe. As depicted in figure~\ref{fig:ppchain}, the $pp$-chain starts out by the basic fusion of two protons ($^1$H$^+$) to a deuteron ($^2$H$^+$), releasing an $\nu_e$ of low energy, $Q_\beta(pp)= 423$\,keV. Less frequently, a three-body interaction $p+e^-+p$ leads to the formation of a deuteron by electron capture, releasing a mono-energetic neutrino of 1.44\,MeV ($pep$). In any case, the new-born deuteron quickly fuses with a proton to form {$^{3}$He} in a strong interaction process.\\
From here, three reactions paths are possible: Either the fusion of two {$^{3}$He} nuclei (pp-I), the fusion of {$^{3}$He} and {$^{4}$He} (pp-II), or the capture of a further proton on {$^{3}$He} leading to the direct formation of {$^{4}$He} (pp-III). From point of view of solar neutrino detection, the pp-II branch is the most interesting. In this case, the newly formed {$^{7}$Be} nucleus will capture either an electron to be converted to {$^{7}$Li} and in the process release a mono-energetic $\nu_e$ of 862\,keV ($^7$Be), or a proton to form $^8$B, a short-lived ($<1$\,s) isotope undergoing $\beta^+$-decay and releasing a high-energy neutrino, $Q_\beta(^8{\rm B})= 14.6$\,MeV \cite{Raffelt:1996wa}.\\
In principle, the highest-energy neutrinos are expected from the direct fusion of $^{3}{\rm He}+p$. However, due to the low fusion rate the corresponding $hep$ neutrinos are outside the reach of current-day detectors \cite{Raffelt:1996wa}.
 \medskip\\
{\bf Bethe-Weizs\"acker cycle.} Alternatively, {$^{4}$He} is formed from protons in a system of several reaction cycles that are based on the elements carbon, nitrogen and oxygen (CNO) \cite{Weizsaecker:1938}. The general layout of the double-cycle is shown in figure~\ref{fig:cnocycle}. Note that the heavier elements (metals) involved serve only as catalysts for hydrogen fusion to helium and are themselves not consumed in the process. As a consequence, the production of {$^{4}$He} involves also in the CNO case the emittance of two neutrinos per full cycle, that populate the solar neutrino spectrum most notably in the (1$-$2)\,MeV region.
 
\begin{figure}[ht]
\centering
\includegraphics[width=\textwidth]{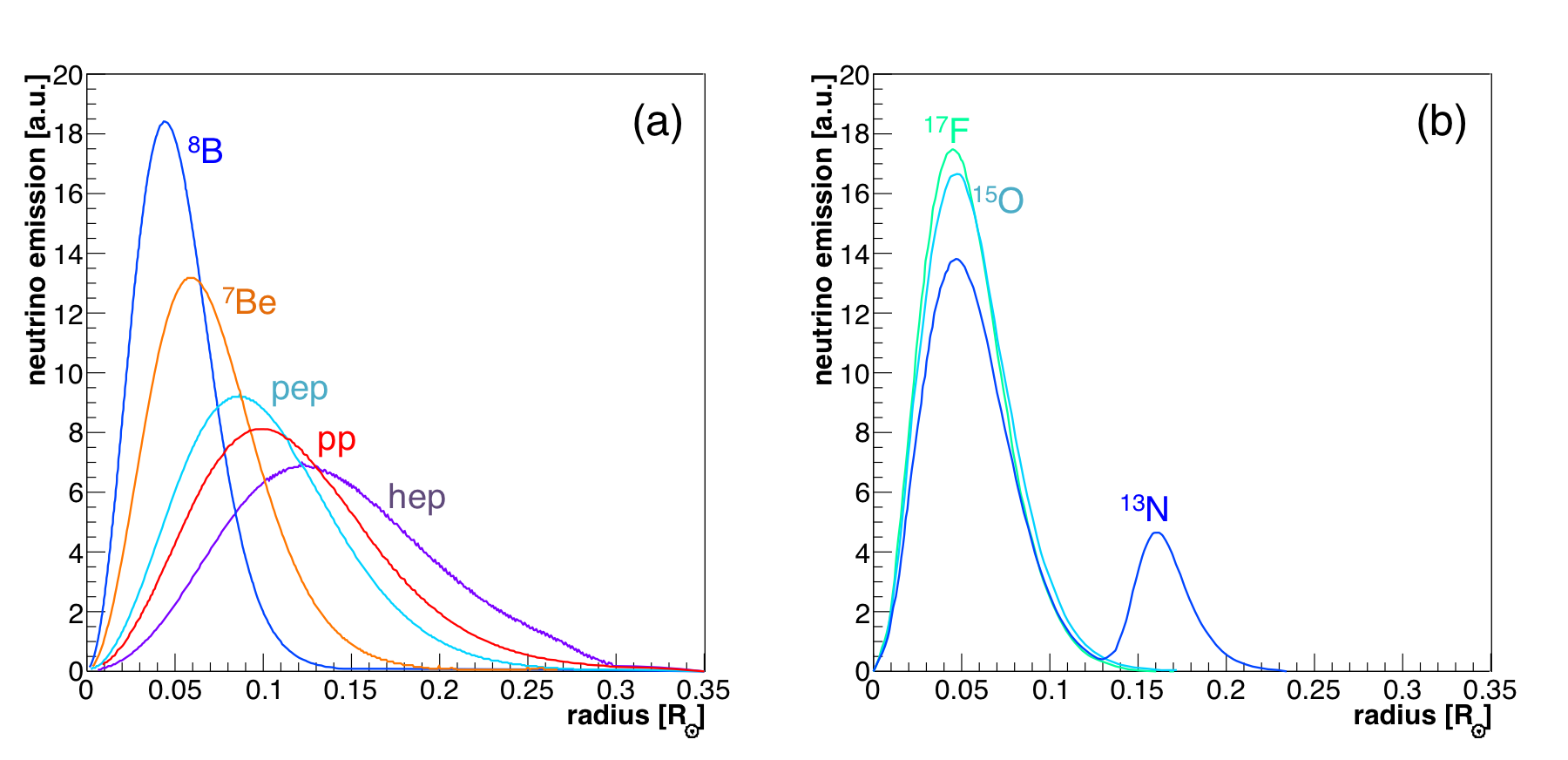}
\caption{Radial profiles of solar neutrino production (neutrinospheres) for the different sub-branches of (a) $pp$-chain and (b) CNO cycle \cite{Gann:2015yta,Lopes:2013nfa}.}
\label{fig:raddists}
\end{figure}

While the CNO cycle features prominently in heavier stars, the Sun as a relatively light star doesn't reach sufficiently high central temperatures (yet) to allow for a significant CNO contribution to the total energy budget. In fact, a steady reaction cycle is only maintained in the very center of the Sun and creates an energy output corresponding to only (1$-$2)\,\% of the total energy budget. At larger radii, the CNO cycle stalls at the reaction $^{14}{\rm N}(p,\gamma)^{15}{\rm O}$ which forms the bottle neck of the whole cycle \cite{Robertson:2012ib}. The experimental uncertainty on the corresponding reaction cross-section directly translates to the CNO reaction rate occurring in the Sun and thus to the expected ratio of $pp$ to CNO neutrinos emitted \cite{Robertson:2012ib, Adelberger:2010qa}.
\medskip\\ 
\noindent{\bf Reaction rates.} Fusion processes are usually initiated by the collision of two reaction partners that are recruited from the constituents of the solar plasma. The fusion rate $\dot n$ can be written as
\begin{eqnarray}
\dot n = n_1 n_2 \langle \sigma v \rangle_{\rm MB}
\end{eqnarray}
where $n_{1,2}$ are the number densities of the reactants and $\langle \sigma v \rangle_{\rm MB}$ corresponds to the reaction cross sections averaged over the velocity profile of the particles involved \cite{Raffelt:1996wa}. As fusion inside the Sun is a thermal process, the kinetic energy of the reactants corresponds to the local plasma temperature that reaches up to $T_{\rm c(ore)} = 1.5 \cdot 10^7$\,K in the solar core. The local velocity ($v$) profile is thus given by the well-known Maxwell-Boltzmann distribution,
\begin{eqnarray}
f_{\rm MB}(v) = \sqrt{\left(\frac{m}{2\pi k_B T}\right)}\cdot 4\pi v^2 \cdot \exp \left( \frac{mv^2}{2k_{\rm B}T} \right)
\end{eqnarray}
corresponding to a mean kinetic energy of $\frac{1}{2}mv^2 \sim 1$\,keV for protons at $T_{\rm c}$.\\
However, the kinetic energy provided is not sufficient to overcome the Coulomb repulsion exhibited by the positive charges of the reactants. The reaction rate is thus governed by the quantum-mechanical probability for tunnelling the Coulomb wall, introducing a steep dependence on the kinetic energy of the reaction partners\footnote{Note that Coulomb repulsion is somewhat reduced by screening effects induced by the electrons of the solar plasma. This can be taken into account by adding a Debye-H\"uckel enhancement factor $f_e(E)$ in the cross-section \cite{Raffelt:1996wa, Adelberger:1998qm}.}. The tunnelling probability is usually factored out by re-writing the cross-section as a product of the Gamow factor and the astrophysical $S$-factor:
\begin{eqnarray}
\sigma(E) = S(E)\frac{\exp(-2\pi\eta)}{E},
\end{eqnarray}
with $\eta=\frac{Z_1Z_2e^2}{4\pi\epsilon_0\hbar v}$ the dimensionless Sommerfeld parameter ($Z_{1,2}$ denoting the electric charges of the reactants) \cite{Adelberger:1998qm}. The $S$-factor $S(E)$ is such reduced to reflect only the nuclear processes underlying the fusion reaction.\\
Regarded individually, the rate at which these reactions occur in the Sun is extremely low. The typical live time of a free proton to undergo $pp$-fusion is of the order of 8\,Gyrs. Therefore, the measurement of the relevant cross-sections and $S$-factors in laboratory environments is notoriously difficult. In the energy range of interest, i.e.~the Gamow peak for which the product of the reactants' velocity distribution and the cross-section becomes maximal, the challenge posed to experimenters is to reach sufficiently high rates while maintaining low background conditions \cite{Adelberger:1998qm}. In fact, the cross-sections of reactions involving weak currents are in some cases too low for a direct measurement, e.g.~for the basic $pp$ fusion rate at $\sim2$\,keV. In these circumstances, the $S$-factors are either extrapolated from higher energies, calculated based on weak interaction and nuclear theory, or using a combination of both \cite{Adelberger:1998qm}.
\medskip\\
{\bf Influence of solar state values.} Reaction cross-sections and thus fusion rates depend significantly on the ambient conditions: Especially the Gamow factor introduces a steep dependence on the kinetic energies of the reactants, that is usually described by a temperature power law. For instance, the $pp$-fusion rate and thus neutrino emission scales with $T^4$ \cite{Raffelt:1996wa}, while the same dependence for {$^8$B}-neutrinos is $T^{24}$ \cite{Bahcall:1996vj}. Given the outward-decreasing temperature profile, thermonuclear burning is restricted to the solar core, the radial extension depending on the specific reaction type (fig.~\ref{fig:raddists}). In this context, solar metal abundances play an important role as the local metallicity largely influences the electromagnetic opacity and thus the temperature gradient \cite{Serenelli:2012zw}.

\begin{figure}[ht!]
\centering
\includegraphics[width=0.4\textwidth]{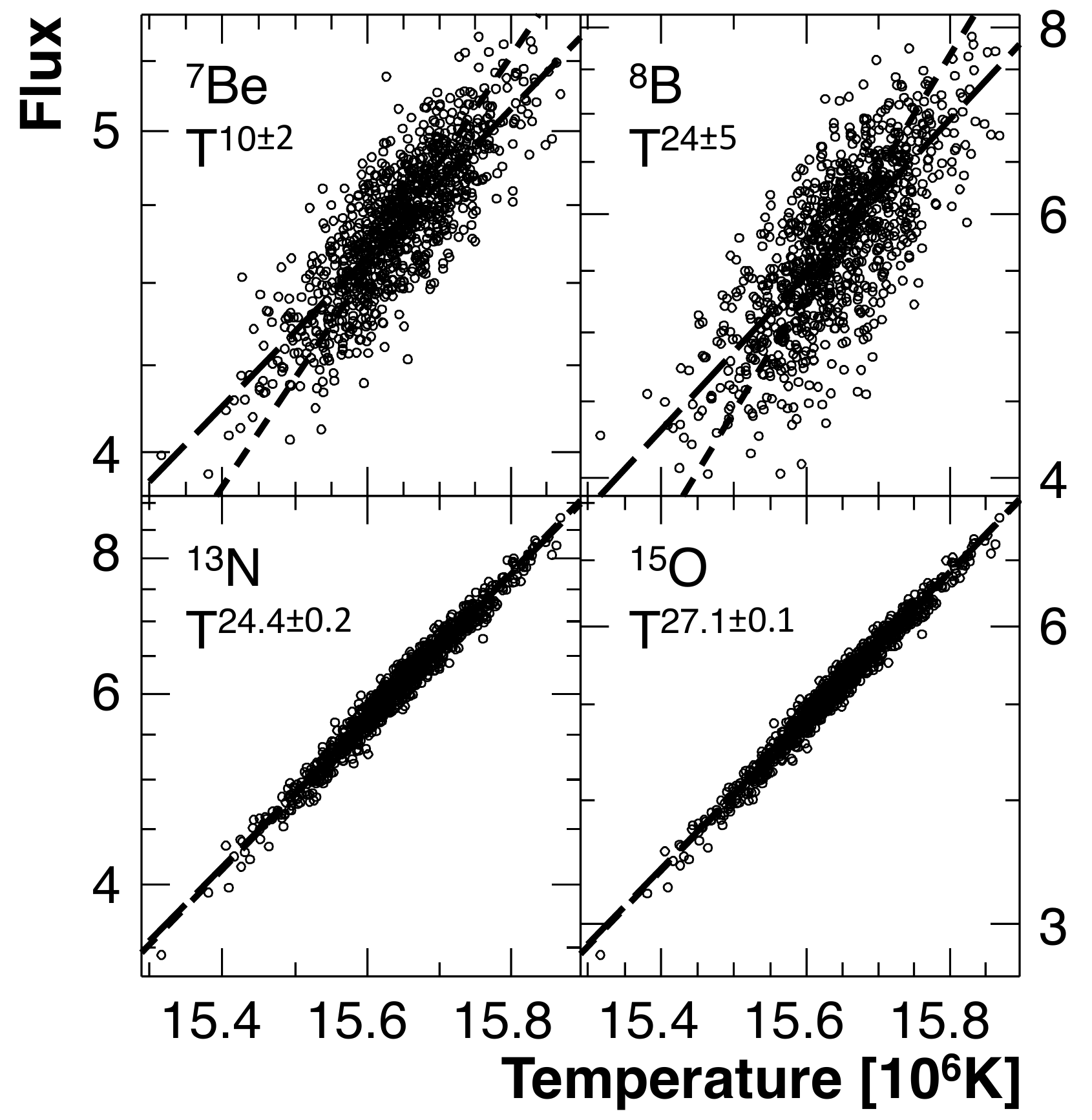}
\caption{SSM calculations of neutrino fluxes from {$^7$Be}, {$^8$B}, {$^{13}$N} and {$^{15}$O} as a function of the solar core temperature. Fluxes are given in units $10^9$, $10^6$, $10^8$ resp.~$10^8$ per cm$^2$s.  The functionality is described by steep power laws \cite{Bahcall:1996vj}.}
\label{fig:nufluxvstc}
\end{figure}

\noindent {\bf Impact of metallicity.} Table \ref{tab:ssm} shows the neutrino rates predicted for the $pp$- and CNO- reactions for two different sets of input metallicities Z (sect.~\ref{sec:ssm}): The former {\it high-Z} model GS98 \cite{Grevesse:1998bj} and the more recent {\it low-Z} model AGSS09 \cite{Asplund:2009fu}. As Rosseland opacities depend on the abundance of metals in the solar plasma, the {\it high-Z} scenario translates to a steeper temperature gradient inside the Sun. The changed temperature profile in turn leads to a slight shift in the relative fusion rates in the different branches of the $pp$ chain and CNO cycle.

\begin{table}
\begin{center}
\begin{tabular}{lccccc}
\hline 
Branch & GS98 & AGSS09 & Unit [cm$^{-2}$s$^{-1}$] & Uncertainty & Difference\\
\hline
$pp$ & 5.98 & 6.03 & $\times10^{10}$ & $\pm$0.6\,\% & +0.8\,\% \\
$pep$ & 1.44 & 1.47 & $\times10^{8}$ & $\pm$1.2\,\% & +2.1\,\% \\
$^7$Be & 5.00 & 4.56 & $\times10^{9}$ & $\pm$7\,\% & $-$8.8\,\% \\
$^8$B & 5.58 & 4.59 & $\times10^{6}$ & $\pm$13\,\% & $-$17.7\,\% \\
$hep$ & 7.91 & 8.22 & $\times10^{3}$ & $\pm$15\,\% & +4.1\,\% \\
$^{13}$N & 2.96 & 2.17 & $\times10^{8}$ & $\pm$15\,\% & $-$26.7\,\% \\
$^{15}$O & 2.23 & 1.56 & $\times10^{8}$ & $\pm$16\,\% & $-$30.0\,\% \\
$^{17}$F & 5.52 & 3.40 & $\times10^{8}$ & $\pm$18\,\% & $-$28.4\,\% \\
\hline
\end{tabular}
\caption{SSM predictions for neutrino fluxes of all relevant $pp$/CNO fusion reactions as observed at Earth. Central values are reported using two different inputs for solar core metallicity: The high-metallicity ({\it high-Z}) GS98 and the more recent low-metallicity ({\it low-Z}) AGSS09 \cite{Robertson:2012ib}}\label{tab:ssm}
\end{center}
\end{table} 

As illustrated by table \ref{tab:ssm}, the discrepancy that arises from the different SSM-input values for GS98 and AGSS09 translates to differing predictions for the emitted neutrino fluxes. Note that some neutrino fluxes ({$^8$B}, {$^7$Be}, CNO) are more affected than others, CNO fluxes featuring the largest differences because the abundances of carbon, nitrogen and oxygen do not only influence opacities but also very directly the amount of catalyst nuclei available for the reaction cycle \cite{Serenelli:2012zw}.
As the discrepancy between {\it high-Z} and {\it low-Z} predictions remains unresolved, a rather pragmatic approach is taken in the following sections: In agreement with the original publications, experimental results will mostly be compared to the {\it high-Z} SSM predictions, pointing out where adaptation of the {\it low-Z} model would affect the interpretation of the individual measurements. As will be discussed in section \ref{sec:statsolcomp}, a combination of precise measurements of several solar neutrino branches (e.g.~{$^8$B} and CNO) has in fact the potential to resolve the controversy over the true metallicity value \cite{Serenelli:2012zw}.

\begin{figure}[ht]
\centering
\includegraphics[width=0.6\textwidth]{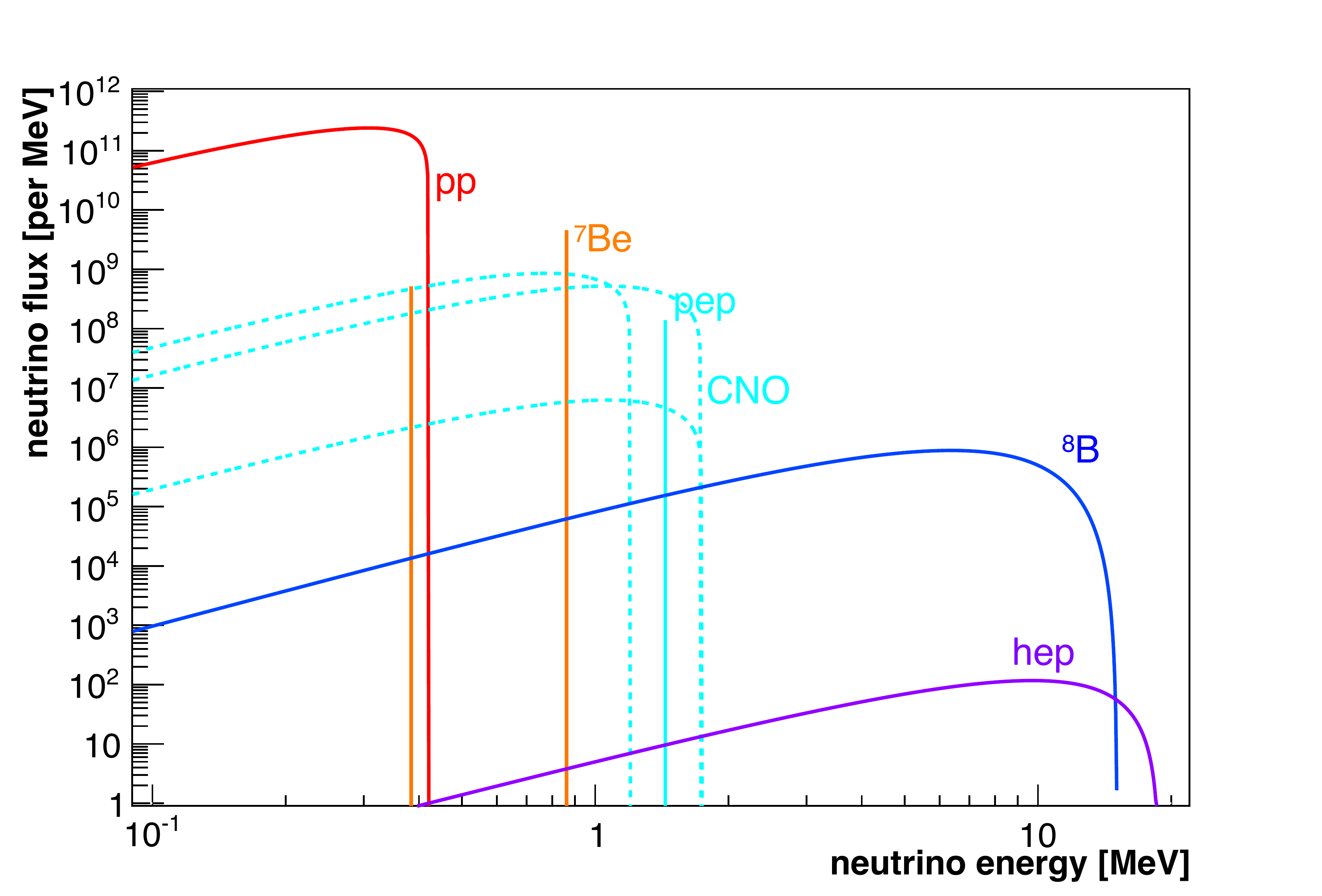}
\caption{The components of the solar neutrino spectrum based on \cite{Bahcall:2004pz}.}
\label{fig:solarnuspectrum}
\end{figure} 

\subsection{Spectral shapes}
\label{sec:solnuspecshape}

The solar neutrino spectrum observed at Earth is a superposition of the partial spectra result- ing from the underlying fusion processes. While only the most basic facts are presented here, the reader is referred to \cite{Raffelt:1996wa} for a more detailed account.
\medskip\\
{\bf Continuous spectra.} Neglecting small corrections based on screening and thermal broadening, the emitted neutrino spectra can be obtained directly from the reaction kinematics: For the three-body final states of $pp$, {$^{13}$N}, {$^{15}$O} and {$^{17}$F} reactions, all super-allowed transitions, the neutrino spectrum is described by
\begin{eqnarray}
\frac{{\rm d}N}{{\rm d}E} \propto (Q+m_e-E)\cdot \sqrt{(Q+m_e-E)^2-m_e^2}\cdot E^2 \cdot F(E),
\end{eqnarray}
where $Q$ is the maximum kinetic energy of the positron/neutrino, while the Fermi function $F(E)$ takes the final state interaction of the positron into account \cite{Raffelt:1996wa}.\\
A notable exception arises for {$^8$B} neutrinos: The decay of the ground state ($J^p = 2^+$) of {$^8$B} to an excited state of $^8$Be (equally $2^+$) is an allowed transition, but the very short half-life of the final state leads to a broad resonance and thus an effective smearing of the allowed decay spectrum. The neutrino spectrum is thus determined from laboratory measurements of the {$^8$B} $\beta^+$-decay. Provided the energies of $\alpha$-particles emitted in the decay of the $^8{\rm Be}^*$ resonance are determined simultaneously, the neutrino spectrum can be obtained by spectral inversion \cite{Raffelt:1996wa, Bahcall:1986qc}.
\medskip\\
{\bf Spectral lines.} On the other hand, both {$^7$Be} and $pep$ neutrinos originate from two-body final states, fixing their energies with respect to the $Q$-values of the reactions. These mono-energetic lines are somewhat broadened due to the thermal motion of the reaction partners, featuring kinetic energies in the keV range. While clearly beyond the resolution of present-day detectors, a measurement of the line width would in principle relate information on the temperature of the solar core \cite{Raffelt:1996wa}.
\medskip\\
{\bf Superimposed spectrum.} Combining the information on the rates of the reactions emitting neutrinos and their corresponding spectral shapes, a sum spectrum can be constructed that reflects the total solar neutrino emission. Figure \ref{fig:solarnuspectrum} displays the SSM prediction for the spectral components \cite{Bahcall:2004pz}, illustrating the contributions of all neutrino-emitting reactions in the $pp$ chain and CNO cycle. While this spectrum holds true for the total neutrino flux disregarding flavors, the electron neutrino spectrum observed by most terrestrial experiments is in addition distorted by neutrino oscillations.

%
%

\section{Solar neutrino oscillations}
\label{sec:solnuosc}

From the very beginning, experiments for the detection of solar neutrinos have been intertwined with the putative occurrence of neutrino flavor oscillations. In the following, the basic concepts of flavor mixing of the three active neutrinos and the resulting oscillations in vacuum and matter. For greater detail and possible extensions of the basic three-flavor picture, the reader is referred to \cite{Bilenky:1998dt, Bilenky:2010zza}.
\medskip\\
{\bf Standard-model neutrinos.} The standard model (SM) of particle physics introduces neutrinos as the partners of the charged leptons in weak isospin doublets \cite{Agashe:2014kda}: In charged-current reactions, a neutrino can be converted to the lepton of the corresponding flavor under emission neutrino flux [per MeV] of an $W^-$ boson or vice versa under $W^+$ emission. Neutrinos therefore come in three flavors associated with the three charged leptons, electron ($\nu_e$ : $e^-$), muon ($\nu_\mu$ : $\mu^-$) and tau ($\nu_\tau$ : $\tau^-$). Moreover, in the basic SM neutrinos are assumed to be massless and to appear always as left-handed particles, while the corresponding antineutrinos ($\bar\nu_e$ etc.) are always right-handed. This reflects the parity-violating (V-A) character of the weak interaction in which neutrinos and antineutrinos are created and destroyed. Flavor family number is conserved at all times.
\medskip\\
{\bf Neutrino mixing.} neutrinos exhibiting finite rest masses create the possibility of a mixing between flavor eigenstates (taking part in weak interaction) and mass eigenstates (propagating in space)\footnote{The direct association of a single flavor state to a single mass state may seem more natural but corresponds only to one specific realization.}. More generally, the three flavor states denoted as $|\nu_\alpha\rangle$ ($\alpha=e,\mu,\tau$) can be described as linear superposition of the three mass eigenstates $|\nu_i\rangle$ ($i=1,2,3$) and vice versa:
\begin{eqnarray}
|\nu_{i}\rangle = U_{\alpha i} |\nu_{\alpha}\rangle;\qquad |\nu_{\alpha}\rangle = U_{\alpha i}^* |\nu_{i}\rangle
\end{eqnarray}
Similarly, the flavor and mass eigenstates of antineutrinos are described by:
\begin{eqnarray}
|\bar\nu_{i}\rangle = U_{\alpha i}^* |\bar\nu_{\alpha}\rangle;\qquad |\bar\nu_{\alpha}\rangle = U_{\alpha i} |\bar\nu_{i}\rangle
\end{eqnarray}
The PMNS mixing matrix\footnote{named in honor of the fathers of the lepton flavor oscillation and mixing theories Pontecorvo, Maki, Nakagawa and Sakata.} $U$ describes the association of flavor to mass eigenstates. To conserve probability, the matrix $U$ is required to be unitary. This concept is in full analogy to the quark sector where the mixing between weak-interaction flavor and strong-interaction mass states is described by the CKM matrix \cite{Bilenky:2010zza}.
\medskip\\
{\bf Parameters of the PMNS matrix.} The neutrino mixing matrix is commonly presented in the parametrization
\begin{eqnarray}
\label{eq:upmns}
U = \left( \begin{array}{ccc}1&0&0\\0&c_{23}&s_{23}\\0&-s_{23}&c_{23}\end{array} \right) \left( \begin{array}{ccc}c_{13}&0&s_{13}e^{-i\delta}\\0&1&0\\-s_{13}e^{-i\delta}&0&c_{13}\end{array} \right) \left( \begin{array}{ccc}c_{12}&s_{12}&0\\-s_{12}&c_{12}&0\\0&0&1\end{array} \right),
\end{eqnarray}
with $c_{ij}=\cos\theta_{ij}$, $s_{ij}=\sin\theta_{ij}$. The mixing is thus represented by a combination of three three-dimensional rotations by the mixing angles $\theta_{12}$, $\theta_{13}$ and $\theta_{23}$. The unitarity requirement allows in addition for a complex phase term $\delta$. The phase introduces the possibility of a CP- asymmetry in the oscillations patterns of neutrinos and antineutrinos. However, this additional degree of freedom plays no role in the interpretation of solar neutrino oscillations \cite{Bilenky:2010zza}.\\
Since the break-through discovery of neutrino flavor transformation by SNO in 2001 (sect.~\ref{sec:b8sno}), by now most of the mixing parameters have been determined in a variety of oscillation experiments. Table \ref{tab:osc_par} lists the current best fit values (and allowed regions) of these parameters based on a recent global analysis of the available oscillation data \cite{Capozzi:2016rtj}. Very different from the quark sector, neutrino mixing angles are large and allow for substantial probabilities of flavor transitions. Most relevant for the solar sector is the so-called solar mixing angle $\theta_{12} \approx 33^\circ$ ("$\nu_e \leftrightarrow \nu_\mu$"), with some second-order corrections from $\theta_{13} \approx 9^\circ$ ("$\nu_e \leftrightarrow \nu_\tau$").
\begin{table}
\begin{center}
\begin{tabular}{lccc}
\hline
Parameter & Central value & $1\sigma$ range & $3\sigma$ range \\
\hline
$\sin^2\theta_{12}$ & 0.297 & $0.281-0.314$ & $0.250-0.354$\\
$\sin^2\theta_{13}$ & 0.0214 & $0.0205-0.0225$ & $0.0185-0.0246$\\
$\sin^2\theta_{23}$ & 0.437 & $0.417-0.470$ & $0.379-0.616$\\
$\delta_{\rm CP}$ & $1.35\pi$ & $(1.13-1.64)\pi$ & $(0-2)\pi$\\
$\Delta m_{21}^2$ [$\times10^{-5}\,{\rm eV}^2$] & 7.37 & $7.21-7.54$ & $6.93-7.97$ \\
$\Delta m_{32}^2$ [$\times10^{-3}\,{\rm eV}^2$] & 2.50 & $2.46-2.54$ & $2.37-2.63$ \\
\hline
\end{tabular}
\caption{Current best fit values for the oscillation parameters provided by a global analysis of all available oscillation data, assuming normal neutrino hierarchy \cite{Capozzi:2016rtj}.}\label{tab:osc_par}
\end{center}
\end{table} 

\subsection{Oscillations in vacuum}
\label{sec:oscvac}

Solar neutrinos are produced in the conversion of protons to neutrons that occur in several weak reactions along the solar $pp$ chain and CNO cycle, e.g. $p + p \to d + e^+ + \nu_e$. Due to energy and flavor conservation, the antilepton-neutrino pair created in these processes are necessarily ($e^+$, $\nu_e$), so all neutrinos emitted in solar fusion are originally of electron flavor.
\medskip\\
{\bf Neutrino production.} Based on neutrino mixing, these $\nu_e$'s are created as a superposition of neutrino mass eigenstates. In case of the low-energy end of the spectrum (as the $pp$-neutrinos), the weak potential created by solar matter can be safely neglected, and the admixture of the three mass eigenstates $|\nu_i\rangle$ to the initial flavor state $|\nu_\alpha(t=0)\rangle$ corresponds to the first row of the PMNS matrix:
\begin{eqnarray}
\label{eq:nueamplitude}
|\nu_\alpha(0)\rangle = |\nu_e\rangle = U_{e1}\cdot|\nu_1\rangle + U_{e2}\cdot|\nu_2\rangle + U_{e3}\cdot|\nu_3\rangle
\end{eqnarray}
{\bf Neutrino propagation.} Different from their production, the propagation of the neutrinos from their point of creation in the solar core is best described in the basis of the associated mass eigenstates. According to Schr\"odinger's equation, the initial mass states $|\nu_i(0)\rangle$ will pick up complex phase factors when developing over time:
\begin{eqnarray}
\label{eq:nuoscphases}
|\nu_\alpha(t)\rangle = \sum_i U_{\alpha i}|\nu_i(t)\rangle = \sum_i U_{\alpha i}|\nu_i(0)\rangle e^{iE_i t}. 
\end{eqnarray}
where $E_i$ denotes the energies of the individual mass eigenstates. Following a commonly used ansatz, we assume that all mass states are produced with identical momenta\footnote{While this standing-wave approximation is for sure inferior to a more correct description based on wave packets \cite{Akhmedov:2010ms}, it returns the correct result for the oscillation formula.}, which will result in slightly different energies $E_i$'s. Based on current upper limits on the minuscule neutrino rest masses ($\sum m(\nu_i) < 0.23$\,eV \cite{Otten:2008zz}), solar neutrinos can be safely assumed as ultra-relativistic ($E \gg m_i$, $E = pc$). Therefore, the energy $E_i$ of the individual mass states can be expanded to $E_i \approx E+\frac{m^2_i}{2E}$. In turn, the complex phase factors of eq.~(\ref{eq:nuoscphases}) will differ after traveling a distance $L\sim t$ (for $c=1$). 
\medskip\\
{\bf Neutrino detection.} Neutrinos can be detected in a selection of weak processes. In case of solar neutrinos, the most important are elastic scattering off electrons and various capture processes on nucleons. As for production, the flavor state $|\nu_\alpha(L)\rangle$ detected at a baseline $L$ is described by the linear superposition of the three time-evolved mass states $|\nu_i(t)\rangle$. As their is no way to differentiate the three mass states in production or detection, the partial amplitudes for all three mass states have to be summed coherently before being squared to obtain the probability for detection in a specific flavor state. However, as the $|\nu_i(L)\rangle$ now feature differing complex phase factors, the quadratic sum for the initial flavor $|\nu_\alpha\rangle$ will no longer add up to unity. Instead the probability to change to any new flavor $|\nu_\beta\rangle$ is given by
\begin{eqnarray}\label{eq:oscformula}
P(\nu_\alpha\to\nu_\beta,L) = |\langle\nu_\beta|\nu_\alpha(L)\rangle|^2=\bigg| \sum_i U^*_{\beta i} U_{\alpha i} e^{i \frac{m_i^2}{2E} L}\bigg|^2=\sum_{i,j} U_{\beta j} U^*_{\alpha j}U^*_{\beta i} U_{\alpha i} e^{i \big(\frac{m_j^2}{2E}-\frac{m_i^2}{2E}\big) L}
\end{eqnarray}
So, the occurrence of non-zero probability for a transition $\nu_\alpha\to\nu_\beta~(\beta\neq\alpha)$  depends on two factors: Firstly, non-zero off-diagonal entries in the mixing matrix $U$, corresponding to $\theta_{ij}\neq 0$ and thus flavor mixing. Secondly, a difference in mass between the eigenstates, i.e.~$m^2_i \neq m^2_j$.\\
The cyclic behavior of the complex phase term in eq.~(\ref{eq:oscformula}) introduces an oscillatory pattern in the probabilities to detect the neutrinos in a given flavor $|\nu_\beta\rangle$. The oscillation frequency depends on the difference of the squares of the masses of the involved eigenstates, $\Delta m_{ji}^2 = m_j^2-m_i^2$, as well as the neutrino energy $E$. Based on these quantities, an oscillation length $\ell_{ij}=\pi\hbar c E_\nu/\Delta m^2_{ji}$ can be defined. On the other hand, the oscillation amplitude depends on the product of mixing matrix elements and thus the size of the mixing angles $\theta_{ij}$.
\medskip\\
{\bf Solar neutrino oscillations.} Solar fusion produces only $\nu_e$, and solar neutrino detectors are mostly sensitive to neutrinos of electron flavor. Therefore, the most interesting quantity for the description of these experiments is the survival probability of $\nu_e$'s, $P(\nu_e\to\nu_e)$ or short Pee. Using the standard parameterization of the mixing matrix (8), Pee at large distances from the production point ($L\gg \ell_{13} = 4\pi\hbar c E_\nu/\Delta m^2_{13}$) can be well approximated by
\begin{eqnarray}
P_{ee}(L,E) &  = & \cos^4\theta_{13}\bigg[ 1-\sin^2(2\theta_{12})\sin^2\bigg(\frac{\Delta m^2_{21}L}{4E_\nu}\bigg)\bigg] + \sin^4\theta_{13}\\
& \approx  &1-\sin^2(2\theta_{12})\sin^2\bigg(\frac{\Delta m^2_{21}L}{4E_\nu}\bigg).
\end{eqnarray}
The second approximation (13) is often applied as $\theta_{13}\ll\theta_{12}$. This effectively allows to describe solar neutrino oscillations in a two-flavor picture, greatly reducing the complexity of the full three-flavor solution. While the term $\sin^2(2\theta_{12})$ describes the oscillation amplitude, the second $\sin^2$-term containing the quotient of distance and neutrino energy, $L/E_\nu$, describes the oscillation phase. As listed in Table~\ref{tab:osc_par}, the relevant $\Delta m^2_{21}\approx 8\cdot 10^{-5}\,{\rm eV}^2$ is small, leading to a typical oscillation length of $\ell_{\rm osc}\approx30$\,km for a neutrino at an energy of 1\,MeV.
\medskip\\
{\bf Mean survival probability.} Mean survival probability. On the scale of $\ell_{\rm osc}$, the propagation distances $L$ from the point of production in the Sun to a neutrino detector localized on Earth are vastly different for individual neutrinos. As illustrated in figure \ref{fig:raddists}, fusion zones extend as far out as $0.25R_\odot$. Moreover, the energy resolution for the neutrino energy $E$ of present-day experiments is limited\footnote{In addition, it can be shown that in the case of solar neutrinos, the superposition of the three mass eigenstates will loose coherence long before reaching Earth, leading to a freeze-out of the oscillatory behavior \cite{Akhmedov:2010ms}}. Therefore, neutrinos will arrive with largely different phase factors, leading to an averaging-out effect in the oscillation probabilities. The effective $\nu_e$ survival probability can be written to
\begin{eqnarray}
\label{eq:peevac}
\langle P_{ee,\rm vac} \rangle = \langle P_{ee}(L,E) \rangle = 1-\frac{1}{2}\sin^2(2\theta_{12}).
\end{eqnarray}
Equation (\ref{eq:peevac}) provides a fairly good approximation of the survival probability of neutrinos in
 the low-energy end of the solar spectrum \cite{Bilenky:2010zza}.

\subsection{Oscillations in matter}
\label{sec:oscmat}

\noindent{\bf Matter potential.} While the assumption of oscillations in vacuum is a good approximation for the low-energy spectrum, the situation changes at higher energies. Starting from $\sim$1\,MeV, the effect of the weak scattering of neutrinos off the solar matter $-$ and especially the electrons of the plasma $-$ can no longer be neglected \cite{Bilenky:2010zza, Rosen:1986jy}. Given the large matter densities of the solar core, $\varrho_{\odot,c}\approx 150\,{\rm g/cm}^3$, the neutrinos begin to 'feel' the effect of a weak matter potential. In turn, this leads to a sizable effect on the propagation of the neutrinos in terms of elastic forward scattering: Like electrons in a solid state body, neutrinos in matter pick up an effective mass.\\
The strength of the potential and thus the magnitude of the mass shift depends on the neutrino flavor: All flavors interact with nucleons and electrons via neutral-current (NC) reactions. However, as the cross-sections are flavor-independent, there is no net effect on oscillations. A difference between flavors arises only when scattering off electrons: Due to the presence of an additional charged-current (CC) channel for electron neutrinos, the corresponding cross-section is about five times larger than for muon and tau neutrinos.
\medskip\\
{\bf Effect on oscillations.} With regard to oscillation, shifts in the absolute mass scale of the participating neutrinos and thus the do not matter. However, the relative increase in the $\nu_e$ mass due to CC scattering on electrons in comparison to $\nu_{\mu,\tau}$ changes the effective mass squared differences between the underlying mass eigenstates. What is more, it introduces off-diagonal terms to the neutrino mixing matrix.
This is best understood when regarding the time evolution of the mass eigenstates using Schr\"odinger's equation in matrix notation. In the vacuum case,
\begin{eqnarray}
\label{eq:oscmatrix}
i\frac{d}{dt}\left(\begin{array}{c}\nu_1 \\ \nu_2 \\Ê\nu_3\end{array}\right) = \underbrace{\frac{1}{2E}\left(\begin{array}{ccc}{m_1}^2 & 0 & 0 \\ 0 & {m_2}^2 & 0 \\Ê0 & 0 & {m_3}^2\end{array}\right)}_{H^{\rm vac}}\left(\begin{array}{c}\nu_1 \\ \nu_2 \\Ê\nu_3\end{array}\right)
\end{eqnarray}
where $H^{\rm vac}$ is the Hamiltonian in mass basis. As the $|\nu_i\rangle$ are eigenstates of the mass basis, $H_i^{\rm vac}$ features only diagonal entries. 

However, for neutrinos propagating in matter, a further matrix term $V$ appears in equation (\ref{eq:oscmatrix}) that represents the weak potential created by the forward-scattering off electrons and nucleons. only the additional potential $V_{\rm CC}$ for $\nu_e$'s has to be regarded because it will induce a divergence in the effective mass squared differences. As the admixture of the $|\nu_e\rangle$ flavor to the respective mass eigenstates $|\nu_i\rangle$ depends on the mixing matrix $U$, the additional term will be of shape $U^\dagger V U$: 
\begin{eqnarray}
\label{eq:mattermatrix}
H^m = H^{\rm vac}+
U^\dagger \left( \begin{array}{ccc}V_{\rm CC}&0&0\\0&0&0\\0&0&0\end{array} \right)U
\end{eqnarray}
where $V_{\rm CC}=2\sqrt{2}G_F N_e$ corresponds to the weak potential only applying to $\nu_e$, featuring the Fermi constant $G_F$ and the electron density of matter, $N_e$. $V_{\rm CC} $ will only play a role if the product $V_{\rm CC}\cdot E$ is comparable or greater in size than the mass squared differences $\Delta m^2$. This condition is in fact fulfilled in the core regions of Sun, where  $N_e$ reaches $\sim10^{26}\,{\rm cm}^{-3}$.
However, a visible effect only comes to bear for higher neutrino energies $E\gtrsim5$\,MeV and the smallest of the three mass splittings $\Delta m^2_{21}$. Again, the oscillation probabilities can be quite comfortably described in a two-flavor picture\footnote{In this case, the mixing matrix is reduced to $U_{2\times2}=\left(\begin{array}{cc} \cos\theta_{12} & \sin\theta_{12} \\ -\sin\theta_{12} & \cos\theta_{12}\end{array}\right)$.}, and we obtain an effective Hamiltonian in matter
\begin{eqnarray}
H^m_{2\times 2} = 
\frac{1}{2E}\left(\begin{array}{cc} m_1^2 + 2\cos^2\theta_{12}V_{\rm CC}E & 2\cos\theta_{12}\sin\theta_{12}V_{\rm CC}E\\  -2\cos\theta_{12}\sin\theta_{12}V_{\rm CC}E &m_2^2 -2\sin^2\theta_{12}V_{\rm CC}E  \end{array}\right) 
\end{eqnarray}
Therefore, the appearance of $V_{\rm CC}$ affects neutrino oscillations in several ways: Firstly, the additional on-diagonal terms will alter the effective mass squared difference and thereby change the oscillation lengths. Secondly, the appearance of off-diagonal terms opens up at the possibility for direct transition between mass eigenstates that will enhance the vacuum flavor mixing. Finally, the vacuum mass eigenstates $|\nu_i\rangle$ are no longer good eigenstates of $H^m$. Instead, neutrino oscillations are best described in terms of the effective matter eigenstates $|\nu_i^m\rangle$, with
\begin{eqnarray}
(m^2_{1,2})^m = \frac{1}{2}\left[m_1^2+m_2^2+V_{\rm CC}E\mp\sqrt{(V_{\rm CC}E-\Delta m^2_{21}\cos2\theta_{12})^2+(\Delta m^2_{21}\sin2\theta_{12})^2}\right],
\end{eqnarray}
the effective mass splitting
\begin{eqnarray}
(\Delta m^2_{21})^{m} = (m^2_2)^m-(m^2_1)^m = \Delta m^2_{21}\cdot \sqrt{(V_{\rm CC}E/\Delta m^2_{21}-\cos2\theta_{12})^2+\sin^22\theta_{12}},
\end{eqnarray}
and the effective mixing angle $\theta^m_{12}$ given by
\begin{eqnarray}
\sin2\theta^m_{12} =\frac{\sin2\theta_{12}}{\sqrt{(V_{\rm CC}E/\Delta m^2_{21}-\cos2\theta_{12})^2+\sin^22\theta_{12}}}
\end{eqnarray}
For $V_{\rm CC}E/\Delta m^2_{21}=\cos2\theta_{12}$, $\sin2\theta^m_{12}$ will become unity, corresponding to maximal mixing between electron and muon neutrino flavors. This resonance effect was first described by Mikheyev, Smirnov and Wolfenstein \cite{Mikheev:1986wj,PhysRevD.17.2369}. For larger densities or energies, $\theta_{12}^m$ can become as large as $90^\circ$, corresponding to a situation in which the flavor eigenstate $|\nu_e\rangle$ is fully associated with the mass eigenstate $|\nu_2\rangle^m$.
\medskip\\
{\bf Solar matter effect.} This condition is in fact realized for B-8 neutrinos of $E\gtrsim10$\,MeV exposed to the matter density of the solar core, which coincides with their emission region (fig.~\ref{fig:raddists}). The $\nu_e$'s emitted in {$^8$B} decays are thus not created as a superposition of mass eigenstates, but are fully associated to the effective mass state $|\nu_2\rangle^m$. When leaving the sun, the matter potential will gradually decrease, and thus also the effective mass of $|\nu_2\rangle^m$. If the transition through the resonance region is sufficiently smooth\footnote{i.e. the adiabacity condition is fulfilled for which the neutrino oscillation length is short compared to the spatial width of the resonance region \cite{Smirnov:2003da}.}, the neutrino will remain in the heavier mass eigenstate and no further flavor conversion takes place. Upon reaching the surface of the Sun, the neutrino will have assumed the vacuum mass eigenstate $|\nu_2\rangle$. However, differently from the solar interior, the vacuum mass state is mostly associated with the flavor state $|\nu_\mu\rangle$ by $U_{22}=\cos\theta_{12}$, while the probability to detect such a neutrino as $|\nu_e\rangle$ is only 
\begin{eqnarray}\label{eq:peemat}
P_{ee}=|U_{12}|^2 = \sin^2\theta_{12}. 
\end{eqnarray}
This leads to the surprising finding that despite the averaging effects discussed in section \ref{sec:oscvac}, the probability to detect these neutrinos in a $\nu_e$ flavor state is suppressed to a value below 50\,\%. The suppression effect would be even more prominent in case of a small vacuum mixing angle $\theta_{12}$. This so-called MSW resonance effect was predicted by Mikheyev and Smirnov well before the discovery of solar neutrino oscillations \cite{Rosen:1986jy, Mikheev:1986wj, PhysRevD.17.2369}.
\medskip\\
{\bf Earth matter effect.} An additional complication is added if solar neutrinos cross terrestrial matter prior to detection (e.g.~\cite{Maltoni:2015kca}): The weak potential generated by Earth matter is sufficient to influence oscillation and thus $\nu_e$ survival probabilities. Depending on neutrino energy and flavor composition this can either result in a further suppression of $P_{ee}$ or in a regeneration of the $\nu_e$ content. The size of the observed effect depends as well on the length of the trajectory through the Earth's matter, that ranges from 1$-$2 kilometers to the full Earth diameter, depending on the rock shielding above a subterrestrial detector and the zenith position of the Sun. In consequence, a day-night asymmetry may be introduced in the neutrino event rates and energy spectra observed of solar neutrino experiments. The reader is referred to \cite{Maltoni:2015kca} for a more complete discussion of the parameter dependences of the $\nu_e$ regeneration factor. 

\subsection{The MSW-LMA solution and solar neutrino survival probabilities}
\label{sec:pee}

\begin{figure}[ht]
\centering
\includegraphics[width=0.6\textwidth]{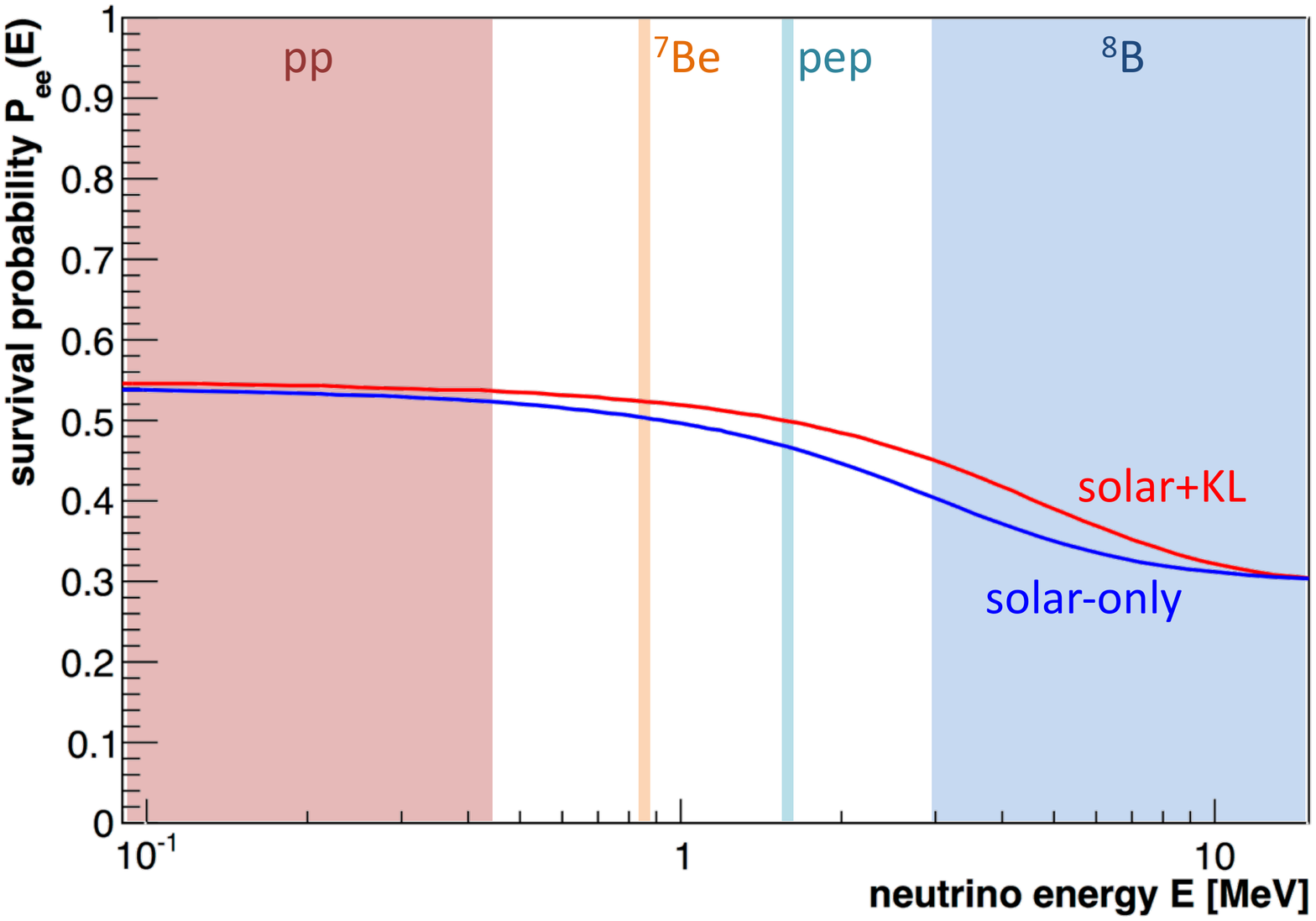}
\caption{The electron neutrino survival probability $P_{ee}$ as a function of solar neutrino energy. While the blue curve ({\it solar-only}) is based on solar neutrino data only, the deviating red curve ({\it solar+KL})  includes as well the KamLAND antineutrino data \cite{Abe:2016nxk}. The shaded regions indicate the corresponding spectra of spectral subbranches.}
\label{fig:solarpee}
\end{figure}

Based on the data collected by solar neutrino detectors (sect.~\ref{sec:solnumeas}) and by a variety of other oscillation experiments, the oscillation scenario firmly established today is the so-called MSW-LMA solution, i.e.~oscillations governed by a large solar mixing angle (LMA) and subject to matter effects (MSW) \cite{Smirnov:2003da}. Given the wash-out of oscillation probabilities, one would generally expect a flat $\nu_e\to\nu_e$ survival probability. However, the presence of matter effects splits the solar neutrino spectrum in three energy regimes as is exemplified by figure \ref{fig:solarpee}.

\noindent{\bf Low energy.} For the region below 1\,MeV, so mostly the $pp$ and {$^7$Be} neutrinos, neutrino energy is too low to create a sizable matter effect even for the maximum density region of the solar core, resulting in pure vacuum oscillations. Because of the vast length and differences of the baselines, the oscillation pattern observed on Earth is washed out, and the $\nu_e$ survival probability corresponds to $P_{ee}^{\rm vac}=1-\frac{1}{2}\sin^2(2\theta)$ given by eq.~(\ref{eq:peevac}).
\medskip\\
{\bf High energy.} On the other end of the spectrum, for energies greater 10\,MeV, matter effects prevail and suppress vacuum oscillations. Instead, electron neutrinos are born in the matter mass state $\nu_2^m$, that is adiabatically converted to $\nu_2$ while traveling through the slowly decreasing matter density of the solar envelopes. The probability for $\nu_e$ detection on Earth is thus given by $P_m =  |U_{12}|^2 = \sin^2\theta_{12}$ [eq.~(\ref{eq:peemat})].
\medskip\\
{\bf Transition region.} There arises an intermediate energy range in which matter effects are present but not fully expressed. Therefore, the survival probability will assume an intermediate value $P_{ee}^m < P_{ee} (E) < P_{ee}^{\rm vac}$ that depends on the exact neutrino energy, and in addition on the exact value of $V_{\rm CC}$ at the position of neutrino emission in the solar core. The MSW-LMA solution predicts a rather gentle transition from one value to the other. However, as will be lined out in section \ref{sec:pee}, the course of $P_{ee}(E)$ in this region is especially sensitive to the occurrence of non-standard interactions, the presence of very light sterile neutrino states or other non-standard effects that would all alter the characteristics of the transition \cite{Friedland:2004pp, Minakata:2010be, deHolanda:2010am}.
\medskip\\
{\bf Day-night asymmetry.} The MSW-LMA solution predicts a small but measurable effect of Earth matter on the high-energy part of the solar neutrino spectrum: Above a neutrino energy of 3\,MeV, the electron flavor will be partially re-generated for neutrinos crossing the Earth's matter potential, leading to an increase of interaction rates measured during night time. The regenerative effect is on the order of 3\,\% and depends on the exact value of the mass squared difference $\Delta m_{21}^2$ \cite{Maltoni:2015kca}.

%
%

\section{Solar neutrino detectors}
\label{sec:detectors}

As the Sun provides the highest natural neutrino flux on Earth, first efforts towards the detection began early on. The pioneering experiment was the Chlorine experiment constructed and operated by Raymond Davis in the Homestake gold mine in South Dakota \cite{Bahcall:1981br}. Neutrino detection relied on a radiochemical method, the signature of interactions provided by the conversion of a handful of the target chlorine-37 atoms into radioactive argon. Despite of the technical difficulties involved, the method proved viable and allowed for a fairly accurate (if low) measurement of solar {$^7$Be} and {$^8$B} neutrinos (sect.~\ref{sec:radchem}) \cite{Davis:1968cp}. As a result, Ray Davis was co-awarded the Nobel Prize in Physics 2002 for "pioneering contributions to astrophysics, in particular for the detection of cosmic neutrinos".

The deficit found in the measured compared to the expected solar neutrino rate triggered the solar neutrino problem \cite{Altmann:2001eu, Haxton:1995hv} and in its wake a series of further experiments trying to reject or to corroborate these findings: On the one hand, the radiochemical method was further refined by choosing gallium as a target, allowing for a lower detection threshold that was now sensitive to the complete solar neutrino spectrum including pp-neutrinos (sect.~\ref{sec:gallium}). Moreover, first attempts at a spectroscopic measurement of the solar neutrino flux were made, with the aim to disentangle the underlying solar neutrino spectrum from the energy-dependent effects of neutrino oscillations or other non-standard processes.

After a short overview of the radiochemical technique in section \ref{sec:radchem}, this section will concentrate on a description of the two main spectroscopic techniques: Huge water Cherenkov detectors (sect.~\ref{sec:wcd}) that sample the {$^8$B} neutrino spectrum with high statistics above an energy of (4-5)\,MeV, and smaller-sized liquid scintillator detectors (sect.~\ref{sec:lsd}) that have the potential to cover the whole spectrum down to the energies of $pp$ neutrinos. The discussion will include the basic principles of detection as well as the respective advantages and challenges of these techniques.

\subsection{Radiochemical experiments}
\label{sec:radchem}

The first generation of detectors to be employed for the measurement of solar neutrinos relied on an experimental technique rather different from current Water Cherenkov and scintillation detectors: Based on a weak charged-current reaction on a specifically chosen target isotope, detection relied not on the final state lepton but on the re-decay of the produced radioactive daughter nucleus. The most famous example of such an inverse beta decay is the reaction
\begin{eqnarray}\label{eq:cl}
{^{37}{\rm Cl}}+\nu_e \to {^{37}{\rm Ar}}+e^-
\end{eqnarray}

\begin{table}
\begin{center}
\begin{tabular}{lccccccc}
\hline 
Reaction & $pp$ & {$^7$Be} & $pep$ & CNO & {$^8$B} & Integral & Unit\\ 
\hline
${^{37}{\rm Cl}}(\nu_e,e^-){^{37}{\rm Ar}}$ & 0 & 1.15 & 0.2 & 0.5 & 5.9 & $7.7^{+1.2}_{-1.0}$ & SNU \\
${^{71}{\rm Ga}}(\nu_e,e^-){^{37}{\rm Ge}}$ & 69.6 & 34.4 & 2.8 & 9.7 & 12.4 & $129^{+8}_{-9}$ & SNU \\
\hline
\end{tabular}
\caption{SSM predictions for the event rates (in SNU) expected in the chlorine and gallium
 radiochemical experiments \cite{Altmann:2001eu}.}\label{tab:radchem}
\end{center}
\end{table} 

\noindent employed in the Homestake experiment (sect.~\ref{sec:homestake}). The minimum energy that the neutrino has to provide in order to induce the reaction corresponds to the mass difference between the reaction partners of the initial and final states, $Q = [m({^{37}{\rm Ar}})+me]-m({^{37}{\rm Cl}}) = 814$\,keV (with $m$ the nuclear masses). Suitable isotopes are selected for low positive $Q$-values, providing a stable target but a low energy threshold for neutrino detection. Contrariwise, the final-state isotope is radioactive and will eventually decay back to the initial isotope. Further considerations are the availability of large amounts of the target isotope and the possibility for a chemical separation of target and daughter elements (see below).
\medskip\\
{\bf Experimental technique.} The setups of the radiochemical experiments did not permit to observe the electrons emitted in the detection reaction but were targeted on the recovery of the produced radioisotope, i.e.~${^{37}{\rm Ar}}$ in the Homestake experiment. Upon start of a specific measurement run, the daughter isotope was slowly enriched due to neutrino-induced conversion. However, there is a maximum concentration of the daughter isotope to be reached as the re-decay to the initial isotope will counteract the neutrino-induced production. When approaching the equilibrium of the two processes, the product isotope is chemically extracted from the neutrino target, the exact technique applied depending on the given experiment. Note that even for detectors featuring tens to hundreds of tons of the target isotope, the number of daughter nuclei produced is vanishingly small. Per single extraction, a mere handful of re-decays has to be counted with high efficiency. The daughter nuclei are usually transferred to a gaseous state and mixed with the counting gas of a miniature proportional counter to detect the isotope's $\beta^+$-decay or electron capture (EC).
\medskip\\
{\bf Background.} Due to the overall low rates involved, the avoidance of additional count rates due to external backgrounds is of paramount importance. This is especially true for the target material, as the nuclear conversions serving as signature of solar neutrino reactions can as well be induced by cosmic radiation. To provide sufficient shielding from the extremely penetrating muon component, the experiments were housed in underground caverns at more than 1000\,m depth, e.g.~the Homestake gold mine in the US or the LNGS in Italy.
\medskip\\
{\bf Integral measurement.} A common characteristic of these measurements is their integrating character: The neutrino events are not detected on an individual event-by event level. Instead, the amount of converted isotopes depends (apart from the measurement time) on the integrated sum spectrum of neutrinos above the detection threshold, convoluted with the energy-dependent cross-sections. As a consequence, only a spectrally averaged value of the survival probability $P_{ee}$ can be determined.

\subsubsection{The Chlorine Experiment at Homestake}
\label{sec:homestake}

The pioneering effort in this field was the Chlorine Experiment that took solar neutrino data from the early 1970's. A neutrino detector based on the reaction (\ref{eq:cl}) on {$^{37}$Cl} was first proposed by Bruno Pontecorvo in 1946 \cite{Dore:2009bq}. In the early 1950's, Raymond Davis began to work on this technique. After first tries to detect antineutrinos at a nuclear research reactor in Brookhaven \cite{Davis:1955bi}, Davis turned to the Sun as a strong natural neutrino source. 

Starting from 1965, the detector was put up in the Homestake mine in Lead, South Dakota \cite{Davis:1968cp}. The experimental hall was located at 1478\,m depth (4400 meters of water equivalent, mwe), providing sufficient shielding from cosmic rays to reduce the cosmogenic conversion of  {$^{37}$Cl}  to  {$^{37}$Ar}  to a level well below the expected signal rate. A stainless steel tank held a target volume of 378,000 liters (615 tons) of prechlorethylene (C$_2$Cl$_4$). Taking into account the natural abundance of  {$^{37}$Cl}  of 24.2\,\%, this corresponds to $2.1\cdot10^{30}$ target nuclei contained by the tank. The energy threshold for the reaction to occur is 814\,keV, limiting detection to solar {$^7$Be}, CNO and {$^8$B} neutrinos. As displayed in table 3, the detector is in fact mostly sensitive to the {$^8$B} neutrino flux because of the steep energy dependence of the cross-section \cite{Davis:1968cp}.

The typical length of a solar run during which  {$^{37}$Ar}  was allowed to accumulate was $\sim$50 days. Thereupon, the  {$^{37}$Ar}  produced was extracted by purging the perchlorethylene target with gaseous helium. The argon was flushed out into an attached gas-handling system where the gas atoms were retrieved by means of a charcoal trap. Circulating the helium for 22 hours through the tank, the resulting collection efficiency reached 95\,\%. The re-decay of  {$^{37}$Ar}  to  {$^{37}$Cl}  ($T_{1/2} = 35$\,d) via electron capture and emission of an Auger electron was counted in miniature low-background proportional counters \cite{Davis:1968cp}.

Operation of the experiment started in 1967. According to Davis, it was clear from the first run that the observed neutrino rate was considerably lower than the SSM predicted \cite{Bahcall:1981br}.
Over the next years, it became apparent that the measured rate was only about 1/3 of the SSM prediction by Bahcall (e.g.~\cite{Bahcall:1968hc}): After 108 solar runs from 1970 to 1994, the final result accounted to an average rate of $(2.56±0.16±0.16)$\,SNU\footnote{NU=solar neutrino unit, i.e.~1 reaction per second and $10^{36}$ target atoms, introduced and used mostly for the interpretation of radiochemical experiments} instead of the predicted $(7.7^{+1.2}_{-1.0})$\,SNU based on the SSM calculations (tab.~\ref{tab:radchem}) \cite{Cleveland:1998nv}. During this time, Davis confirmed in a row of experimental tests that the collection efficiency for the  {$^{37}$Ar}  produced inside the tank was indeed close to 100\,\% and the detection efficiency of the proportional counters correctly estimated \cite{Cleveland:1998nv}. The verification of the applicability of the experimental method dispersed the widely felt disbelief that had first greeted the unforeseen result. In the follow-up, the attention shifted away from detector effects towards plausible flaws in the prediction of the solar {$^8$B} neutrino rate (e.g.~\cite{Bahcall:1998wm}).

\subsubsection{Gallium experiments}
\label{sec:gallium}

In the aftermath of the Homestake result, a second generation of radiochemical experiments was designed to try and disentangle whether solar or neutrino physics were at the basis of the Solar Neutrino Problem. The main difference lay with the chosen target isotope,  {$^{71}$Ga} . As  {$^{37}$Cl} , it offers the possibility to detect electron neutrinos by the inverse beta decay
\begin{eqnarray}
{^{71}{\rm Ga}}+\nu_e \to {^{71}{\rm Ge}}+e^-
\end{eqnarray}
However, the detection threshold of 233\,keV is considerably lower than for  {$^{37}$Cl} , and most importantly below the end-point of the $pp$-neutrino spectrum at 422\,keV \cite{Altmann:2001eu, Bahcall:1997eg}. As displayed in table 3, the event rate is thus expected to be dominated by the $pp$ neutrino signal. Due to its fundamental position in the reaction network of the $pp$-chain the $pp$ flux is closely linked to the well-known solar electromagnetic luminosity (sect.~\ref{sec:sollum}). Including this constraint, the SSM prediction for the expected pp-neutrino flux thus features only minimal uncertainties ($\pm0.6$\,\%). A comparison of measured and predicted rates in a gallium experiment was therefore regarded as a substantially more reliable test from point of the SSM than the earlier chlorine measurement that was primarily sensitive to the much less certain {$^8$B} neutrino flux.

In the early 1990s, two experiments based on the new technique were launched: The GALLium EXperiment (GALLEX, later GNO) \cite{Hampel:1998xg, Altmann:2005ix} at the LNGS and the Soviet American Gallium Experiment (SAGE) at the Baksan underground laboratory \cite{Abdurashitov:1999zd}. While the basic measurement principle was still close to the Homestake experiment, the two new detectors deferred in many technical details. For brevity, we regard here only the GALLEX experiment to exemplify the technique.

GALLEX was operated from 1991 to 1997 at the LNGS underground lab at a depth of 1.4\,km (3,500\,mwe) \cite{Kaether:2010ag}. The detector tank held a liquid target of 101 tons of aqueous GaCl$_3$ solution acidified in HCl. The natural abundance of the target isotope is 39.9\,\%, amounting to $\sim10^{29}$ target nuclei.

Due to the relatively short half-life of  {$^{71}$Ge}  of 11.2\,d, the standing time between two extractions was only 30 days, in which time on the order of $5-10$  {$^{71}$Ga}  atoms were converted to  {$^{71}$Ge} . The emerging germanium atoms formed a new compound, GeCl$_{4}$, that was highly volatile in the acidic environment of the target. After removal from the target by nitrogen purging, the molecule was chemically transformed to the gas GeH$_4$. As in the Homestake experiment, this gas was inserted into a low-background proportional counter to measure the EC re-decay to  {$^{71}$Ga} .

As in the Homestake experiment, the GALLEX measurements showed a significant rate deficit compared to SSM prediction. Combining the data sets of GALLEX and its successor GNO, the measured rate can be determined to $74\pm7$\,SNU, more than 40\,\% short of the expectation of $129^{+8}_{-6}$\,SNU (tab.~\ref{tab:radchem}) \cite{Altmann:2005ix}. This result was in full agreement with the findings of the SAGE experiment (still operational at the time of writing) \cite{Abdurashitov:1999zd}. As in Homestake, the experiments were subject to a long row of calibration measurements to exclude a systematic bias of the results. Maybe most remarkably, this included several calibrations with high-intensity neutrino sources, that relied on the close positioning of 51Cr and  {$^{37}$Ar}  sources in the immediate proximity of the detectors (e.g.~\cite{Kaether:2010ag}). The emitted neutrino flux was determined based on the source activities (overall on the order of 1\,MCi) and compared to the measurements of additional  {$^{71}$Ga}  conversions induced in the target, reaching an agreement between measurement and expectation on the level of $(86\pm5)\,\%$. While this constitutes a $2.8\sigma$ deficit that is today discussed as possible evidence for disappearance oscillations from electron to eV-mass sterile neutrinos \cite{Giunti:2010zu}, it is by no means sufficient to explain the much larger deficit observed in solar neutrino data.

By linking the measured neutrino rate to the best understood part of the $pp$-chain and the SSM, the gallium results virtually excluded the possibility that the observed deficit was based either on flaws in the detection or on inadequate modelling of the solar interior \cite{Bahcall:1994xv}. In the following, non-standard properties of the neutrinos shifted now into focus. However, the integrated rate measurement of radiochemical experiments gave only very limited information on the energies of the detected neutrinos and were only sensitive to electron flavor, providing no handle to investigate the mechanism underlying the apparent neutrino deficit. Therefore, the next generation of experiments incorporated either the possibility of a spectral measurement or sensitivity for other neutrino flavors.

\subsection{Water-Cherenkov detectors}
\label{sec:wcd}

The central characteristic of the radiochemical experiments is the fact that they perform a time and energy-integrated measurement of the neutrino flux: The measured time-averaged rate only reflects the convolution of cross-sections and solar neutrino spectra above energy threshold. However, from the point of view of both solar and neutrino physics, an energy-resolved measurement is much more attractive: Ideally, it allows an unambiguous assortment of the detected neutrino events to the individual subbranches of the spectrum (sect.~\ref{sec:solnuspecshape}). Moreover, the energy dependence of the survival probability $P_{ee}(E)$ (sect.~\ref{sec:pee}) can be accessed directly via spectroscopy \cite{Altmann:2001eu}.

The advent of time and energy resolved or so-called real-time detection of solar neutrinos was marked by the arrival of the first kiloton-scale Water-Cherenkov detectors, most prominently the KamiokaNDE experiment starting operation at the Kamioka mine in the Japanese Alps in 1983 \cite{Hirata:1989zj}. The primary intention of this first generation of experiments was not the search for neutrinos but for nucleon decay. In particular, the proton decay $p \to \pi^0 e^+ \to \gamma\gamma e^+$ provides a distinct event signature in the 1\,GeV range that can be easily identified in Cherenkov detectors. Solar neutrinos on the other hand can only be detected via the elastic scattering on electrons, providing a single recoil electron with energies not greater than $\sim$15\,MeV. So while only the high-energy end of the solar spectrum was above detection threshold, Kamiokande was the first experiment to demonstrate that the observed neutrino signal indeed originated from the Sun, relating the directional distribution of recoil electrons to its position in the sky (sect.~\ref{sec:wcdtech}).

The experimental focus shifted thereafter, mostly due to an anomaly in atmospheric neutrino fluxes hinting towards muon neutrino disappearance by neutrino oscillations \cite{Hirata:1992ku}. In relation to this, solar neutrino detection was getting more into focus, so that the second generation of large water Cherenkov detectors was much better adapted for detection in the low-energy range. In the following, these detectors proved incredibly successful: the 25-kt Super-Kamiokande (SK) detector allowed for the discovery of atmospheric neutrino oscillations \cite{Fukuda:1998mi} and performed amongst many other important results precise measurements of the flux, shape and time variation of the solar {$^8$B} neutrino spectrum \cite{Abe:2016nxk}. On the other hand, the SNO experiment employing a 1-kt target of heavy water was able to provide an unambiguous confirmation that flavor oscillations from $\nu_e$ to $\nu_{\mu,\tau}$ are the cause for the solar neutrino deficit \cite{Ahmad:2002jz}.

\subsubsection{Detection technique}
\label{sec:wcdtech}

{\bf Neutrino electron scattering.} The detection channel common to all water Cherenkov detectors is via neutrinos scattering elastically off electrons in the target material \cite{Raffelt:1996wa}. In accordance with the characteristic interaction strength of the weak force, the total cross section for $\nu_ee$-scattering is only $\sigma_{\nu_ee}(E) \approx 9.0\cdot10^{-44}(E/10 MeV)\,{\rm cm}^{2(}$\footnote{In the range below 1\,MeV, true cross-sections are even lower when suitable radiative correction terms are taken into account \cite{Bahcall:1995mm}}$^)$ \cite{Bahcall:1995mm}. Therefore, very large volumes on the scale of 1,000 tons (equalling $43.3\cdot 10^{32}$ electrons) are required to obtain neutrino rates of  at least a few {$^8$B} neutrino events per day. Moreover, the cross sections of elastic muon and tau neutrino scattering are significantly lower, $\sigma_{\nu_\mu e}(E) = \sigma_{\nu_\tau e}(E) \approx 1.6 \cdot 10^{-44}\,(E/10 MeV)\,{\rm cm}^2 \approx 0.2\sigma_{\nu_ee}$ \cite{Bahcall:1995mm}. Therefore, flavor oscillations $\nu_e \to \nu_{\mu,\tau}$ reduce the detected event rate with respect to the SSM expectation.

\begin{figure}[ht]
\centering
\includegraphics[width=0.6\textwidth]{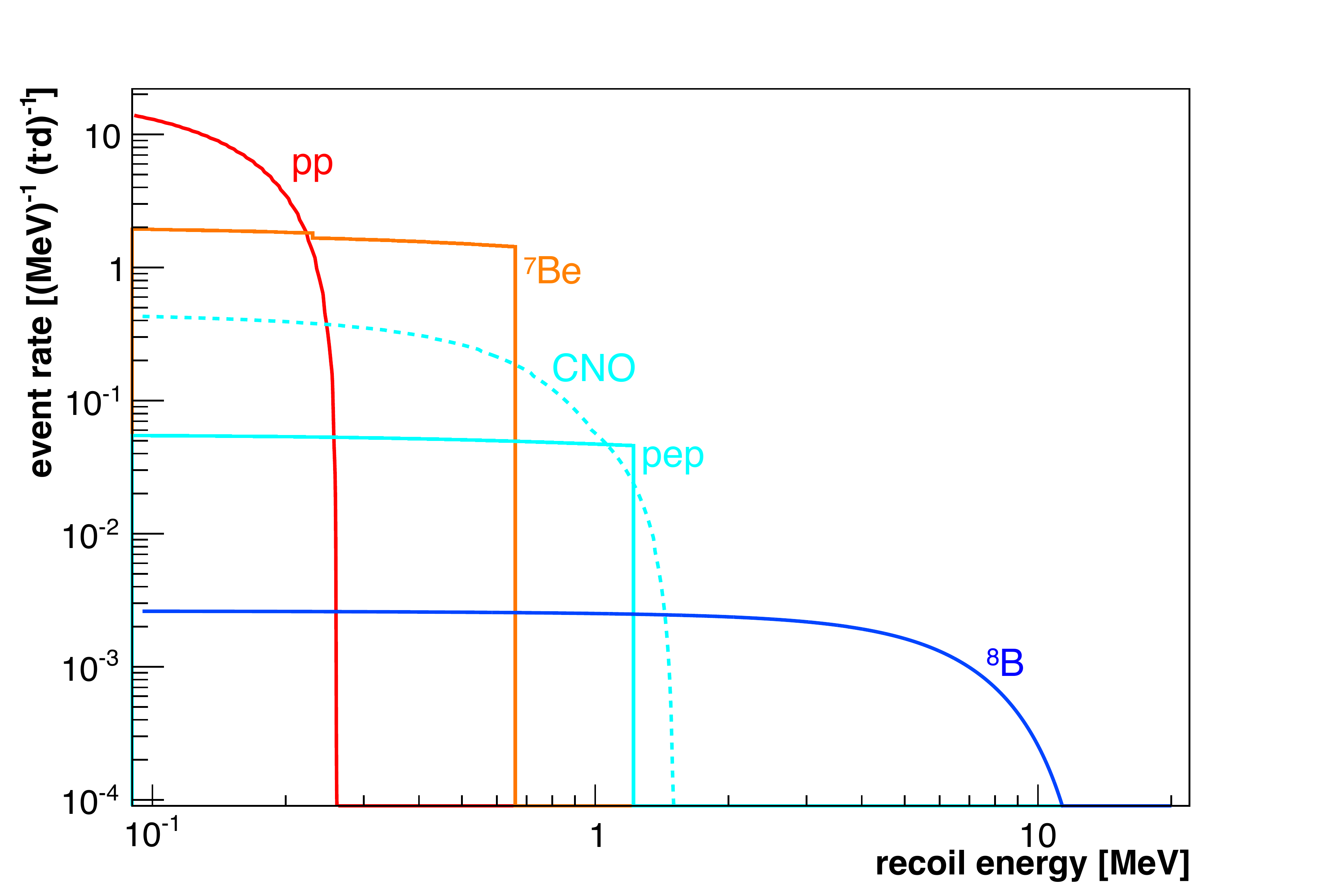}
\caption{Kinetic energy spectra of recoil electrons from elastic scattering of electron neutrinos. Spectra are scaled for relative fluxes on Earth (without oscillations).}
\label{fig:solarrecoilspectrum}
\end{figure} 

The kinetic energy transferred to the recoil electrons depends on the scattering kinematics, i.e. on the scattering angle $\vartheta$. For $m_\nu\ll E_\nu$, the kinetic energy $T_e$ of the recoil energy is given by the Compton formula,
\begin{eqnarray}
E_e(\vartheta,E_\nu) = E_\nu\left(1-\frac{1}{1+m_e(1+\cos\vartheta)}\right)
\end{eqnarray}
the maximum transfer is realized in backscattering where $T_e(\vartheta=180^\circ) = 2m_eE_\nu/(1+2m_e)$ \cite{Raffelt:1996wa}. Therefore, electron recoils from mono-energetic neutrinos ({$^7$Be}, $pep$) form a characteristic shoulder in the recorded event energy spectrum. Recoils from the continuous neutrino spectra will result in a smooth event spectrum, as illustrated in figure \ref{fig:solarrecoilspectrum}. The underlying spectral shape of the incident neutrino flux can be reproduced by a de-convolution of the complete recoil spectrum and the differential cross-section
\begin{eqnarray}\label{eq:diffx}
\frac{{\rm d} \sigma}{{\rm d} T_e} = \frac{2G_F^2m_e}{\pi E_\nu^2}\left[\varepsilon_-^2E_\nu^2+\varepsilon_+^2(E_\nu^2-T_e^2)-\varepsilon_-\varepsilon_+m_eT_e\right]
\end{eqnarray}
(with $\varepsilon_- = 1-\sin^2\theta_W$, $\varepsilon_+ = -\sin^2\theta_W$ , and $\theta_W$ the Weinberg angle) from the overall signal \cite{Raffelt:1996wa}. This allows to test the agreement between the neutrino spectrum recorded and the spectrum predicted from nuclear theory, thereby providing sensitivity for energy-dependent effects like neutrino oscillations (sect.~\ref{sec:solnuosc}).
\medskip\\
{\bf Event energy.} The expected track length of a 10\,MeV electron is $\sim$5\,cm, much smaller than the dimensions of the detectors. Thus, the tracks will originate and end in the bulk of the detector, most of its kinetic energy lost by ionization of the medium according to the Bethe-Bloch formula. Only a minor part of the energy loss ($\sim$0.14\,\%) is converted into Cherenkov
light, which amounts to about 350 photons per MeV of deposited energy. The wavelength spectrum of the Cherenkov light produced can be approximated by the formula 
\begin{eqnarray}
\frac{{\rm d} I}{{\rm d} \lambda} =\frac{e^2L\sin^2\theta_c}{\lambda c^2}
\end{eqnarray}
where $\lambda$ is the wavelength of the emitted photons, $L$ is the path length, and $\theta_c$ the Cherenkov angle. As $L$ is closely related to the electron energy, the total amount of Cherenkov light emitted is a good estimator for the recoil energy \cite{Radel:2012kw, Hosaka:2005um}. 
 \medskip\\
{\bf Directionality} The light is emitted in the characteristic Cherenkov cone centered on the electron track and featuring an opening angle $\cos\theta_c= \frac{1}{\beta n}$. Most photons are emitted under the minimum Cherenkov angle in water, $\theta_c = \arccos(1/n) \approx 41^\circ$. The orientation of the resulting cone is reconstructed based on the shape of the Cherenkov ring projected on the PMTs.

The cone axis coincides with the momentum direction of the recoil electron, which in turn is closely related to the incident direction of the solar neutrino by scattering kinematics. Regarding a 10\,MeV neutrino, the recoil electron will keep the direction of the incident neutrino within $\sim18^\circ$ \cite{Raffelt:1996wa}. However, detector effects partially smear out this alignment: The tracks of the recoil electrons are not perfectly straight as they undergo multiple Coulomb scattering processes on the electrons of the water target. As a consequence, the observed Cherenkov ring will in fact be an overlay of multiple rings of slightly different alignment. This results in an effective angular resolution of $\sim18^\circ$ for a 10\,MeV electron, the resolution decreasing at lower energies \cite{Abe:2016nxk}. As a consequence, the resulting directional resolution for the recoil electron is not sufficient for a full reconstruction of the scattering kinematics that, together with the known incident direction of the neutrino, would provide an event-by-event reconstruction of the neutrino energy.

However, the remaining correlation between true neutrino and reconstructed recoil-electron direction is sufficiently pronounced to employ it very effectively for background discrimination: As demonstrated in figure \ref{fig:skangular}, electron recoils aligned with the Sun's position in the sky stick out over a constant background of randomly oriented radioactive decays \cite{Abe:2010hy}.
 \medskip\\
{\bf Light propagation.} The initial Cherenkov spectrum peaks in the UV, but will be modified by propagation through the water of the target volume. For most of the spectral range, the transparency of the water is primarily defined by Rayleigh scattering of the water molecules. light absorption on impurities playing only a minor role. The overall light attenuation can be described in one dimension by the interplay of scattering and absorption processes, quantified by the length scales of the Beer-Lambert-law
\begin{eqnarray}
I(x) = I_0\exp(-x/\ell);\qquad \frac{1}{\ell}=\frac{1}{\ell_{\rm abs}}+\frac{1}{\ell_{\rm scat}},
\end{eqnarray}
with $\ell_{\rm abs}$ the absorption and $\ell_{\rm scat}$ the scattering length, and l describing the combined overall attenuation length. For Cherenkov detectors in enclosed water tanks, the transparency can be significantly enhanced by constant re-purification of the target medium. In SK, the water is circulated continuously through an external multi-step system that includes ultra-filtration, UV sterilization, reverse osmosis and degasification \cite{Fukuda:2002uc}. As a consequence, a value of $\ell \approx 100$\,m is reached for the blue range of the spectrum.

In accordance with the Rayleigh scattering cross section, light attenuation is more prominent at short wavelengths, $\ell_{\rm scat} \propto \lambda^4$. Especially in large detectors, this leads to a relative suppression of short UV wavelengths compared to the blue component of the Cherenkov spectrum \cite{Fukuda:2002uc}.
\medskip\\
{\bf Light detection.} In experiments for MeV neutrinos, the light is detected by a dense array of photomultiplier tubes (PMTs) located at the verge of the detection volume. Standard bi-alkali PMTs are almost uniquely suited to the detection of the Cherenkov light \cite{Brack:2012ig}: They are sensitive to low light intensities, featuring a 20$-$30\,\% probability to detect single photons. The peak sensitivity is around 400\,nm in wavelength, which is an almost ideal match with the effective Cherenkov spectrum. PMTs can cover large areas with a relatively low amount of read-out channels required (the largest PMTs used in neutrino experiments are 20" in diameter). The time resolution is in the order of few nanoseconds and thus provides sensitivity to the minuscule time differences resulting from the inclination of incident light cone relative to the surface formed by the PMT photocathodes ($\sim$1\,ns per 25\,cm difference in way length). Note, however, that PMT dark noise rates are on the order of 5\,Hz per cm$^2$ \cite{Brack:2012ig}. Given the large photoactive areas of neutrino detectors, random coincidences of dark hits set the instrumental trigger threshold, while their overlap with the nanosecond-long neutrino signal has to be taken into account in event reconstruction (see below) \cite{Hosaka:2005um}. 

Not to insert the sensors directly inside the detection volume $-$ as is the case for high-energy neutrino telescopes $-$ is recommendable for two reasons: Standard PMTs feature an angular acceptance of no more than $2\pi$, so shadowing effects are bound to occur for direct photon tracks. Much more importantly, the low neutrino interaction rates require correspondingly low background levels. The relatively large contents of radioactive contaminants in the PMT glass and dynode chain would raise background rates in the vicinity of the PMTs substantially \cite{Boger:1999bb}.

\begin{figure}[ht]
\centering
\includegraphics[width=0.9\textwidth]{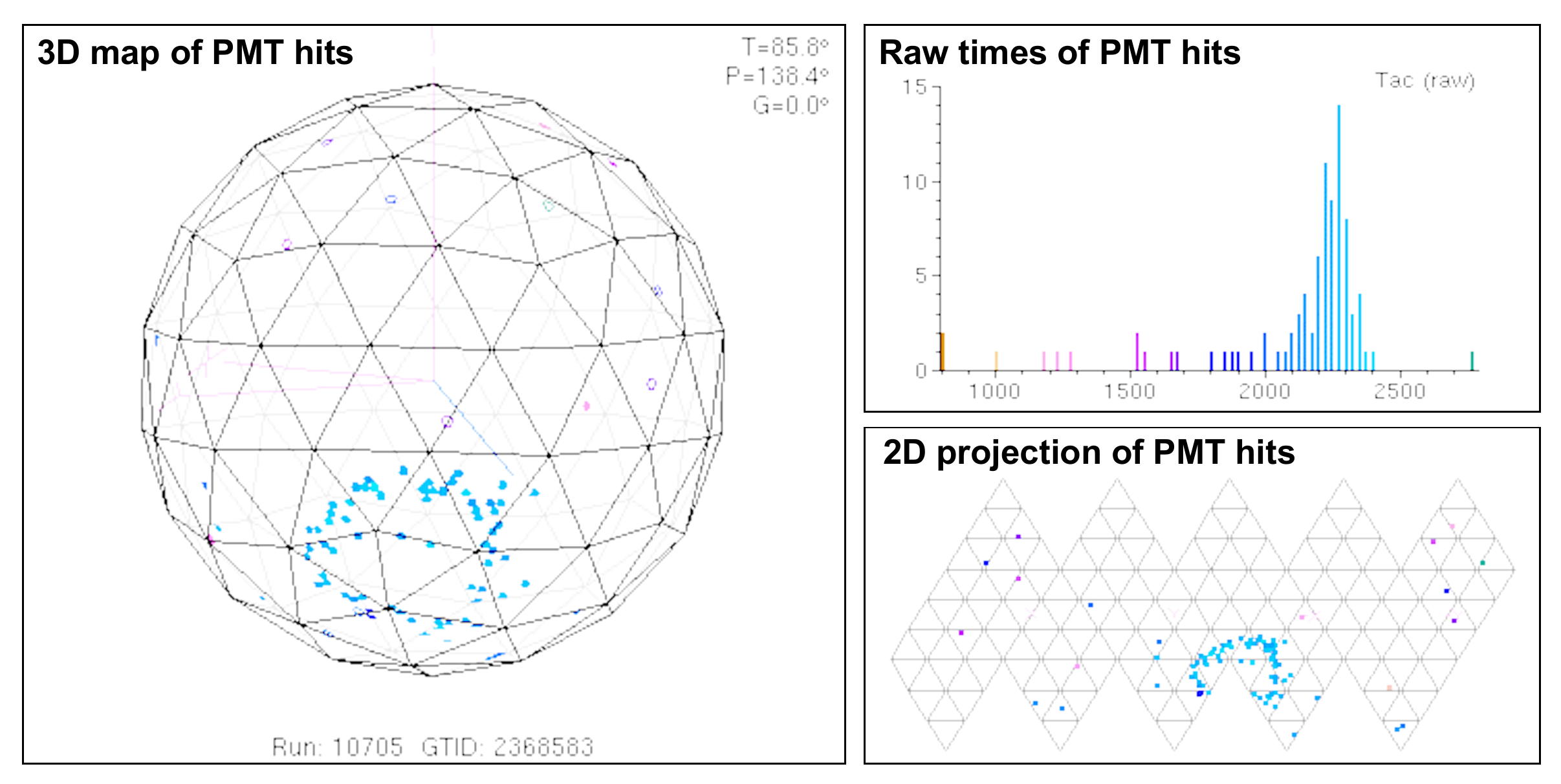}
\caption{Event display of SNO showing a {$^8$B} neutrino-electron scattering event: The large panel to the left shows the 3D position of hit PMTs on the spherical holding structure, while the lower right panel shows a 2D projection. The distribution of (raw) hit time values is shown on the upper right panel and indicated in all panels by the color coding.}
\label{fig:snob8event}
\end{figure}

\noindent{\bf Event reconstruction.} When arriving at the detector walls, the Cherenkov cone is projected as a ring-shaped pattern of PMT hits. Usually, the ring will spread over a large area and hit multiple PMTs. Many (but not all) of these will detect only a handful or single photons for a given event, while the vast majority of PMTs in the detector will register no light at all. Due to multiple scattering of the recoil electron in the medium (see above), the projected ring image will be somewhat washed out. Further signals created by photons scattered in the water or reflected from the mirroring photocathodes as well as a majority of PMT dark hits can be removed by a timing cut of late hits (e.g.~50\,ns after the first photon hits in Super-Kamiokande) \cite{Hosaka:2005um}.

Based on this ring image (exemplified in figure \ref{fig:snob8event}), vertex position, track direction and electron energy can be reconstructed: The size of the ring and the relative differences in photon arrival times will be proportional to the distance of the vertex from the wall, while timing and distortion of the ring is sensitive to the inclination angle of the incoming light cone relative to the normal vector of the PMT surface. The number of photons detected is nearly proportional to the energy of the initial electron. Based on the distance the photons had to travel from the vertex before reaching the PMTs, a correction taking into account light attenuation in the liquid has to be applied \cite{Hosaka:2005um}.

The (electron) energy resolution depends on the number of photoelectrons detected. To first order, it will depend on the ratio of attenuation length to detector size, the photoactive area covered by PMTs compared to the total geometric surface of the target volume, and the detection efficiency of the individual PMTs. Typical effective values for the photoelectron yield are of the order of 7 photoelectrons (pe) per MeV of deposited energy, corresponding to an effective light detection efficiency of $\sim$2\,\%. Residual dark hits falling within the reconstruction window can be statistically corrected for \cite{Hosaka:2005um}. 
\medskip\\
{\bf Detection threshold.} As no additional particles are created in the final state, there is no intrinsic energy threshold for $\nu_e$-scattering (see above). A relatively weak threshold is imposed by the Cherenkov effect itself as a velocity $v > c/n \approx 0.75c$ is required of a particle in order to emit Cherenkov light in the water \cite{Radel:2012kw}. This corresponds to a minimum kinetic energy $T_e \geq 262$\,keV for the recoil electron, which is fulfilled for most solar neutrino interactions.

More stringent is the effective threshold set by PMT dark noise levels. Each individual PMT produces dark pulses due to the thermal emission of electrons from the photocathode that are intrinsically indistinguishable from single photon hits. Typical dark rates for 8"$-$20" PMTs are of the order hundreds to thousands per second. To identify physical events, a trigger is issued only if a close coincidence of several hits in different PMTs is observed within a narrow time window of 10$-$100 nanoseconds. In Super-Kamiokande IV, the instrumental threshold for low-energy events is set to the level of 47 hits in 200\,ns \cite{Abe:2016nxk}.

In practice, the neutrino detection threshold is imposed by the background from radioactive decays inside the detection volume. Depending on the purity of the target water and the surrounding detector components, this will set the energy threshold to 3.5$-$5\,MeV.
\medskip\\
{\bf Radioactive background.} The detector layout of water Cherenkov (and liquid-scintillator) detectors is dictated by the necessity to reduce radioactive backgrounds as far as possible. As solar neutrino event rates are on the level of $\leq$1 event per day and ton (e.g.~\cite{Bellini:2014uqa}), the required radiopurity is far beyond natural levels. The following sources have to be regarded:
\medskip\\
{\it Internal radioactivity.} In water Cherenkov detectors, the dominant internal background arises from the radon isotope  {$^{222}$Rn}  dissolved in the water. The radioactive noble gas enters into the water by emanation from detector surfaces and decays with a half life of 3.8 days. The primary background does not originate from the radon decay itself\footnote{ {$^{222}$Rn}  decays to 218Po under the emission of an $\alpha$-particle of 5.6\,MeV well below the Cherenkov threshold} but from the subsequent $\beta^-$-decay of  {$^{214}$Bi}  ($Q_\beta = 3.3$\,MeV) several steps along in the decay chain. In both SK and SNO, inverse osmosis is part of the multi-step water purification system to remove the  {$^{214}$Bi}  ions before decaying. However, the residual  {$^{214}$Bi}  $\beta$-activity dominates the electron recoil spectrum up to $\sim$4\,MeV. This is due to the relatively low photon statistics that smears the detected visible energy to values above the $\beta$-endpoint \cite{Abe:2016nxk}.
\medskip\\
{\it External background.} While the materials used in state-of-the-art neutrino detectors are screened and selected for radioactive contaminations and meticulously cleaned during construction and commissioning, the solid materials (metals, glass, etc.) cannot be purified to the level of the target liquid. An important concept to obtain a clean neutrino data sample is thus the definition of a fiducial volume (FV) for neutrino detection. During data analysis, only events featuring vertices reconstructed within the FV are regarded for analysis. Events at short distance from the target surface (typically 2$-$3 meters) are rejected (e.g.\cite{Abe:2016nxk}), thereby removing the very short-ranged $\alpha$- and $\beta$-activity but also a major part of the $\gamma$-rays emitted by the external materials. Due to absorption, the $\gamma$-flux decreases exponentially with the width of the self-shielding layer.
\medskip\\
{\it Cosmogenic background.} Even in a radiopure detector, cosmic radiation poses a major source of background that makes neutrino detection at the Earth's surface virtually impossible. Experiments are therefore performed deep underground, shielding all of the cosmic-ray components except the most penetrating highly energetic muons. Depending on the specific site, the residual cosmic muon flux is usually orders of magnitude lower than on the surface (e.g.~\cite{Aglietta:1998nx}).

While the muons by themselves are easily identified with the help of either a veto zone defined within or a subdetector system surrounding the neutrino target volume, muon spallation produces radioactive isotopes from the oxygen (or carbon) atoms contained in the detection medium. In Cherenkov detectors, the $\beta$-decays of these isotopes pose the most severe background above the detection threshold \cite{Abe:2016nxk}. Starting from the most abundant oxygen isotope  {$^{16}$O} , a selection of lighter isotopes with large proton-neutron asymmetries is produced, either by direct photo-nuclear interactions of the muon or by secondary processes in hadronic showers caused by the muons. The most important short-lived isotopes featuring high production cross- sections as well as high beta-decay endpoints are the radioisotopes  {$^{12}$B}  and  {$^{12}$N}  with half-lives of $\sim$10\,ms, while  {$^{16}$N}  with a half-life of 7\,sec dominates the long-lived component \cite{Super-Kamiokande:2015xra}. The individual signals of electrons and positrons emitted in the decays cannot be distinguished from electron recoils induced by solar neutrinos. However, the time and space correlation with the parent muon can be employed for a significant reduction of this background \cite{Hosaka:2005um} (sect.~\ref{sec:b8sk}).

\subsubsection{Super-Kamiokande (SK)}
\label{sec:sk}

\begin{figure}[ht]
\centering
\includegraphics[width=0.55\textwidth]{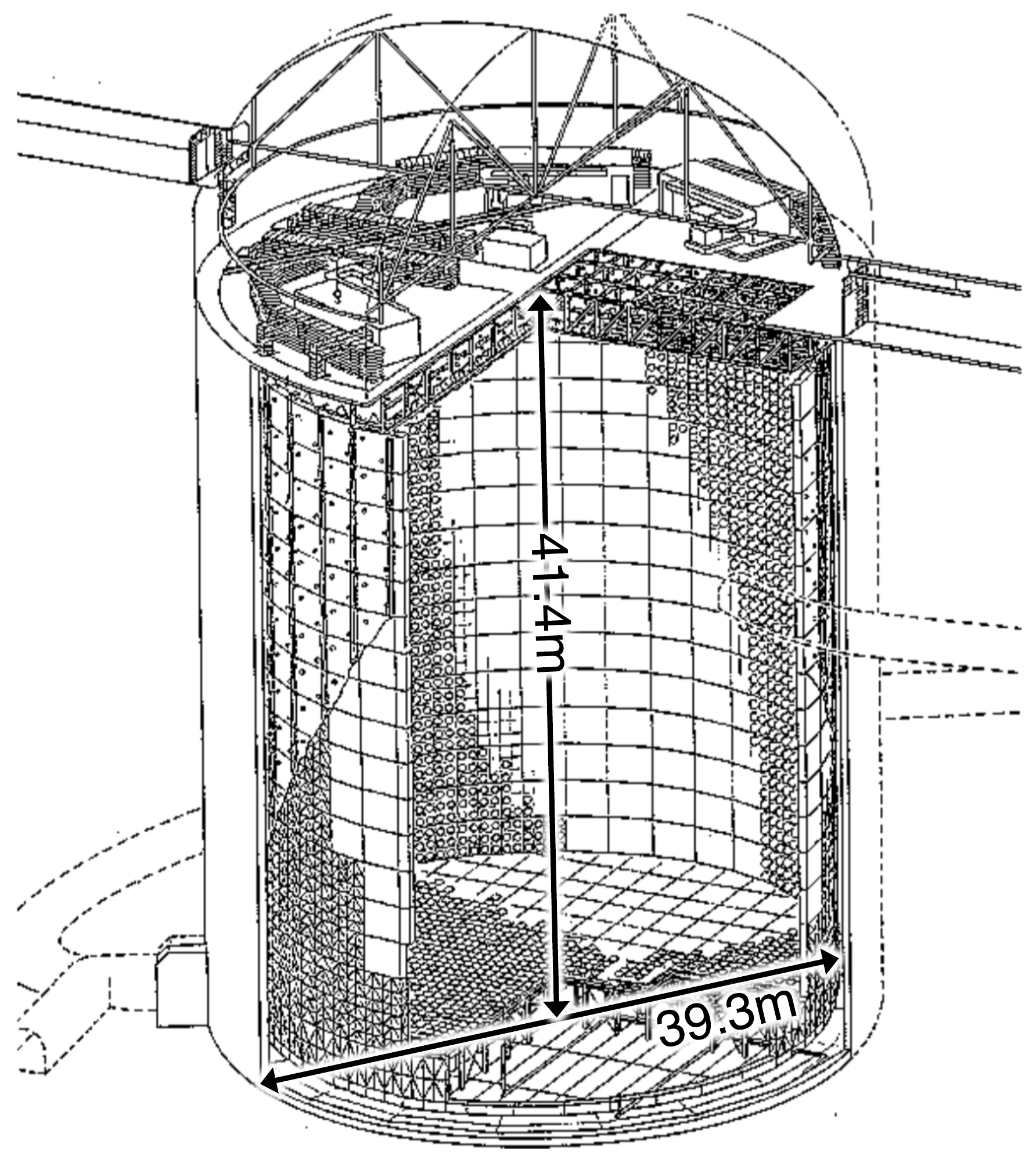}
\caption{Layout of the Super-Kamiokande detector \cite{Ashie:2005ik}.}
\label{fig:sklayout}
\end{figure}

In the mid 1990s, the Super-Kamiokande (SK) started operation in the Kamioka Observatory in Gifu prefecture (Japan) \cite{Fukuda:2002uc}. SK is successor to the previous Kamiokande experiment, that started out as a nucleon decay experiment but instead found first hints for atmospheric neutrino oscillations, performed the first real-time observation of solar neutrinos and observed a dozen neutrino events from the close-by Supernova SN1987A \cite{PhysRevLett.58.1490}. As a consequence, SK was designed with the primary purpose to observe atmospheric and Supernova but also solar {$^8$B} neutrinos. After an upgrade of the read-out electronics, the experiment is now in its fourth data taking phase 'SK-IV' since September 2008 \cite{Abe:2016nxk}. 
\medskip\\
{\bf Detector layout.} The SK detector tank measures 41.4 meters in height and 39.3 meters in diameter, corresponding to a total volume of 50 kilotons of water (fig.~\ref{fig:sklayout}). The Cherenkov light emitted by neutrino events is detected by 11146 inward-looking PMTs of 20" diameter, mounted on a scaffolding running along the inside of the tank walls. The photoactive detection area corresponds to 40\,\% of the tank surface. Surrounding the main detection volume, a gap of 2\,m width is left on the backside of the scaffolding, optically separated from the main tank by black plastic sheets and equipped with 1885 PMTs (8"). This auxiliary detector is used to identify events entering the detection volume from the outside, thus discriminating intruding cosmic muons from genuine neutrino events contained inside the target. The Kamioka laboratory, formerly a mine, provides a rock overburden of 1\,km thickness, corresponding to 2700 meters of water equivalent (mwe). Compared to the surface, the muon flux is reduced by five orders of magnitude, corresponding to $\sim$2 muon events per second in the SK detector \cite{Fukuda:2002uc}.

Great care has been given to reduce the radioactivity levels in the detection volume, most of all radon emanating from the laboratory rock and detector materials. All rock surfaces in the SK dome-shaped cavern are covered with radon-resistent plastic sheets. Special radon-reduced air ($<$3\,mBq/m$^3$) is used as a blanket above the water surface inside the tank. The target is continuously purified running in loop-mode through a dedicated water purification system, removing radioactive and optical impurities. The water transparency is dominated by the influence of Rayleigh scattering (with some additions from Mie scattering and absorption). During the operation, the attenuation lengths has been measured to range between 73\,m to 98\,m. The reader is referred to \cite{Fukuda:2002uc} for a more detailed description.
\medskip\\
{\bf Neutrino detection.} At the time of writing, SK is the largest solar neutrino detector world-wide. The fiducial volume (FV) for solar neutrinos detection is defined by a distance cut 2\,m inwards from the PMT photocathodes. The corresponding target mass is 22.5 kilotons, providing for the detection of $\sim$19 neutrino scattering events per day in SK-IV \cite{Abe:2016nxk}.

Despite the great size, SK is able to collect about 7 photoelectrons (p.e.) per MeV of deposited energy. Based on this p.e.~yield, a trigger is issued when about 30 PMTs are firing within a sliding time window of 200\,ns. In that incidence, the timing and charge information of all PMTs is digitized. The first phase of the experiment (SK-I) started with an instrumental trigger threshold of 5.2\,MeV and a trigger rate of $\sim$10\,Hz, while the offline analysis threshold was set to 6\,MeV in electron recoil energy\footnote{Note that SK publications quote total event energies that include the electron rest mass \cite{Abe:2016nxk,Hosaka:2005um}, while here kinetic/visible energy values are used for consistency with other experimental results.} to assure 100\,\% detection efficiency \cite{Hosaka:2005um}. In the most recent phase SK-IV, this hardware threshold was lowered. Now, a trigger efficiency of 84\,\% is achieved for visible event energies as low as 3.5\,MeV, reaching 99\,\% at 4\,MeV \cite{Abe:2016nxk}.
\medskip\\
{\bf Event reconstruction} relies on the spatial and timing patterns of PMT hits. Even at a neutrino energy of 15\,MeV, the length of the recoil-electron track is below 10 centimeters and thus can be treated as a point-like vertex. Due to the low number of photoelectrons, the majority of PMTs is hit only by a single photon. Therefore, the vertex reconstruction relies on the differences in the absolute photon arrival times at the PMTs. The time resolution of the 20"-PMTs is $\sim$3\,ns for single photons \cite{Hosaka:2005um}. Throughout SK-II to IV, the reconstruction algorithm performs a maximum likelihood fit to the timing residuals of the Cherenkov signal as well as the dark noise background for each testing vertex \cite{Abe:2010hy}. The vertex resolution is a function of energy, improving from 1\,m at 5\,MeV to 50\,cm at 12\,MeV. This is sufficient to apply a fiducial volume cut and to associate cosmogenic decay events to their parent muons (see below) \cite{Abe:2016nxk, Hosaka:2005um}.

The track orientation is found by a maximum likelihood method, that identifies the characteristic Cherenkov ring pattern: Starting from the vector connecting the event vertex with the cluster of hit PMTs, most photons are located on a surrounding cone with an opening angle of $42^\circ$, smeared out by multiple scattering of the electron and Rayleigh scattering of the photons. By optimizing the fit to reproduce the ring under the correct angle, the angular resolution obtained is about $25^\circ$ for 10\,MeV electrons \cite{Abe:2016nxk, Hosaka:2005um}.

The event energy is roughly proportional to the number of produced Cherenkov photons, which is in turn proportional to the number of detected photoelectrons. Small corrections have to be applied to take into account variations in the water transparency, the geometric acceptance of the PMTs hit, and the random contribution of PMT dark noise rate ($\sim$2 dark noise hits in the 50-ns time window associated with an event) \cite{Abe:2016nxk, Hosaka:2005um}.
\medskip\\
{\bf Calibration.} The event reconstruction is tested and calibrated by several low-energy particle sources that can be introduced into the FV: The most important and unique to SK is a low-energy electron LINAC featuring a long evacuated beam pipe with a 0.1\,mm thick titanium window. Able to reach nine distinct positions, the LINAC produces single electrons of defined direction and energy, covering an energy range from 5 to 16 MeV. This device is complemented by an artificial  {$^{16}$N}  source that provides electrons emitted with random orientation and a visible energy spectrum peaking at 7\,MeV. The short-lived isotope  {$^{16}$N}  is produced by $(n,p)$-reactions on  {$^{16}$O} , the neutrons provided by a DT-generator. In addition, a Ni($n,\gamma$)Ni reaction has been used to produce high-energy gammas for estimating systematic shifts in the spatial reconstruction \cite{Fukuda:2002uc, Hosaka:2005um}.

The purity of the water target is continuously monitored using the signals of Michel electrons and positrons produced by cosmic muons decaying inside the fiducial volume. Identification of these events is based on the fast coincidence of the muon stopped inside the target and its decay signal \cite{Fukuda:2002uc, Hosaka:2005um}.

\subsubsection{The SNO Experiment}
\label{sec:sno}

From 1999 to 2006, the Sudbury Neutrino Observatory (SNO) was operated in the Creighton Mine in eastern Canada, 2000\,m below surface. While SNO relied like the substantially larger SK detector on the water Cherenkov technique, the two experiments differed in many respects besides size. Most significantly, SNO used 1\,kton of heavy water (D$_2$O) as detection medium, providing deuterons ($^2$H) as additional reaction target for solar neutrinos \cite{Boger:1999bb}.

\begin{figure}[ht]
\centering
\includegraphics[width=0.65\textwidth]{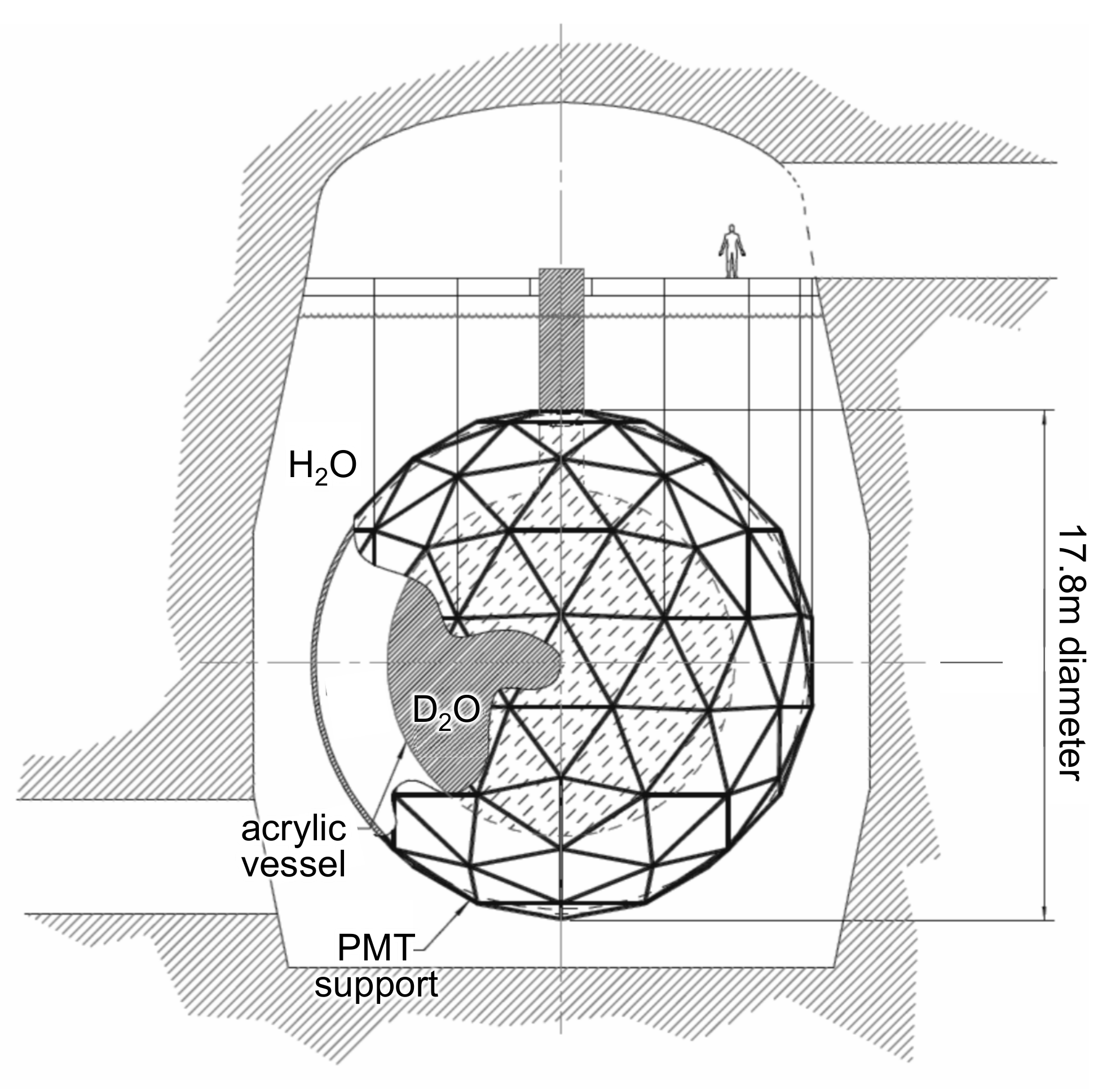}
\caption{Layout of the SNO detector \cite{Boger:1999bb}.}
\label{fig:snolayout}
\end{figure}

\noindent{\bf Neutrino detection in heavy water.} In heavy water, solar neutrinos undergo three different kinds of interactions with the constituents of the detection medium \cite{Chen:1985na, Aharmim:2006kv}:
\begin{eqnarray}
\rm (ES) & \nu_x+e^-\to \nu_x+e^- \nonumber \\
\rm (CC) & \nu_e+d\to p+p+e^- \nonumber\\
\rm (NC) & \nu_x+d\to p+n+\nu_e \nonumber
\end{eqnarray}
{\it Elastic scattering (ES)} of neutrinos on electrons is equivalent to the detection process used in SK. The channel features no intrinsic energy threshold but is mostly sensitive to $\nu_e$ because of the enhanced cross section compared to $\nu_{\mu,\tau}$, and in addition provides some energy and directional information for the neutrinos (sect.~\ref{sec:wcd}), the latter giving evidence for the solar origin of the neutrino signal and used in event discrimination (sect.~\ref{sec:b8sno}).
\medskip\\
The {\it charged-current (CC)} reaction produces a final-state electron and breaks up the deuteron when converting the bound neutron to a proton. The reaction features a threshold of $Q_{\rm CC} = 1.44$\,MeV, and a cross-section for {$^8$B} neutrinos that is about an order of magnitude larger than for ES. While the final-state protons remain undetected below the Cherenkov threshold, the electron produces a visible Cherenkov signal. Because of the heavier reaction partners, the final-state electron's kinetic energy is much more closely correlated to the incident neutrino energy than in the case of ES, providing a more precise measurement of the neutrino energy spectrum. Most importantly, the channel is accessible exclusively for neutrinos of electron flavor because solar neutrino energies are not sufficient to create muons or taus in the final state.
\medskip\\
The {\it neutral-current (NC) reaction} results in the break-up of the target deuteron, requiring a minimum of 2.22\,MeV to overcome the weak bond of the nucleus. The cross-section is of the same order of the (CC) reaction. The interaction is detected by the subsequent signal from neutron capture (see below). It provides no energy information but instead offers equal cross sections to neutrinos of all flavors, i.e.~it can be used to measure the total solar neutrino flux disregarding flavor oscillations.
\medskip\\
As a consequence, the SNO experiment featured the unique capability to measure not only the pure $\nu_e$ flux but also the combined flux of all neutrino flavors emitted by the Sun. By comparison of the event rates in the three detection channels, it could be clearly demonstrated that the solar $\nu_e$ flux is not simply smaller than expected but that the missing $\nu_e$ are converted into the $\nu_{\mu,\tau}$ flavors, providing the key for resolving the solar neutrino puzzle \cite{Ahmad:2002jz}.
\medskip\\
{\bf Detector layout.} SNO contained 1\,kt of D$_2$O in a transparent acrylic sphere of 12\,m diameter (fig.~\ref{fig:snolayout}). The Cherenkov light produced inside the target was detected by an array of 9456 PMTs (8") mounted to a spherical open support structure of stainless steel that itself was located 2.5\,m from the acrylic surface. Each PMT was equipped with a light concentrator, i.e.~a close-to conical metal reflector that increased the total photon collection area to nearly 55\,\% of the solid angle. The whole setup was constructed in a considerably larger underground cavern. The interspace between sphere and cavern walls was filled with 7.4 kiloton of light water. The inner 1.7\,kt shielded the target from the radioactivity of the PMTs and the support structure, while the outer 5.7\,kt were viewed by 91 outward-facing PMTs to identify through-going cosmic muons by their Cherenkov light emission \cite{Boger:1999bb}.
\medskip\\
{\bf Detection phases.} The experiment was conducted in three phases that mostly differed by the methods employed for the detection of the final-state neutrons from the (NC) reaction \cite{Aharmim:2011vm}: In {\it phase I} (1999$-$2001), the neutrino target contained only heavy water, and neutrons were identified by their delayed capture on another deuteron in the target medium. The capture releases a single $\gamma$-ray of 6.25\,MeV, that in turn produced a visible Cherenkov signal either by Compton scattering on an electron or by $e^+e^-$-pair production.\\
In the later {\it phase II} (2001$-$2003), 2 tons of NaCl salt were added to the water. The isotope  {$^{35}$Cl}  offers a much larger neutron capture cross-section than deuterons and emits upon capture a $\gamma$-cascade of a total energy of 8.6\,MeV. The resulting larger Cherenkov signal greatly enhanced the detection efficiency.\\
In the final {\it phase III} (2004$-$2006), the salt was removed and  {$^{3}$He}  proportional counters introduced into the main detector volume. These were arrayed in 36 strings consisting of 2\,m long nickel tubes. Neutron capture on  {$^{3}$He}  is highly efficient and easy to identify on an event-by-event basis, thereby further reducing the systematics for the overall measurement.
\medskip\\
{\bf Cherenkov signals.} Contrary to the situation in SK, the dimensions of the SNO detector were considerably smaller than the light attenuation lengths of both heavy and light water (order of 100\,m). In addition, the photoactive coverage was larger. Therefore, the resulting photoelectron yield is somewhat higher than in SK, on the order of 9\,pe/MeV. The low background levels in the detector permitted to set the instrumental trigger threshold to 16 PMTs firing within a time-window of $\sim$93\,ns in {\it phase I}, corresponding to a visible event energy of $\sim$2\,MeV. The resulting trigger rate from physics events is found to be roughly 5\,Hz \cite{Aharmim:2006kv}. For comparison, the event rate originating from neutrino interactions is on the order of 10 per day \cite{Aharmim:2006kv}.

As in SK, event reconstruction algorithms relied on the arrival time distributions of the detected photons at the PMTs to obtain vertex position, electron direction and event energy from a likelihood fit. Energy was derived in two different fashions: A simple estimator counting the number of PMT hits, and a more sophisticated energy reconstructor considering as well the optical model of the detector. In this way, the considerable position (and direction) dependence of the detector response could be taken into account, translating the number of hits observed for a given event to an electron-equivalent energy. The energy resolution scaled with the event energy, amounting to $\sim$14\,\% at an electron energy of 10\,MeV. The angular resolution was evaluated to $\sim$27$^\circ$, the spatial resolution to $\sim$20\,cm \cite{Aharmim:2009gd}.
\medskip\\
{\bf Backgrounds.} In the initial analysis of  {\it phase I} data, the analysis threshold was set to $\sim$5\,MeV, largely to avoid the background created by radioactive contaminants in both water volumes and in the acrylic vessel \cite{Ahmad:2002jz}. Like in SK, this effectively limited solar neutrino detection to the {$^8$B} spectrum. Even above this threshold, the analysis was beset by various backgrounds, mostly intrinsic radioactivity from elements of the natural U/Th chains dissolved in the water. Differently from SK, background from muon spallation products played only a minor role as the great depth of the detector cavern, corresponding to 6000\,mwe, reduced the muon rate to only 3 events per hour in the D$_2$O target volume \cite{Aharmim:2009gd}.
\medskip\\
{\bf Calibration.} To characterize the detector response, radioactive and optical sources were inserted into both the heavy and light water volumes. Optical parameters and PMT timing were determined with the help of an isotropic light source that could be placed at various positions in the detector, realized in the form of a diffusor ball fed by a multi-wavelength laser covering the relevant range of the Cherenkov spectrum. The quality as well as the systematics of event reconstruction were evaluated by calibration runs with weak radioactive sources inside the target volume: The stability of the energy scale was monitored by a pure  {$^{16}$N}  $\gamma$-source at 6.1\,MeV, that allowed as well to determine the angular resolution based on the reconstructed signals. The reconstruction of electrons was tested with the high-endpoint (14\,MeV) $\beta$-emitter  {$^{8}$Li} . Neutron capture signals and efficiencies were calibrated with the help of a 252Cf source \cite{Aharmim:2009gd}.

\subsection{Liquid-scintillator detectors}
\label{sec:lsd}

Liquid scintillators (LS) have been used as a target medium for neutrino detection since the very beginnings of experimental neutrino detection. In the 1950s, F.~Reines and C.~Cowan chose LS as a target material for the famous Poltergeist experiment that was their first try for detection of antineutrinos from the Hanford research reactor and predecessor to the famous Savannah River experiment that finally gave conclusive evidence for the existence of the neutrino \cite{PhysRev.90.493, Reines:1953pu}. In the following decades, the technique matured to provide ever greater neutrino detection volumes. By now, LS detectors are at the forefront of low-energy neutrino experiments, the scientific program ranging from solar neutrinos over reactor and geo antineutrinos to the search for neutrino-less double-beta decay \cite{Alimonti:2008gc, Markoff:2003tg, KamLAND-Zen:2016pfg}.

\subsubsection{Detection technique}

Organic liquid scintillators (LS) are widely used as detection media in nuclear and particle physics. Compared to the Cherenkov effect, scintillation is a much more efficient mechanism to convert the energy deposited by ionizing particles in the medium to visible light: About $10^4$ blue or near-UV photons are created per MeV, providing for superior energy resolution \cite{PhysRev.90.493, Birks:1964zz}. As in light water, solar neutrino detection is by elastic scattering off electrons. The scintillation signal is caused by the short, quasi point-like tracks of ionization induced by the recoil electrons \cite{Bellini:2013lnn}.
\medskip\\
{\bf Scintillation mechanism.} Organic LS are usually composed of (at least) two hydrocarbon components \cite{Alimonti:2008gc, Buck:2015jxa}: The bulk of the volume is made up by an aromatic solvent, doped with low amounts (gram per liter) of another aromatic molecule, commonly referred to as fluorophore or, for brevity, fluor. When an ionizing particle traverses the medium, the delocalized $\pi$-bound electrons of the aromatic rings are brought into an excited state with an excitation energy of about 4$-$5\,eV, corresponding to the UV range of the photon spectrum. By non-radiative transfer (e.g.~by F\"orster interactions), the excitations are passed on to the fluor molecules. The combination of materials is chosen to optimize the overlap of the absorption band of the fluor to the emission band of the solvent, thereby maximizing the efficiency of the excitation transfer. What is more, the fluor induces a large Stokes' shift and thus photons are emitted at consider- ably larger wavelengths, e.g.~in the violet end of the visible spectrum\footnote{This effect is sometimes enhanced by adding a secondary fluor of even longer absorption and emission wavelengths, cascading the light to the blue regime.}. As the transparency of the solvent improves substantially with wavelengths, the red-shifted scintillation photons thus effectively avoid self-absorption and can travel for macroscopic distances through the LS $-$ an important precondition for the construction of very large detectors \cite{Alimonti:2008gc}.
\medskip\\
{\bf Light emission.} Photons are emitted isotropically, so that all directional information on the initial particle is lost (unless its track length exceeds the position resolution of the event reconstruction) \cite{Bellini:2013lnn}. The time profile of the light emission is a superposition of several exponentially decaying components with typical fluorescence time scales ranging from nanoseconds to microseconds, corresponding to the excitation life times of short-lived singulet states and long-lived triplet states \cite{Buck:2015jxa, MarrodanUndagoitia:2009kq}. The relative strengths at which the different contributions are present depend on the ionization density and thus on the type of the particle, providing a handle for particle identification \cite{Birks:1964zz}.
\medskip\\
{\bf Detection threshold.} Finally, there is a variety of standard purification procedures that have proven themselves as very effective in removing radioactive contaminants of all kinds \cite{Alimonti:2009zz}. As a consequence, detection thresholds are generally much lower than in water Cherenkov detectors, reaching down to the level of several 100\,keV for KamLAND or even 50\,keV in the singular case of Borexino. This renders LS an especially suitable medium for the detection of solar neutrinos, including the sub-MeV energy range \cite{Bellini:2014uqa}. Purification also is used to remove organic impurities from the solvent, greatly increasing the transparency of the liquid. In many experiments, LS purity is close to optimum, so that scintillation light transport in the blue regime is solely defined by the Rayleigh scattering off the solvent molecules. Scattering lengths of 10\,m and more can be reached depending on the solvent \cite{Djurcic:2015vqa}.
\medskip\\
{\bf Basic detector layout.} For most of the currently running and proposed LS detectors, the target liquid is contained in a single, unsegmented volume holding tens to hundreds of tons of LS. The target is located in the center of the detector, contained in an optically transparent vessel that is either made from ultra-thin flexible nylon (for ultra-low background) or a rigid acrylic tank (for sturdiness). The vessel is emerged in a further bath of inactive organic liquid contained in a stainless steel tank. This layer of buffering liquid shields the target volume from external radiation that would otherwise overwhelm the data taking rates. The scintillation light produced within the target volume is registered by inward-looking PMTs that are mounted along the inner walls of the steel tank. Standard bi-alkali PMTs are widely used because they offer large photocathode areas and peak sensitivity in the near-UV/blue spectrum, matching the emission spectrum of the fluors (e.g.\cite{Alimonti:2008gc}).

As for water, cosmic radiation poses an important source of background. All operating large-scale detectors feature an overburden of at least 100\,mwe, while the two experiments discussed here, KamLAND and Borexino, are located deep underground with a rock overburden of at least 1000\,m. In addition, both detectors feature a muon veto encompassing the central LS detector. The inner detector is contained in a second steel tank, surrounding it with a layer of 2$-$3 meters of water equipped with a low number of PMTs to register the Cherenkov light of through-going cosmic muons. In addition, the water serves as a passive shielding against external radioactivity \cite{Alimonti:2008gc}.
\medskip\\
{\bf Energy reconstruction.} The relation between the kinetic energy deposited by an electron and the amount of photons emitted by scintillation is virtually linear. Energy resolution is largely dependent on the statistical fluctuation of photoelectrons collected by the surrounding PMTs. Typical photon detection efficiencies are in the order of 5\,\%, corresponding to a yield of $\sim$500 photoelectrons per MeV of deposited energy, or an energy resolution of (5$-$7)\,\% at 1\,MeV (photon statistics only) \cite{Bellini:2013lnn}.
\medskip\\
{\bf Spatial reconstruction.} At solar neutrino energies, the short ionization tracks of recoil electrons can be regarded in good approximation as point-like vertices. As light emission is isotropic, there is no possibility to infer the track orientation and thus the incident neutrino direction. However, the absolute location of the interaction vertex inside the LS volume can be determined with an accuracy of $\sim$10\,cm. Spatial reconstruction is usually based on the relative time-of-flight differences from the event vertex to the PMTs surrounding the detection volume. In practice, the vertex is found by a likelihood minimization of the arrival time pattern, the {\it pdf}'s depending on the time profile of the scintillation light emission, the distance from vertex to PMT and the number of photons detected per tube \cite{Bellini:2013lnn}.
\medskip\\
{\bf Background discrimination.} Spatial reconstruction is employed for two purposes: Firstly, for the definition of a fiducial volume by a radial cut on vertex positions. By regarding only events in the innermost region of the LS volume for solar neutrino analysis, external backgrounds (especially $\gamma$-rays) can be largely suppressed. Secondly, to correct individual photon arrival times by the time of flight in order to obtain the initial, undistorted light emission profile of the scintillation event. Due to the differences in the light emission profiles that depend on the local ionization density (see above), pulse shape analysis can be used to distinguish electron-like events ($\nu$-induced recoil electrons, but also $\beta$ and $\gamma$ decays) from $\alpha$-decays\footnote{Note that the light output of $\alpha$-particles is quenched compared to electrons by a factor $\sim$10.}. Discrimination of positrons (from $\beta^+$ decays) and electrons is much more demanding but has been employed successfully in the $pep$/CNO neutrino analysis of Borexino (sect.~\ref{sec:pepcno}) \cite{Birks:1964zz, Bellini:2013lnn}.
\medskip\\
{\bf Radioactive background.} Regarding solar neutrinos, the primary advantage of LS over water Cherenkov detectors is provided by the greater light yield, offering the possibility to set significantly lower detection thresholds. However, a considerable challenge is set by the natural radioactivity of target and detector materials: Below 3\,MeV, the background from $\beta^\pm$ and $\gamma$ decays rises exponentially, providing a vast number of background events largely indistinguishable from $\nu$-induced recoil electrons. The great achievement of Borexino is therefore the unprecedented radiopurity level of less than 1 event per day and ton in the central region of the LS target that is the most vital precondition for measuring solar neutrinos at sub-MeV energies (sect.~\ref{sec:bx}).

Background events inside the fiducial volume can be caused by a number of radioactive impurities (e.g.~\cite{Bellini:2013lnn}):
\medskip\\
The {\it radioactive carbon isotope}  {$^{14}$C}  is intrinsic to the hydrocarbons of the LS. Due to the low spectral end point of 156\,keV, the $\beta$-decay affects mostly $pp$ neutrino detection (sect.~\ref{sec:pp}). The {$^{14}$C} contained in the LS corresponds to the tiny residual left in the raw oil used for the production of solvent and fluor. Produced by the interactions of cosmic rays with  {$^{14}$N}  in the Earth's atmosphere, it was incorporated 200 million years ago by the living organisms forming the base of the mineral oil. Due to the long time of underground storage, the initial {$^{14}$C} concentration is reduced considerably. Levels of $10^{-18}$ are reached, depending on the age and physical conditions of the oil well and the treatment of the raw materials during LS manufacture.
\medskip\\
{\it Metals.} Elements of the long-lived {$^{238}$U} and {$^{232}$Th} decay chains can be dissolved in LS either as ions or in organic complexes. Their introduction is by two primary ways: Both chains contain radon isotopes ({$^{222}$Rn}  resp.~{$^{220}$Rn}) that are emanated from the rock walls of underground laboratories but also from detector materials like the PMT glass or cables. Once dissolved in the LS, the radon isotopes decay and their metallic daughter nuclei remain suspended in the liquid. The ions can as well be picked up directly from the inner surfaces of pipes and storage vessels if these have been contaminated with radon and its daughter nuclei, e.g.~by contact with air. The same is true of the long-lived and abundant isotope {$^{40}$K}. Prior cleaning of all surfaces coming into contact with the LS and the employment of nitrogen atmospheres during storage, filling and operation of the detectors are therefore mandatory. Fractional distillation (under reduced atmosphere) has proven as the probably most effective purification technique to remove these impurities from the LS and can be employed either before detector filling or in loop mode during operation \cite{Bellini:2013lnn, Gando:2014wjd}.
\medskip\\
{\it Noble gases.} In addition to radon, exposure of the LS to air may lead to the introduction of further radionuclides. The noble gas isotopes  {$^{85}$Kr}  and  {$^{39}$Ar}  are both $\beta$-emitters of intermediate half-life and spectral end points of several hundred keV. Therefore, they potentially pose a background to {$^7$Be} neutrino detection. Borexino has demonstrated that  {$^{85}$Kr}  can be removed very efficiently from the LS by purging with ultrapure nitrogen \cite{Bellini:2014uqa}.
\medskip\\
{\it Cosmogenic isotopes.} In addition to radionuclides dissolved in the LS, cosmic muon spallation on carbon is an important background factor. The break-up of the dominant  {$^{12}$C}  produces a variety of lighter radioactive nuclei. The most prominent background for solar neutrino spectroscopy is  {$^{11}$C} . The decay spectrum of this relatively long-lived ($T_{1/2}\approx 20$\,min) $\beta^+$-emitter falls into the region of interest for $pep$/CNO neutrino detection (1$-$2\,MeV) \cite{Collaboration:2011nga}. What is more, the production cross-section is quite high, and the decay signal will cover the solar neutrino spectrum for all but the deepest labs, making elaborated veto schemes a necessity (sect.~\ref{sec:pepcno}). Further cosmogenic backgrounds of relevant rates are posed by the $\beta$-decays of  {$^{10}$C}  and {$^{11}$Be} for {$^8$B} analysis (sect.~\ref{sec:b8leta}) and {$^{9}$Li}/{$^{8}$He} for antineutrino detection \cite{Bellini:2008mr, Bellini:2013nah}.

\subsubsection{The Borexino experiment}
\label{sec:bx}

\begin{figure}[ht]
\centering
\includegraphics[width=0.65\textwidth]{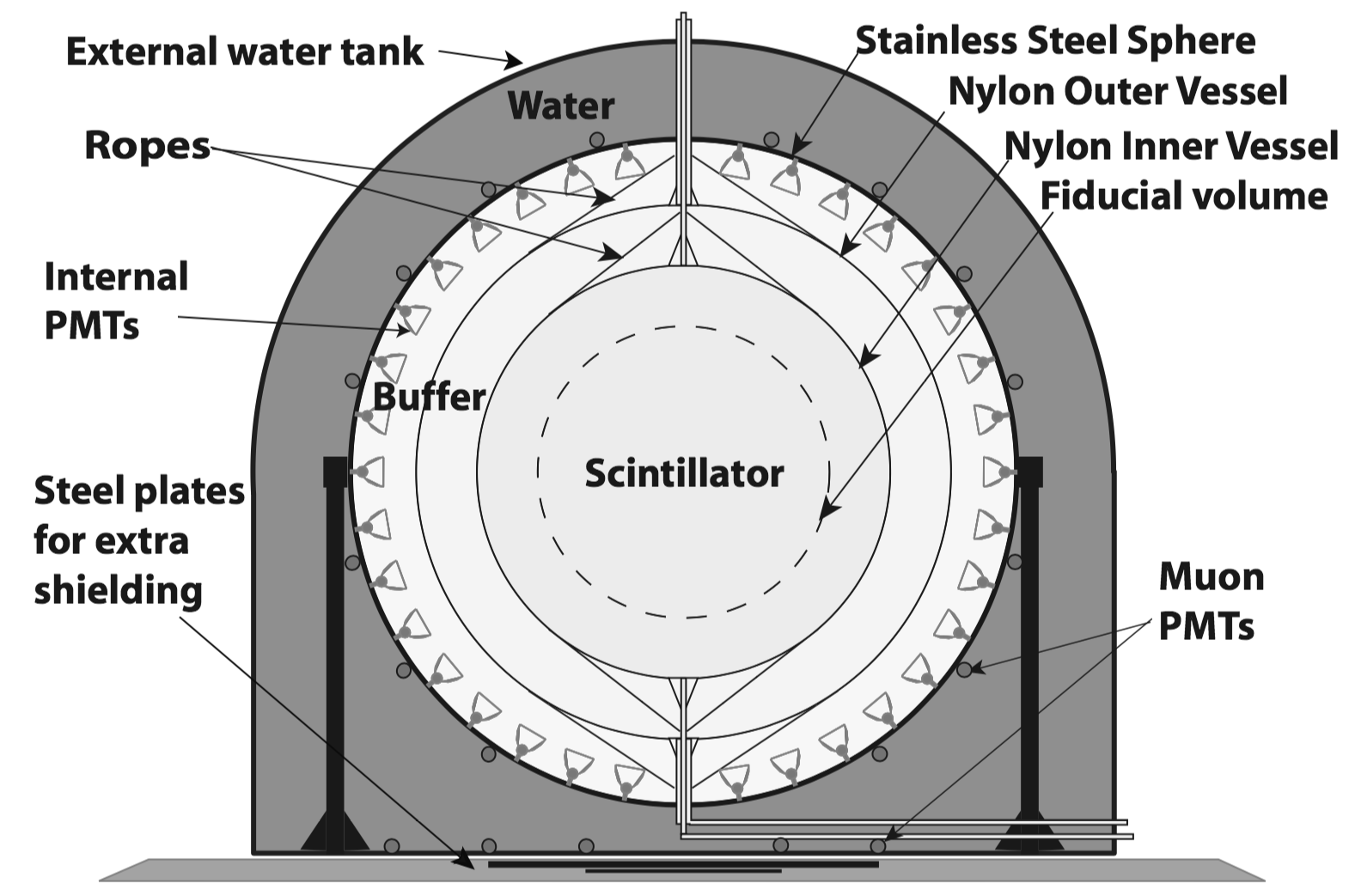}
\caption{Layout of the Borexino detector \cite{Alimonti:2008gc}.}
\label{fig:bxlayout}
\end{figure}

Borexino started data taking in 2007. The liquid-scintillator detector is located at the Laboratori Nazionali del Gran Sasso (LNGS). The underground laboratories are located adjacent to a highway tunnel crossing the Abruzzese mountains from Rome to the Adriatic coast. The corresponding rock overburden is $\sim$1.5\,km (or 3800\,mwe) \cite{Alimonti:2008gc}.
The primary purpose of the experiment was a measurement of the low-energy {$^7$Be} neutrino flux by elastic electron scattering. As a consequence, the detector was designed to meet the requirements of this measurement: A high light yield to be able to discern the $\nu$-induced electron recoil shoulder (sect.~\ref{sec:be7bx}), sufficient shielding from external $\gamma$-rays and, most importantly, the purification of the target LS to a level of $\leq 1$ background event per day and ton in the energy region of interest \cite{Bellini:2011rx}.
\medskip\\
{\bf Detector layout.} The basic layout of Borexino is presented in figure \ref{fig:bxlayout} \cite{Alimonti:2008gc}. The Inner Detector (ID) is confined by a spherical steel tank of 6.85\,m radius. At its very center, the ID contains the neutrino target of 278 tons of ultrapure LS (see below). The target is surrounded by two concentric layers of inactive buffer liquids that provide shielding from external $\gamma$-rays. The sub-volumes are separated by two ultra-thin (125\,m) balloon-shaped nylon vessels \cite{Benziger:2007iv}. The Inner Vessel (IV) of 4.25\,m radius separates the target volume from the Inner Buffer, while a second balloon at 5.50\,m prevents convective motion between inner and outer buffer and impedes radon diffusion towards the target. Both vessels are held in place (and shape) by a system of nylon ropes. The scintillation light from the target is detected by an array of 2212 8" PMTs that are in their majority equipped with aluminum light concentrators (conical mirrors) to increase the light collection efficiency \cite{Oberauer:2003ac}. 371 PMTs are left without cones to assist the Cherenkov light collection and tagging of cosmic muons passing only the buffer region \cite{Alimonti:2008gc}.
The ID is surrounded by a dome-shaped water tank of 18\,m diameter. This Outer Detector (OD) contains 2.4\,kt of water and is equipped with 208 PMTs. The OD provides a coincidence veto for through-going cosmic muons and assists the muon track reconstruction in the ID. The residual flux of muons at LNGS is only $\sim$1.2\,m$^{-2}$h$^{-1}$, corresponding to 4300 muons per day crossing the Inner Detector of Borexino \cite{Bellini:2011yd}.
\medskip\\
{\bf Liquid scintillator.} The bulk material of the LS is the solvent pseudocumene (PC, 1,2,4- trimethylbenzene). PPO (2,5-diphenyloxazole) acts as the light-emitting fluor and features a concentration of 1.5\,g/l. This mixture was selected for the high light yield, good transparency at the emission wavelength of PPO (about 5.5\,m at 400\,nm) and, most importantly, the excellent radiopurity levels obtained in preceding test runs \cite{Alimonti:2000wj}. To arrive at this choice, different LS materials and purification techniques were investigated with the 4-ton prototype Counting Test Facility (CTF) that was operated from 1995 to 2006 at the LNGS \cite{Alimonti:1998nt}.

PC was used as well as basic solvent for the inactive buffer liquid. Here, a scintillation quencher dimethylphthlate (DMP) was added instead of PPO in order to suppress light emission. The initial concentration of 5\,g/l was later on lowered to 2\,g/l to decrease the density difference between buffer and target scintillator and thus the buoyancy forces acting on the IV \cite{Alimonti:2008gc}.

The solvent was produced under very clean conditions and specifically for Borexino at a production plant in Sardinia owned by Polimeri Europa (Sarroch-IT). It was shipped to LNGS in custom-built clean transport tanks. Arriving at LNGS, the PC underwent distillation before being mixed either with PPO for the target liquid or DMP for the buffer. All tank and piping surfaces to come into contact with the liquid were meticulously cleaned. An atmosphere of ultrapure nitrogen was maintained in the liquid handling system during filling to avoid contamination with  {$^{85}$Kr}  and  {$^{39}$Ar} . All ID volumes were first filled with ultrapure water. In a second step, the lighter LS was inserted from the top while the water was drained at the bottom of the ID, creating an increasing layer of LS floating on top of the receding water. This allowed constant monitoring of the LS for the appearance of an access in radioactive background \cite{Alimonti:2008gc}.
\medskip\\
{\bf Neutrino detection.} Solar neutrinos are detected by their recoil electrons, the kinetic energies ranging from $\sim$100\,keV to 10\,MeV. Starting from an initial light yield slightly above $10^4$ photons per MeV, the ID provides a geometric coverage of 30\,\% from PMT photocathodes and cones and an average detection efficiency of 20\,\%. Including small absorption losses in the LS, the photoelectron yield is on the order of 550\,p.e.~per MeV, corresponding to an energy resolution of $\sim$5\,\% and a position resolution of $\sim$10\,cm for electron-like events at 1\,MeV. The instrumental threshold for detection is set to 25$-$30\,p.e.~or $\sim$50\, keV. The trigger rate of 30\,Hz is dominated by the low-energy decay of the  {$^{14}$C}  intrinsic to the LS hydrocarbons \cite{Alimonti:2008gc, Bellini:2014uqa}.

In all solar neutrino analyses, not the entire LS volume can be used for neutrino detection because external $\gamma$-rays and radioactive contaminants on the nylon vessel exceed the rate levels of the neutrino signals. Instead, reconstruction of the event vertex positions is employed to impose a radial cut that defines an inner fiducial volume (FV). The exact FV chosen depends on the neutrino species analyzed. Usually, about 70$-$100 tons are selected by a combination of radial cut and top-bottom cuts that take into account the larger background rates close to the vessel end-caps \cite{Bellini:2014uqa, Bellini:2013lnn}.
\medskip\\
{\bf Background levels.} The materials of the internal detector components (stainless steel, photomultiplier glass and dynode chain, cables, light concentrators, nylon vessel) were selected for extremely low radioactive background levels (typically by Ge spectroscopy) and were manufactured in ultra-clean processes (e.g.~the nylon vessel was assembled and glued under clean-room conditions \cite{Benziger:2007iv}). As a consequence, the external $\gamma$ background accounts for only a few events per day in the FV set for neutrino detection (sects.~\ref{sec:be7bx}, \ref{sec:pepcno}, ~\ref{sec:pp}).

These efforts culminated in unprecedented radiopurity levels: During Phase I (2007-2011), {$^{238}$U} and {$^{232}$Th} concentration in the target LS were on the level of a few times $10^{-18}$ gram per gram LS. However, secular equilibrium was broken in several places along the {$^{238}$U} decay chain.  {$^{210}$Pb},  {$^{210}$Bi} and  {$^{210}$Po} levels proved to be comparable to or above the solar neutrino rates. In addition,  {$^{85}$Kr} was found to contribute sizably to the low energy spectrum \cite{Bellini:2013lnn}. Following up on first analyses of the {$^7$Be}, $pep$ and {$^8$B} (sects.~\ref{sec:be7bx},  \ref{sec:pepcno},  \ref{sec:b8leta}), an online purification campaign based on water extraction and nitrogen purging of the LS took place in 2011/12. As a result, background levels of the follow-up Phase II (from 2012) were significantly reduced, removing  {$^{85}$Kr}  almost completely and reducing the  {$^{210}$Pb}/{$^{210}$Bi}  rates \cite{Bellini:2014uqa}. The first direct measurement of $pp$ neutrinos based on Phase II data is described in sect.~\ref{sec:pp}.
\medskip\\
{\bf Calibration} of the detector response, i.e.~vertex position and energy reconstruction of events, is performed with weakly-radioactive sealed sources directly inserted into the target volume. A rod system featuring a rotatable, bendable arm can position the calibration source with $\sim$1\,cm accuracy inside the scintillator volume. A camera system is used for affirmation. For purposes of solar neutrinos, the calibration source contains a combination of several $\gamma$-emitters selected to cover the energy range from about 0.1 to 1.4\,MeV. The $\gamma$-rays deposit their complete energy in the scintillator, providing events at fixed energies for the calibration of energy and position reconstruction. In addition, a 2.6\,MeV {$^{208}$Tl} $\gamma$-source can be inserted into the outer buffer by a pipe system, permitting to calibrate spatial distribution and spectral shape of external backgrounds \cite{Back:2012awa}.

\subsubsection{The KamLAND experiment}
\label{sec:kl}

\begin{figure}[ht]
\centering
\includegraphics[width=0.6\textwidth]{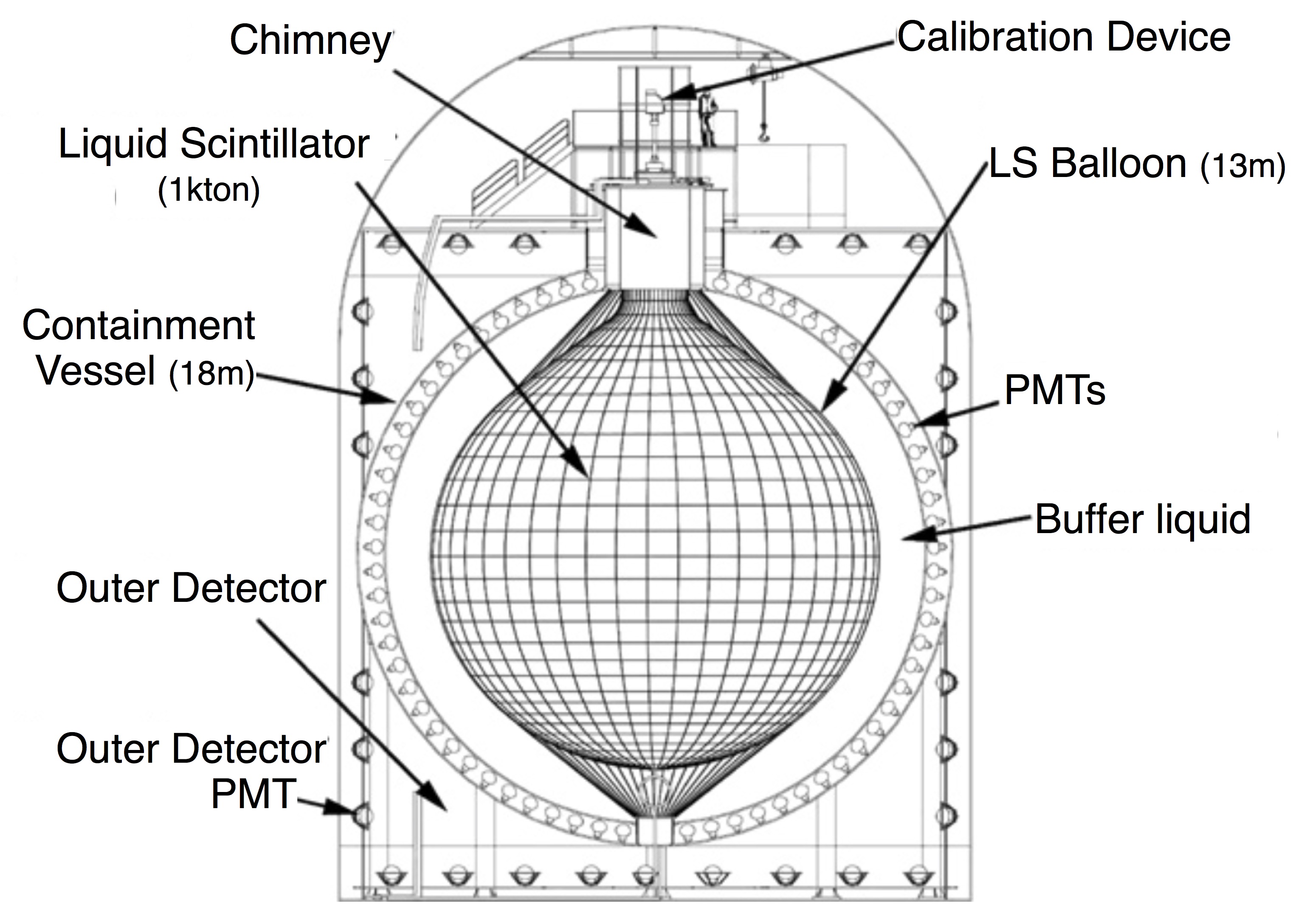}
\caption{Layout of the KamLAND detector \cite{Markoff:2003tg}.}
\label{fig:kllayout}
\end{figure}

The KamLAND (Kamioka Large Anti-Neutrino Detector) experiment started operation at the underground Kamioka Observatory in 2002 \cite{Markoff:2003tg}. The initial phase (2002$-$2007) of the experiment was dedicated to the search for long-baseline reactor antineutrino oscillations. Almost immediately, KamLAND discovered a deficit in the detected $\bar\nu_e$ rate \cite{Eguchi:2002dm}. In the follow-up, the experiment performed a measurement of the oscillation parameters $\theta_{12}$ and especially $\Delta m^2_{21}$, providing complementary information to the $\nu_e$ oscillation data from the solar neutrino experiments \cite{Capozzi:2016rtj}. After an analysis of high-energy {$^8$B} neutrinos (sect.~\ref{sec:b8kl}) and an extensive purification campaign of the target LS, KamLAND performed a two-year measurement (2009- 2011) of the solar {$^7$Be} neutrino flux \cite{Gando:2014wjd, Abe:2011em} that is described in section \ref{sec:be7kl}. In its most recent incarnation, the KamLAND-ZEN detector has loaded a sub-volume of the LS target with several 100\,kg of the xenon-isotope {$^{136}$Xe}, searching for the neutrino-less double-beta decay \cite{KamLAND-Zen:2016pfg}.
\medskip\\
{\bf Detector layout.} The basic structure of KamLAND very much resembles the Borexino de- tector (fig.~\ref{fig:kllayout}) \cite{Markoff:2003tg}: A stainless steel sphere of 18\,m diameter separates the inner LS detector (ID) from a surrounding water-Cherenkov veto. The neutrino target consists of 1\,kt of LS (see below), confined by a 135\,\textmu m thin spherical balloon of 13\,m diameter, made from several layers of nylon and EVOH (ethylene vinyl alcohol). The target is floating in an inactive buffer liquid, held in place by Kevlar ropes. The scintillation light is detected by 1879 20"-PMTs mounted to the inside of the steel sphere. For about 2/3 of the PMTs, the photo cathode is masked down to 17" diameter to improve timing and single photo-electron resolution.
The outer tank holds 3.2\,kt of water and is equipped with PMTs to provide a muon anti- coincidence veto for the ID. The muon rate traversing the ID is $\sim$0.2\,s$^{-1}$. Due to the lower rock shielding at Kamioka compared to the LNGS, cosmogenic background rates in KamLAND are on average a factor 3$-$4 higher than in Borexino (sect.~\ref{sec:bx}) \cite{Markoff:2003tg}.
\medskip\\
{\bf Liquid scintillator.} During its reactor antineutrino measurements and also later on in its solar phase, KamLAND has been using a three-component scintillator: The bulk material consists of a mixture of 20 volume percent pseudocumene (PC, as in Borexino) and 80\,\% Dodecane (C12), a normal paraffin featuring 12 carbon atoms. Although the addition of the C12 decreases the initial light yield by $\sim$20\,\%, it greatly improves the transparency of the LS, accommodating the large detector diameter, and increases the hydrogen content of the LS that is important for $\bar\nu_e$ detection via the inverse beta decay on protons. As in Borexino, 1.5\,g/l of PPO are added as a fluor.

The buffer liquid is composed of two paraffins, Dodecane (40\,\%) and Paraol-250 (60\,\%). This mixing ratio was chosen to reproduce the density of the target LS to mitigate buoyancy forces \cite{Markoff:2003tg}.
\medskip\\
{\bf Neutrino detection} is by elastic scattering on electrons. The photoelectron yield achieved is $\sim$300\,pe/MeV, corresponding to an energy resolution of (6.9$\pm$0.1)\,\% and a position resolution of $\sim$13\,cm at 1\,MeV that has been evaluated by calibrations (see below). In the initial antineutrino phase, the background levels from intrinsic radioactivity limited solar neutrino detection to the region above the spectral endpoint of  {$^{208}$Tl}  ($Q_{\beta\gamma} = 5.0$\,MeV). After the purification campaign, overall event rates were greatly reduced and the trigger level could be reduced from 180 to 70 coincident PMT hits, corresponding to an energy threshold of 0.4\,MeV \cite{Gando:2014wjd}.
\medskip\\
{\bf Background levels.} LS raw materials, nylon and all other inner detector components were selected for low radioactivity. However, the initial requirements for radiopurity were considering antineutrino detection only and were thus not sufficient for a low-energy solar neutrino measurement. In the first data taking phase, radioisotopes dissolved in the LS, primarily  {$^{85}$Kr} ,  {$^{210}$Bi}  and  {$^{210}$Po}  caused a background level of $8\cdot10^7$ decays per day, by far overwhelming the solar neutrino signal. Solar neutrino detection was thus confined to the {$^8$B} spectrum above a threshold of 5.5\,MeV caused by  {$^{208}$Tl}  decays (sect.~\ref{sec:b8kl}) \cite{Abe:2011em}.
In order to reach the required radiopurity for sub-MeV neutrino detection, two subsequent purification campaigns were carried out in the years 2007-2009. By distillation and nitrogen purging of the LS, a substantial improvement of the background conditions was achieved. Final rate reduction factors were between $6\cdot10^{-6}$ for  {$^{85}$Kr}  and $5\cdot10^{-2}$ for  {$^{210}$Po} . Despite this huge improvement, the signal-to-background ratio for {$^7$Be} and  {$^{210}$Bi}  $\beta$-decays was still at 1:4. This situation permitted a positive measurement of the {$^7$Be} flux although at lower accuracy than Borexino (sect.~\ref{sec:be7bx}) \cite{Gando:2014wjd}.
\medskip\\
{\bf Calibration} of energy and position resolution was performed in regular intervals lowering weakly radioactive $\gamma$-sources into the target LS along the central axis. After LS purification, an additional off-axis system was used to check for deviations of the detector response from rotational symmetry \cite{Gando:2014wjd}.

%
%

\section{Solar neutrino measurements}
\label{sec:solnumeas}

From the first evidence of neutrinos from solar fusion in the Homestake experiment \cite{Davis:1968cp} to the latest {$^8$B}-neutrino measurement in SK-IV \cite{Abe:2016nxk}, all solar neutrino experiments have added incremental amounts of information, often relying on earlier results for the best interpretation of their data. However, the current understanding of solar neutrino oscillations and the underlying neutrino spectrum relies mainly on the data of three experiments: For energies above 3.5\,MeV, the most accurate data is provided by the SK and SNO experiments (sects.~\ref{sec:sk}+\ref{sec:sno}). Below, the Borexino (sect.~\ref{sec:bx}) results dominate, by now equalling or surpassing the accuracy of all former measurements from radiochemical experiments.
In the following section, the spectroscopic measurements performed by the Water Cherenkov and LS detectors will be reviewed, giving primary attention to the results most relevant to flavor oscillation and SSM physics. Ordered by the spectral components investigated and proceeding from the highest ({$^8$B}) to lowest ($pp$) energies, the account follows roughly chronological order. A more global interpretation of the results will be given in section \ref{sec:status}.

\subsection{$^8$B-neutrinos}
\label{sec:b8}

First observed by the Homestake experiment (sect.~\ref{sec:homestake}), {$^8$B} neutrinos have been measured by all subsequent solar neutrino experiments. Based on the {$^8$B} neutrino flux, the presence of neutrino flavor oscillations was impressively demonstrated by the SNO experiment (sect.~\ref{sec:b8sno}). The combination of SNO and SK data provides the tightest constraints on the solar mixing angle $\theta_{12}$. To date, SK has performed the most accurate measurement of the shape of the {$^8$B} recoil spectrum and of the tiny day-night asymmetry in {$^8$B} neutrino oscillation probabilities expected from Earth matter effects (sect.~\ref{sec:b8sk}). In the energy range below 5\,MeV that is of special interest for oscillation physics (sect.~\ref{sec:statsolnuosc}), specific low-threshold analyses have been performed by both SNO and SK. However, the lowest energy data point at 3\,MeV electron recoil energy is provided by Borexino (sect.~\ref{sec:b8leta}). In 2011, KamLAND released a {$^8$B} result compatible with all earlier measurements (sect.~\ref{sec:b8kl}).

\subsubsection{SNO: Flavor conversion}
\label{sec:b8sno}

\begin{figure}[ht!]
\centering
\includegraphics[width=0.49\textwidth]{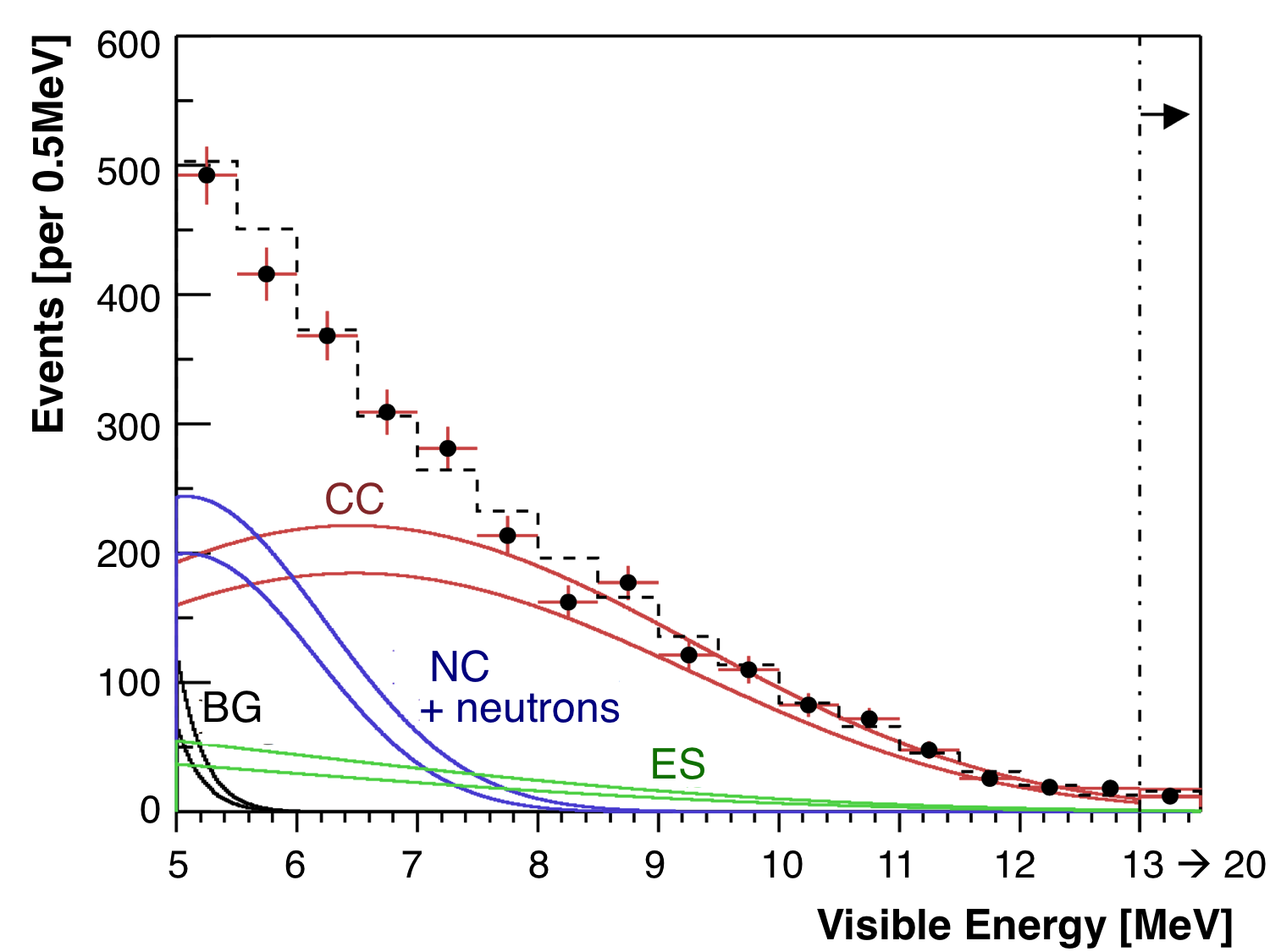}
\includegraphics[width=0.49\textwidth]{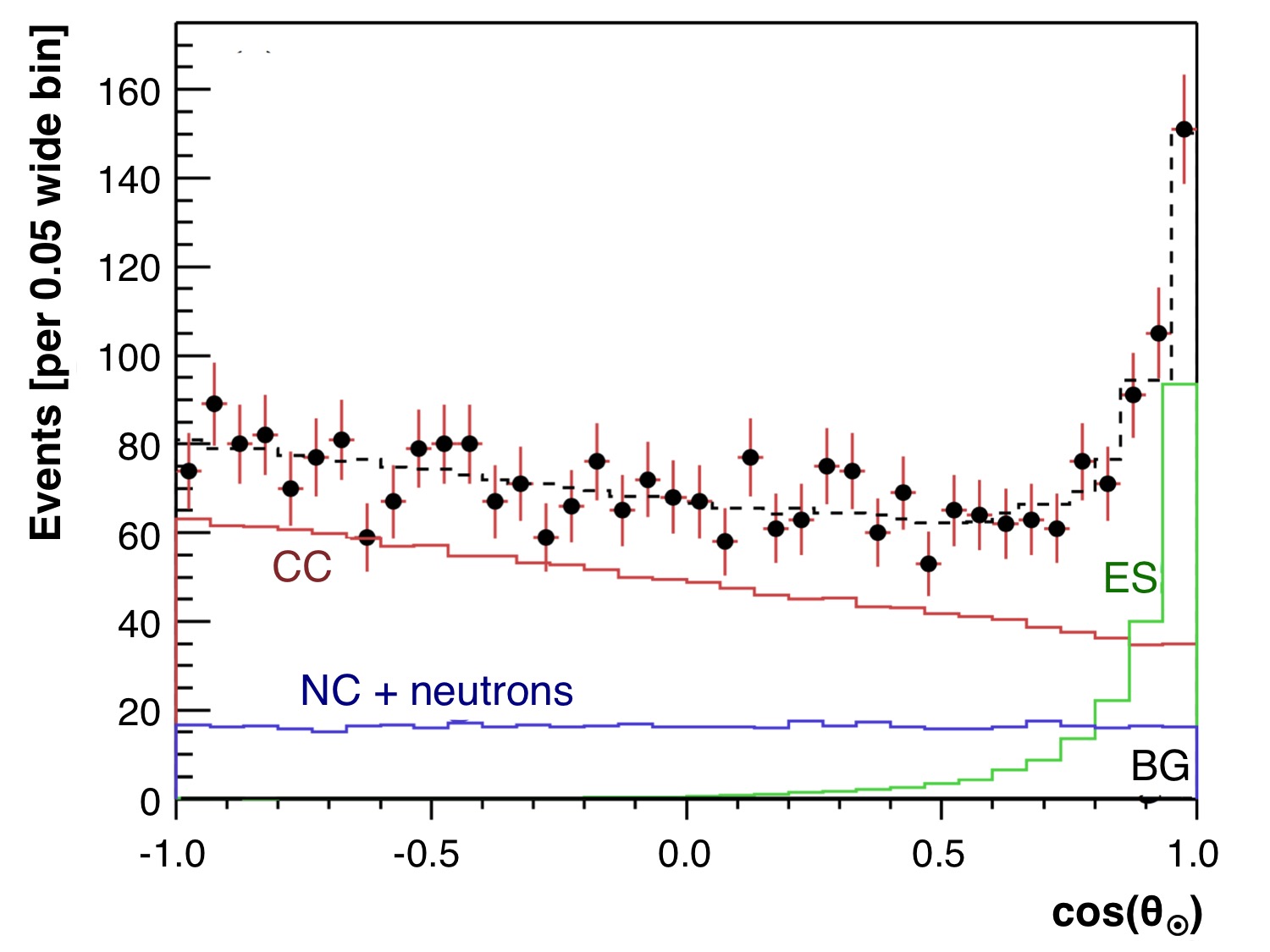}
\includegraphics[width=0.49\textwidth]{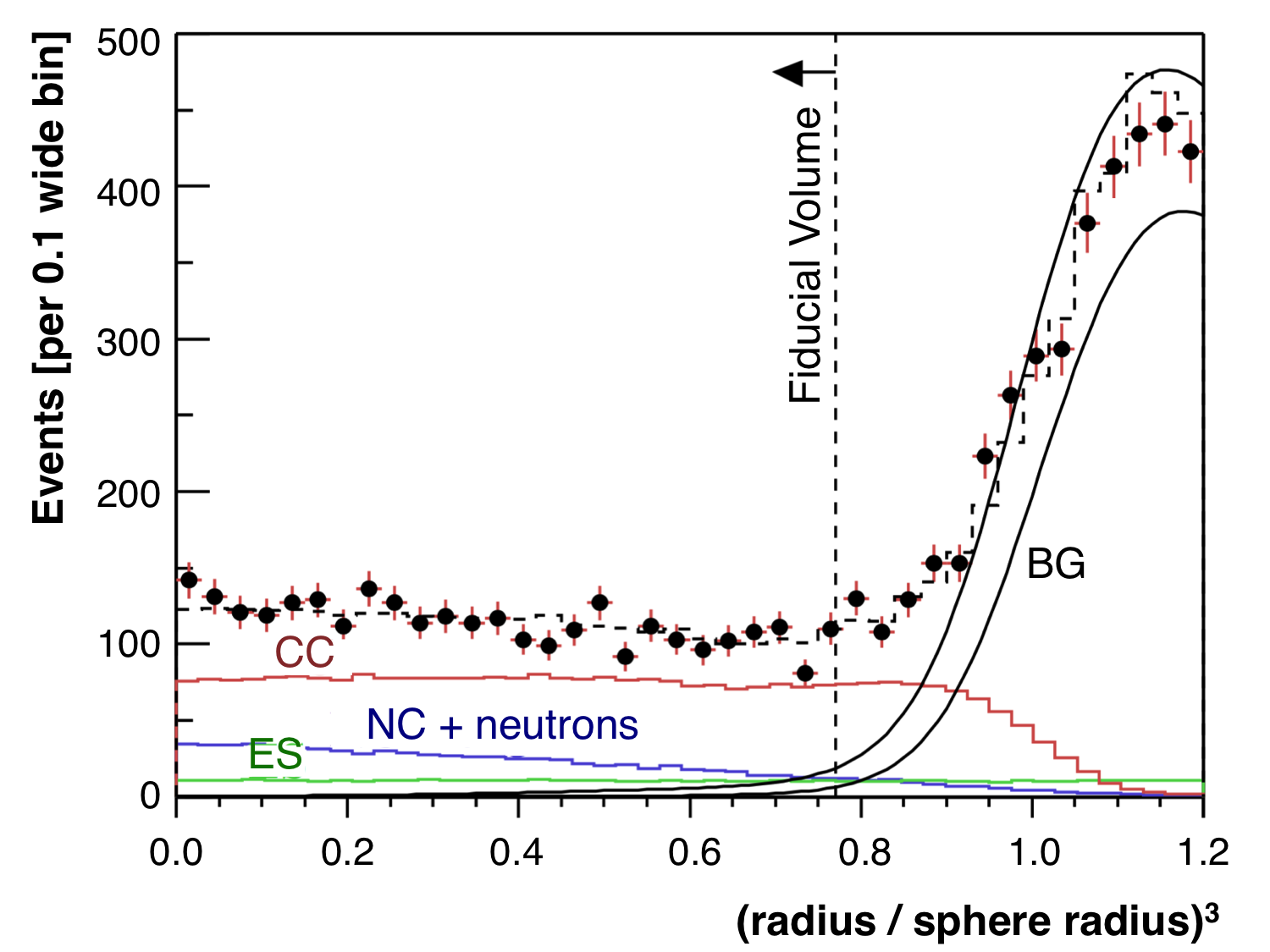}
\caption{Event distributions and Monte-Carlo fits to the energy spectrum, recoil direction and radial distributions of neutrino (CC,NC,ES) and background (BG) events in SNO-I \cite{Aharmim:2006kv}.}
\label{fig:snofit}
\end{figure}

The discovery of flavor conversion in the {$^8$B} neutrino flux by SNO is considered a mile stone of neutrino physics and one of the central building blocks for establishing neutrino flavor oscillations. The great novelty of the SNO measurement was the availability of three detection channels in heavy water sensitive to different neutrino flavors: Elastic scattering off electrons (ES) as well as charged current (CC) and neutral current (NC) reactions on deuterons. The latter permit to distinguish the $\nu_e$ component in CC reactions from the total {$^8$B} neutrino flux determined by the NC channel. In the following, the final analysis of the Phase-I data set is re- viewed \cite{Aharmim:2006kv}. A combined low-threshold analyses \cite{Aharmim:2009gd} of Phase-I and II data is shortly discussed in sect.~\ref{sec:b8leta}, while the final analysis of all three phases was published in \cite{Aharmim:2011vm}.
\medskip\\
{\bf Data analysis strategy.} During SNO's first data taking phase, the three event categories could not be distinguished on an event-by-event basis \cite{Aharmim:2006kv}. Instead, their respective contributions to the total solar neutrino signal in the detector were extracted based on a combined fit to the distributions of three main event parameters: the energy spectrum, the directional orientation of the Cherenkov cone, and the position distribution of the event vertices (fig.~\ref{fig:snofit}).

\noindent{\it Energy spectra.} Both ES and CC reactions produce continuous visible energy spectra corresponding to the final state electron recoil energies. While ES is strictly decreasing, CC features a maximum around 7\,MeV. The NC channel produces an energy line corresponding to the 6.25\,MeV $\gamma$-rays from neutron capture on deuterons (for phase I).
\medskip\\
{\it Direction.} Event kinematics dictate the orientation of the recoil-electron tracks: While in case of ES, the momenta of electrons are closely aligned with the incident neutrinos, an opposite (but weaker) dependence is found for the electrons from the CC reaction. The $\gamma$-rays emitted from neutron capture following NC are fully isotropic.
\medskip\\
{\it Radial distribution.} The vertices of neutrino interactions are uniformly distributed over the entire detection volume. However, the differences in detection efficiency can be exploited to distinguish the reaction channels: As expected, CC interactions are confined to the D2O target, while ES reactions extend uniformly in the light water region. On the other hand, the NC tagging efficiency drops towards the edge of the D2O volume as neutrons spill-out into the surrounding light water. The effect is pronounced as the diffusion length of thermal neutrons in heavy water is $\sim$1.2\,m and the neutron capture cross-section of {$^{1}$H} is much larger than in {$^{2}$H}. Neutron capture on {$^{1}$H} releases a merely 2.2\,MeV $\gamma$-ray below detection threshold.
\medskip\\
Figure \ref{fig:snofit} displays the distributions for these three parameters derived from data, as well as a multi-dimensional likelihood fit with the ES, NC and CC {\it pdf}'s derived from simulations.
\medskip\\
{\it Background rejection.} As laid out in section \ref{sec:sno}, several sources of background have to be taken into account: Events from external $\gamma$-rays were removed by defining a fiducial volume (FV) for neutrino detection at 5.5\,m radius. The otherwise overwhelming background from radioactive contaminants dissolved in the heavy water was suppressed by the lower energy threshold of 5\,MeV. Cosmogenic radioisotopes with high spectral endpoints were removed by a 20\,s time cut following each muon, introducing negligible dead time due to the exceptionally low muon rate of 3\,h$^{-1}$ at 6000\,mwe depth. The residual background rates were found to be 0.23\.d$^{-1}$ for neutrons from photo-disintegration of target deuterons and 0.15\,d$^{-1}$ for $\beta$-$\gamma$-events from the decay of  {$^{208}$Tl}  dissolved in the heavy water, both featuring uncertainties on the level of 0.04\,d$^{-1}$. These backgrounds were included in the fits shown in figure \ref{fig:snofit}, using {\it pdf}'s derived from the detector MC and constraining there amplitude based on auxiliary measurements of contamination levels.
\medskip\\
{\bf Results.} Based on the complete phase-I data set \cite{Aharmim:2006kv}, the equivalent fluxes observed in the three channels were
\begin{eqnarray}
\phi_{\rm CC} &=& 1.76^{+0.06}_{-0.05}{\rm(stat)}^{+0.09} _{-0.09}{\rm(syst)}\cdot 10^6\,{\rm cm}^{-2}{\rm s}^{-1}\nonumber\\
\phi_{\rm ES} &=& 2.39^{+0.24} _{-0.23}{\rm(stat)}^{+0.12} _{-0.12}{\rm(syst)}\cdot 10^6\,{\rm cm}^{-2}{\rm s}^{-1}\nonumber\\
\phi_{\rm NC} &=& 5.09^{+0.44} _{-0.43}{\rm(stat)}^{+0.46} _{-0.43}{\rm(syst)}\cdot 10^6\,{\rm cm}^{-2}{\rm s}^{-1}\nonumber
\end{eqnarray}
While $\phi_{\rm NC}$ is in good agreement with the SMM expectation from {$^8$B} neutrino production, $\phi_{\rm SSM}=5.05^{+1.01}_{-0.81} \cdot 10^6\,{\rm cm}^{-2}{\rm s}^{-1}$, both $\phi_{\rm CC}$ and $\phi_{\rm ES}$ are in comparison substantially suppressed. Using the known correlations between the cross-sections of the three channels, the corresponding neutrino flavor fluxes can be inferred:
\begin{eqnarray}
\phi_{e} &=& 1.76^{+0.05} _{-0.05}{\rm(stat)}^{+0.09} _{-0.09}{\rm(syst)}\cdot 10^6\,{\rm cm}^{-2}{\rm s}^{-1}\nonumber\\
\phi_{\mu,\tau} &=& 3.41^{+0.45} _{-0.45}{\rm(stat)}^{+0.48} _{-0.45}{\rm(syst)}\cdot 10^6\,{\rm cm}^{-2}{\rm s}^{-1},\nonumber
\end{eqnarray}
i.e.~roughly 2/3 of the original $\nu_e$ flux is converted to $\nu_{\mu,\tau}$ flavors, providing a natural explanation of the suppressed $\nu_e$ rates measured by all other solar neutrino experiments beforehand. This defining result was confirmed and refined in all the later phases \cite{Aharmim:2011vm}. Based on the full data set and the flux measurements in the three channels, the electron neutrino survival probability was determined to $P_{ee}^m=0.317\pm0.016_{(stat)}\pm0.009_{(syst)}$.

\subsubsection{Super-Kamiokande: Energy spectrum and day-night asymmetry}
\label{sec:b8sk}

The SNO result demonstrated the presence of flavor conversion in {$^8$B} neutrinos beyond doubt. However, the best measurement of the underlying mixing angle $\theta_{12}$ is obtained by combining the results of SNO with the very accurate electron-scattering (ES) measurement of the {$^8$B} neutrino flux by Super-Kamiokande (SK). SK has accumulated more than $7\cdot10^4$ ES events in its four data taking phases (sect.~\ref{sec:b8sk}). This high-statistics sample enables a very precise analysis of the {$^8$B} induced recoil spectrum at high-energies $E_{\rm vis} \geq 5$\,MeV. The result corresponds to the shape predicted by nuclear theory, which is in agreement with the flat course of $P^m_{ee}$ expected from the MSW-LMA solution at these energies \cite{Abe:2016nxk}.

\begin{figure}[ht!]
\centering
\includegraphics[width=0.5\textwidth]{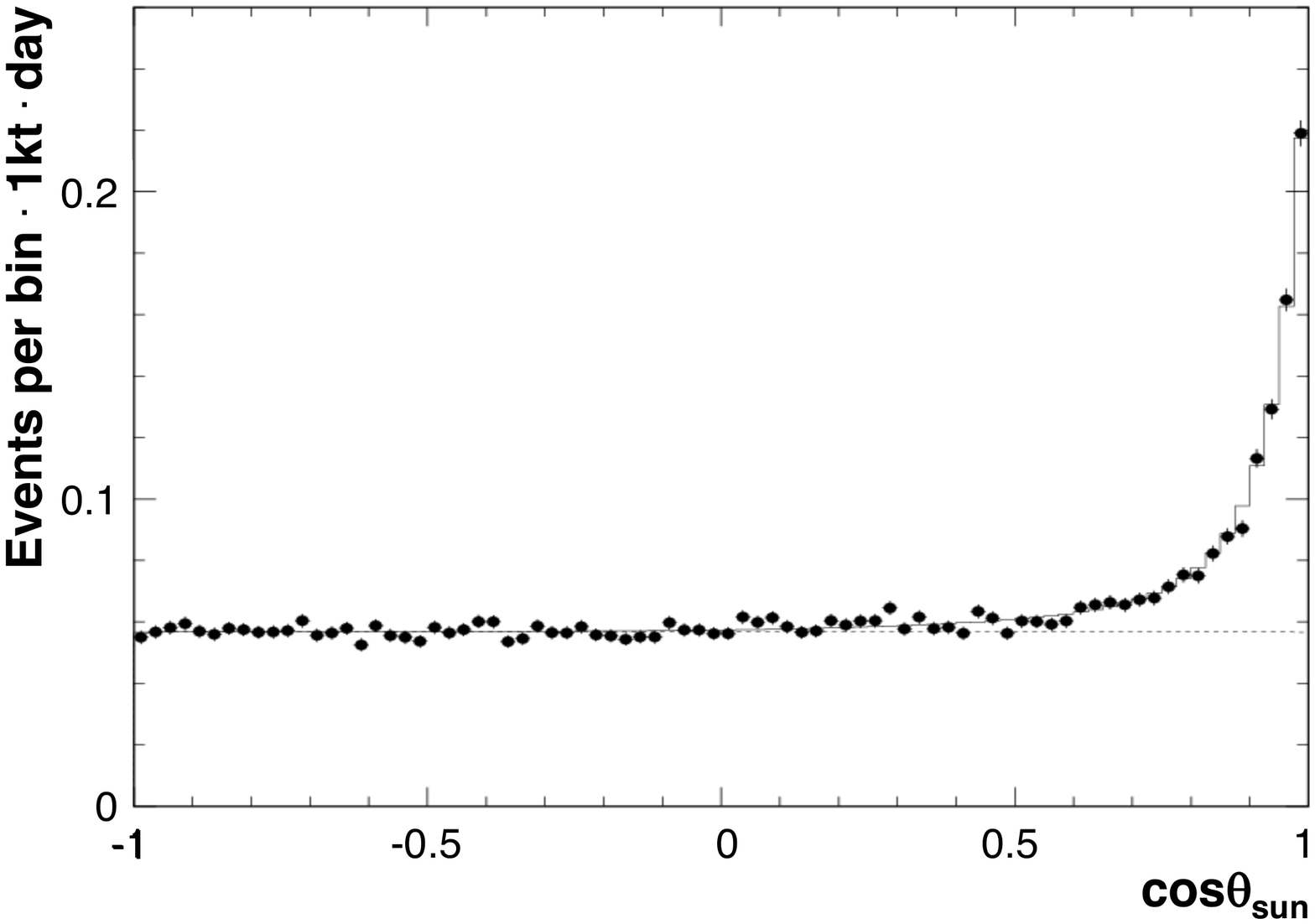}
\caption{The angular distribution of low-energy electron-like events acquired in SK-III \cite{Abe:2010hy}. The correlation with the Sun's position in the sky is used to discriminate aligned {$^8$B} recoil electrons from isotropic background.}
\label{fig:skangular}
\end{figure}

The large statistics offer as well the possibility to subdivide the data set into time bins, searching for variations in the signal rate: The slight eccentricity of the Earth's orbit is expected to induce a 6.6\,\% alteration in the neutrino rates of terrestrial detectors, the minimum expected in summer when the Earth reaches the aphelion. An annual modulation compatible with this has been observed by both SNO and SK \cite{Abe:2016nxk, Aharmim:2011vm}. More interesting from the point of view of neutrino oscillations is a diurnal modulation of the survival probability $P_{ee}$ introduced by terrestrial matter effects: {$^8$B} neutrinos crossing the Earth matter experience a regeneration effect increasing the effective value of $P_{ee}$. A corresponding analysis in SK provides a strong indication for the presence of this effect \cite{Abe:2016nxk}.
\medskip\\
{\bf Data analysis strategy.} As a light water detector, SK detects solar neutrinos solely by elastic scattering off electrons. The analysis proceeds in two steps \cite{Hosaka:2005um}: Firstly, a number of selection cuts is applied to reduce the background levels in the residual data set as far as possible. Sec- ondly, the remaining events are characterized by their vertex properties.
\medskip\\
{\it Rejection cuts.} To suppress background events arising from radioactive impurities in the surrounding detector materials, the SK analysis uses a dynamical fiducial volume cut that rejects all events reconstructed closer than 2\,m to the PMT photocathodes. In addition, the orientation of the reconstructed Cherenkov cones is taken into account, rejecting events featuring inward-pointing tracks even at further distances from the PMTs as they are more likely to result from external $\gamma$-rays. Moreover, the $\beta(\gamma)$-decays of  {$^{208}$Tl}  and  {$^{214}$Bi}  dissolved in the water dominate the recoil spectrum at low energies. As described in section \ref{sec:sk}, this background effectively set the detection threshold in SK-I to a minimum electron recoil energy of 6.5\,MeV, while the latest SK-IV achieved a threshold of 3.5\,MeV based on improved event identification algorithms (sect.~\ref{sec:b8leta}).
\medskip\\
{\it Cosmogenic background.} At energies above 6.5\,MeV, the dominating background sources are high-endpoint $\beta$-emitters induced by cosmic muon spallation on oxygen. The most important isotopes in this respect are  {$^{12}$B}  ($T_{1/2}=20$\,ms, $Q_\beta=$13.4\,MeV),  {$^{12}$N}  ($T_{1/2}=11$\,ms, $Q_\beta=17.3$\,MeV) and  {$^{16}$N}  ($T_{1/2}=7.1$\,s, $Q_\beta=$10.4\,MeV) \cite{Super-Kamiokande:2015xra}. As the SK site is relatively shallow, the cosmic muon rate in the detector amounts to 2\,s$^{-1}$. Therefore, A SNO-type veto of the whole detector for 20 seconds after each muon event is not practicable. Instead, additional information on the parent muon is exploited to assign a likelihood that a given high-energy electron event is of cosmogenic origin: Criteria used are the spatial distance to a potential parent muon track, the time elapsed since the passing of the muon and the muon's visible light output. An increased value marks the production of a hadronic shower inside the detector, corresponding to a larger probability that spallation isotopes were produced. Applying an event selection based on this likelihood removes a considerable fraction of cosmogenic background events while introducing only a $\sim$20\,\% loss in exposure \cite{Hosaka:2005um}.
\medskip\\
{\it Angular selection.} From the remaining event sample, the solar neutrino signal is extracted based on the angular distribution of the reconstructed Cherenkov cones. The tracks of recoil electrons from elastic neutrino scattering point away from the Sun, while the tracks of background events from radioactive decays are randomly orientated. Fig.~\ref{fig:skangular} displays the corresponding distribution for the third data taking phase SK-III \cite{Abe:2010hy}.

\noindent{\bf Results.} Based on the combined analysis of all data sets SK-I to SK-IV, the value of  $P_{ee}^m=0.334^{+0.027}_{-0.023}$, slightly higher than SNO and slightly worse in measurement precision \cite{Abe:2016nxk}. The spectral shape observed above 5\,MeV (and actually also at lower energies, sect.~\ref{sec:b8leta}) is fully consistent with the expectation from the {$^8$B} decay spectrum and thus with the constant expectation of $P^m_{ee}$ in this energy regime (fig.~\ref{fig:peeesplot}).

The day-night asymmetry of $P_{ee}$ has been investigated by grouping the data in time bins, evaluating the {$^8$B} neutrino event rates observed during day-time (D) and night-time (N). The selection is performed based on the position of the Sun relative to the horizon. The corresponding asymmetry factor has been determined to
\begin{eqnarray}\label{eq:daynight}
A=\frac{2(D-N)}{D+N} = -(3.3\pm1.0_{(stat)}\pm0.5_{(syst)})\,\%,
\end{eqnarray}
providing a $2.8\sigma$ indication for the presence of an asymmetry. As the amplitude of the asymmetry depends on Earth matter effects (sect.~\ref{sec:pee}), the most recent SK result sets by now the tightest limit on $\Delta m^2_{21}$ from solar neutrino oscillations \cite{Abe:2016nxk}. Note that the systematic uncertainty of this relative rate measurement is significantly smaller than that of the absolute {$^8$B} rate measurement as many normalization uncertainties can be omitted.

\subsubsection{Low-threshold analyses: SNO, SK, Borexino}
\label{sec:b8leta}

As illustrated in figure \ref{fig:solarpee}, the MSW-LMA solution predicts a slow transition from $P^{\rm vac}_{ee}$ to $P^m_{ee}$ in the energy range from 1 to 10\,MeV, the exact course depending on the values of $\theta_{12}$ and $\Delta m^2_{21}$ as well as on the radius of the neutrinospheres and the solar matter profile \cite{Smirnov:2003da}. As the transition region is especially sensitive to new physics (sect.~\ref{sec:statsolnuosc}), testing the exact dependence of $P_{ee}(E)$ is of great interest. However, the access to the {$^8$B} neutrino spectrum in this low energy range is very demanding: For a given electron recoil energy $T_e$, the contribution of lower-energy neutrinos $E_\nu \leq T_e$ will be minor because of the low d$\sigma$/d$T_e$ [eq.~(\ref{eq:diffx})]. Instead, the recorded recoil spectrum will be dominated by higher-energy neutrinos ($E_\nu > T_e$) undergoing forward-scattering. A determination of the neutrino spectral shape must thus rely on a spectral de-convolution of the energy-dependent cross-section \cite{Mollenberg:2014mfa}.

\begin{figure}[ht!]
\centering
\includegraphics[width=0.6\textwidth]{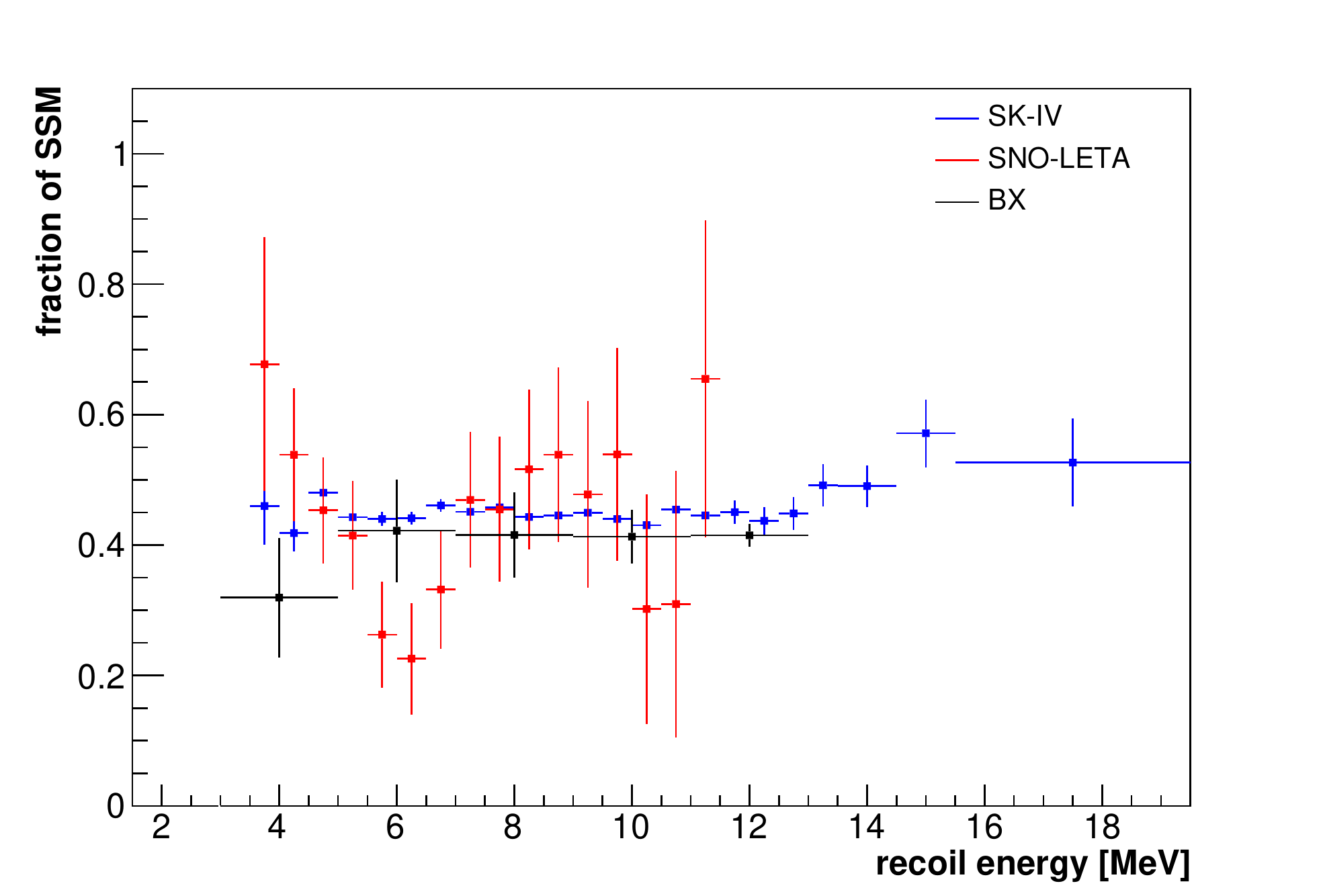}
\caption{Recoil energy spectra from the elastics scattering of {$^8$B} neutrinos as recorded by SNO, SK and Borexino \cite{Abe:2016nxk, Aharmim:2009gd, Bellini:2008mr}. Spectral bins are normalized to the SSM expectation of the {$^8$B} spectrum and show no significant deviation from a constant value of $P_{ee}$.}
\label{fig:peeesplot}
\end{figure}

When accessing the low-energy regime $E_{\rm vis} \approx T_e \leq 5$\,MeV, the radioactive contamination of the target material immediately plays a much more important role. Especially in water Cherenkov detectors low-energy background events from radioactive decays are smeared out to higher energies. Here, the low-threshold analyses of the SNO and SK experiments are shortly presented \cite{Abe:2016nxk, Aharmim:2009gd}, followed by a more elaborate description of the corresponding analysis of Borexino that for now provides the lowest-energy data point \cite{Bellini:2008mr}.
\medskip\\
{\bf SNO LETA.} The SNO Low-Energy Threshold Analysis (LETA) is based on the combined data sets of phases I and II \cite{Aharmim:2009gd}. The starting point is provided by the analysis presented in section \ref{sec:b8sno}: The signal channels are extracted based on a simultaneous fit of energy, radius and direction profiles of the events. Moreover, a fourth variable is included that reflects the isotropy of photon hit positions on the spherical surface that is formed by the PMTs. This is especially valuable in the salt phase-II of the experiment, where the new variable allows to discriminate the asymmetric hit distributions resulting from the Cherenkov cones of CC and ES events from the more isotropic photon pattern resulting from the $\gamma$-cascade caused by neutron capture on  {$^{35}$Cl} . Due to this additional handle, the spectral shapes of ES and CC channels can be partially unconstrained in the fit, providing sensitivity to the energy-dependence of $P_{ee}(E)$.

\begin{figure}[ht!]
\centering
\includegraphics[width=0.6\textwidth]{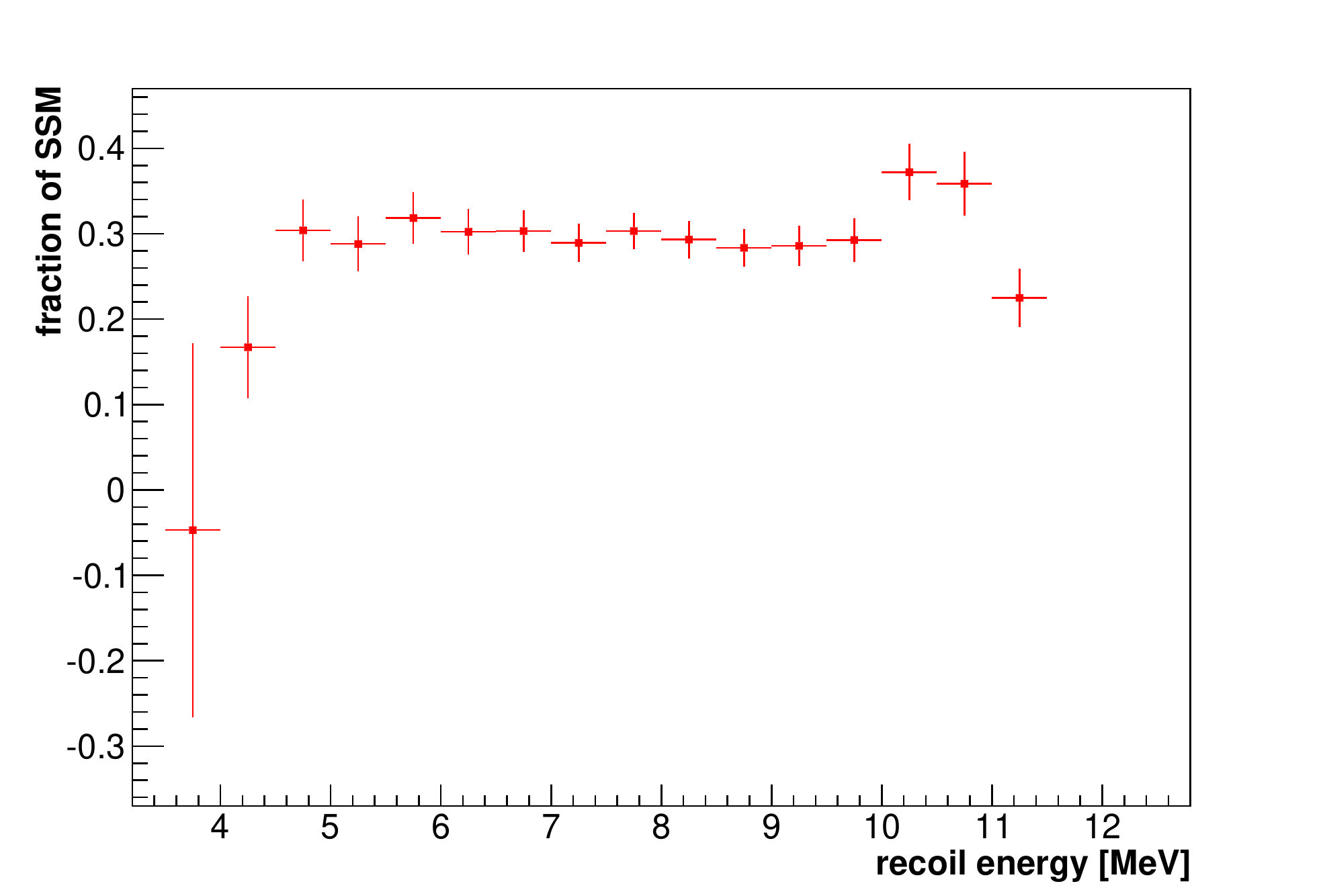}
\caption{Electron energy spectrum of the CC interactions of {$^8$B} $\nu_e$'s in SNO \cite{Aharmim:2009gd}. Spectral bins are presented as a fraction of the SSM expectation. The two lowest energy bins are somewhat low compared to a flat resp. rising expectation from the MSW-LMA model.}
\label{fig:peeccplot}
\end{figure}

Further improvements of the analysis encompassed the inclusion of new background components in the fit: internal ( {$^{214}$Bi}  and  {$^{208}$Tl} ) decays and external $\gamma$-rays; the energy reconstruction was refined to reduce the spectral overlap of signal and internal background spectra; an external background veto similar to that of the SK analysis (sect.~\ref{sec:b8sk}) as well as an energy-dependent fiducial volume cut were applied; and finally, runs known for high radon content in the heavy water target were removed from the data set.
As a consequence, the overall analysis threshold could be reduced to a visible energy of 3.5\,MeV. Furthermore, spectral shapes for both CC and ES channels were extracted from the data to determine $P_{ee}(E)$ from the comparison to the SSM prediction: The result for the ES spectrum that has been included in figure \ref{fig:peeesplot} is inconclusive, while the CC data shown in figure \ref{fig:peeccplot} show a moderate tension ($\sim2\sigma$) with the SSM prediction for the two lowest-energy data points. Note that the CC spectrum is more sensitive to low-energy neutrinos in the vacuum-matter transition region because in this case the electron recoil energy $T_e$ is directly related to the neutrino energy $E_\nu$ via $E_\nu = T_e + 1.44$\,MeV.
\medskip\\
{\bf SK-IV.} The solar analysis of the most recent Super-Kamiokande data taking phase IV includes an effort to lower the energy threshold for the detection of ES events \cite{Abe:2016nxk}. Several improvements in the experimental hardware and in the analysis of the data indeed allow for a lower energy threshold of now only 3.5\,MeV:

Firstly, the water circulation speed of the online purification system has been lowered to reduce the entry of radioactive radon and thus the amount of low-energy background in the target volume. This mostly regards the $\beta$-decay of  {$^{214}$Bi}  ($Q_\beta=2.8$\,MeV). that is smeared out to visible energies of up to 5 MeV due to the comparatively poor energy resolution.

Secondly, a new discrimination technique has been developed that permits the suppression of  {$^{214}$Bi}  based on the acquired signal shape: Based on an upgrade of the front-end electronics that acquire now the full signal waveforms of the PMTs, the new method enables the discrimination of upward-fluctuated low-energy background and larger-energy neutrino events based on the fuzziness of the reconstructed Cherenkov ring: Generally speaking, the lower the energy of a $\beta$-like event, the larger the chance for multiple scattering of the decay electron on electrons in the water target and thus the more pronounced the fuzziness of the ring. This basic property persists even if an upward-fluctuation of photon hits caused by Poissonian statistics leads to a considerably larger visible energy. The new technique is successfully employed to discriminate $\nu$-induced electron recoils from  {$^{214}$Bi}  $\beta$-decays.
Based on the these improvements, an extraction of the directional forward-scattering peak of ES recoil electrons becomes possible even at energies as low as 3.5\,MeV (sect.~\ref{sec:b8sk}) \cite{Abe:2016nxk}. The resulting ES spectrum is included in figure \ref{fig:peeesplot}.
\medskip\\
{\bf Borexino.} Finally, also Borexino has performed a measurement of the {$^8$B} neutrino spectrum based on ES off electrons. Its liquid scintillator target provides two advantages compared to water: The substantially greater light yield provides a sizable and easy to detect signal even for the low-energy range of the {$^8$B} spectrum, allowing for a better reconstruction of the event characteristics. Moreover, the neutrino target and surrounding buffer volumes feature substantially lower radioactive background levels. As a consequence, the Borexino {$^8$B} analysis features the lowest energy threshold of only 3.0\,MeV. On the other hand, an important disadvantage of Borexino is the relatively small target size of the detector: In \cite{Bellini:2008mr}, the spherical fiducial volume (FV) is defined by a radial cut at 3\,m, leaving only the innermost 100 tons for neutrino analysis. The measurement accuracy is thus limited primarily by the comparatively low event statistics.

Inside the FV, the lower detection threshold is imposed by external background: The $\gamma$-rays from the $\beta\gamma$-decay of  {$^{208}$Tl}  in the PMT glass feature an energy of 2.6\,MeV and are thus the most penetrating component of external background. The self-shielding by buffer and outer scintillator layers is not sufficient to reduce this background to a rate lower than the expected {$^8$B} ES signal. Taking into account a modest energy smearing, the ES detection threshold has thus been set to $T_e > 3.0$\,MeV \cite{Bellini:2008mr}.

Table \ref{tab:b8} lists the residual signal and background rates for two energy ranges: Once above the detection threshold of 3\,MeV, once for $T_e \geq 5$\,MeV for comparison with earlier SNO, SK results (sects.~\ref{sec:b8sno}+\ref{sec:b8sk}). The main background between 3 and 5\,MeV is the $\beta\gamma$-decay of  {$^{208}$Tl}  ($Q_{\beta\gamma} = 5.0$\,MeV) dissolved in the target scintillator. As  {$^{208}$Tl}  is in itself a decay product of  {$^{220}$Rn} , the rate has been determined {\it in-situ} based on the decay of its immediate parents 2 {$^{12}$B} i and 212Po, which can be tagged via a fast coincidence ($\tau(^{212}{\rm Po}) = 431$\,ns). Based on this indirect rate information, the  {$^{208}$Tl}  background spectrum can be statistically subtracted from the ES signal.

At higher energies, cosmic muons and their spallation products pose the most dangerous background: Muons crossing the Inner Detector (ID) in the inactive buffer region produce a sizable Cherenkov signal that overlaps with the {$^8$B} signal spectrum. However, this direct muon background is efficiently suppressed by the anti-coincidence veto with the Outer Detector and pulse shape discrimination in the ID. Following each muon identified, a 6.5-second veto of the full detection volume is applied, thereby removing $\gamma$-ray background from captures of spallation neutrons on carbon ($E_\gamma \sim 4.5$\,MeV) and the decays of a number of short-lived spallation isotopes (e.g.~{$^{12}$B}  with $Q_{\beta} = 13.4$\,MeV). Finally, there are two long-lived isotopes with end-points above threshold:  {$^{10}$C}  ($Q_{\beta^+} = 3.6$\,MeV, $\tau = 28$\,s) is efficiently rejected by a three-fold-coincidence veto based on the signals of the muon, the capture of one or two neutrons knocked out from  {$^{12}$C}  to produce  {$^{10}$C} , and the $\beta^+$ decay itself. {$^{11}$Be} ($Q_{\beta^-} =11.5$\,MeV, $\tau = 20$\,s) is estimated based on a fit to the time-correlation with the preceding parent muons and again subtracted statistically \cite{Bellini:2008mr}. The remaining signal and background rates are displayed in table \ref{tab:b8}.
\medskip\\
{\bf Results.} The {$^8$B} ES recoil spectra of SNO, SK-IV and Borexino are presented in figure \ref{fig:peeesplot}. In addition, SNO provides a measurement of the pure $\nu_e$ spectrum based on the CC reaction channel on deuterons (fig.~\ref{fig:peeccplot}). In both plots, the spectra are presented as fractions of the SSM prediction. While figure \ref{fig:peeesplot} shows an approximately constant fraction of $\sim$0.45 for ES, the CC measurement returns $\sim$0.3 including an apparent decrease for the lowest energy bins. These findings are in moderate ($2\sigma$) tension with the predicted course of $P_{ee}(E)$ in the MSW-LMA scenario that would favor a spectral upturn. Possible implications will be discussed in section \ref{sec:statsolnuosc}.

\begin{table}
\begin{center}
\begin{tabular}{lccc}
\hline
 & \multicolumn{3}{c}{Rate [(d$\cdot$kt$^{-1}$)]}\\
Contribution & Borexino (3\,MeV) & Borexino (5\,MeV) & KamLAND (5.5\,MeV) \\
\hline
External $\gamma$-rays & $(6.4\pm0.2)\cdot10^{-2}$ & $(3\pm11)\cdot10^{-5}$ & $(2.0\pm1.0) \cdot 10^{-1}$ \\
 {$^{208}$Tl}  & $(8.4\pm0.2)\cdot10^{-2}$ & 0 & 0 \\

Muons & $(4.5\pm0.9)\cdot10^{-3}$ & $(.53\pm0.8)\cdot10^{-3}$ & $-$ \\

Fast cosmogenic & $(1.7\pm0.2)\cdot10^{-2}$ & $(1.3\pm0.2)\cdot10^{-2}$ & $(3.3\pm4.9) \cdot 10^{-3}$ \\

{$^{11}$Be} & $(3.2\pm0.6)\cdot10^{-1}$ & $(2.3\pm0.4)\cdot10^{-1}$ & $(7.2\pm1.6) \cdot 10^{-1}$ \\
 {$^{10}$C}  & $(2.2\pm0.2)\cdot10^{-2}$ & 0 & 0 \\
 {$^{8}$Li}  & $-$ & $-$ & $(1.7\pm0.3) \cdot 10^{-1}$ \\
\hline
Total background & $1.27\pm0.63$ & $0.25\pm0.44$ & $1.28\pm0.19$ \\
{$^8$B} $\nu$ES signal & $2.17\pm0.38\pm0.08$ & $1.34\pm0.22\pm0.08$ & $1.49\pm0.14\pm0.17$ \\
\hline
\end{tabular}
\caption{Signal and background rates for the {$^8$B} neutrino analyses in Borexino and KamLAND. The signal rates include both statistical and systematic uncertainties \cite{Bellini:2008mr, Abe:2011em}.}\label{tab:b8}
\end{center}
\end{table} 

\subsubsection{KamLAND analysis}
\label{sec:b8kl}

\begin{figure}[ht!]
\centering
\includegraphics[width=0.55\textwidth]{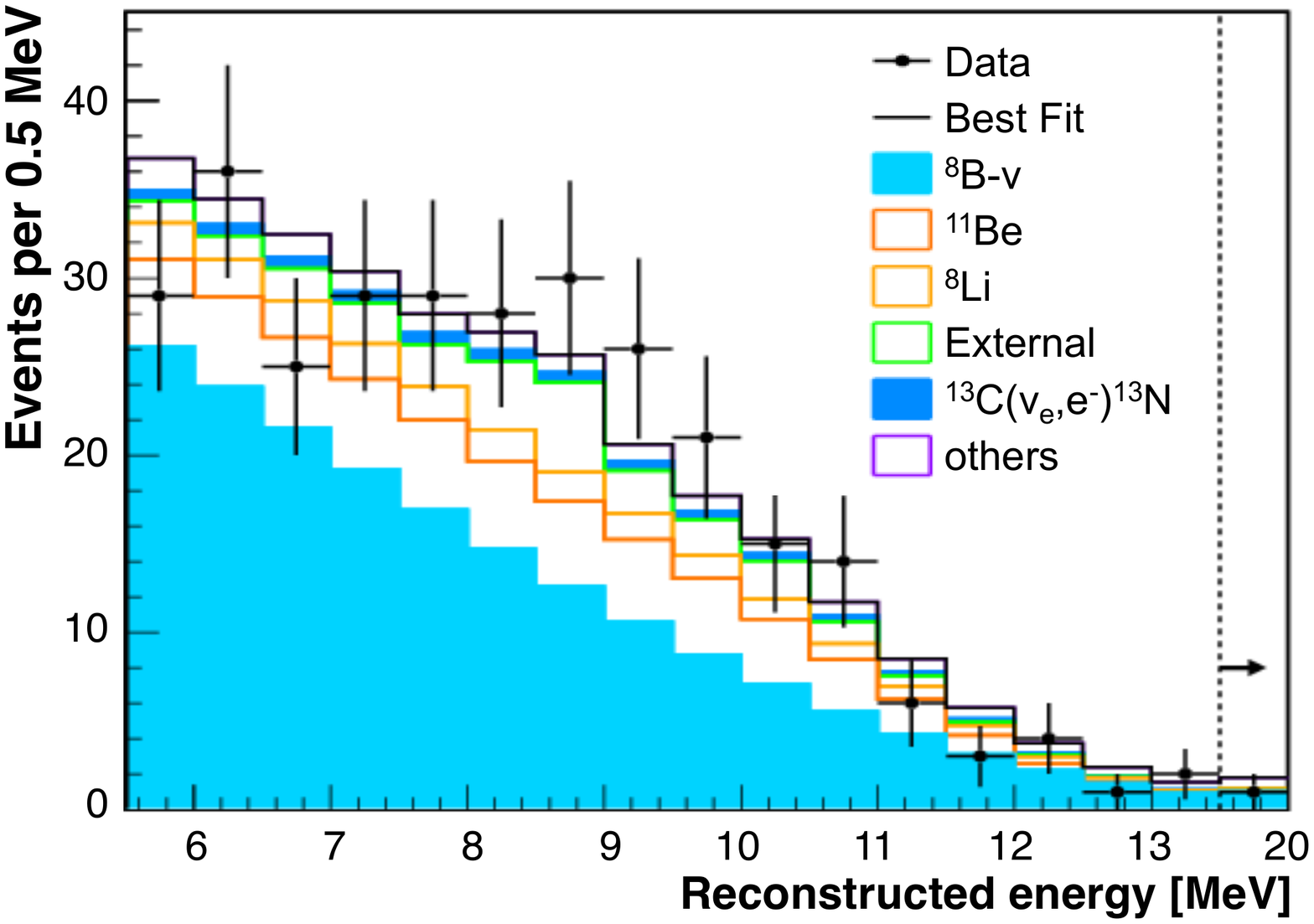}
\caption{Signal and background spectra in the {$^8$B} neutrino energy region above 5\,MeV as reported by KamLAND \cite{Abe:2011em}.}
\label{fig:kamlandb8}
\end{figure}

In 2011, the KamLAND collaboration published an analysis of the ES signal from {$^8$B} neutrinos based on the first reactor antineutrino data taking phase (2002$-$2007), corresponding to an overall run time of 1432 days \cite{Abe:2011em}.
\medskip\\
{\bf Analysis overview.} The general analysis strategy is quite similar to that of Borexino de- scribed in section \ref{sec:b8leta}. As background conditions in KamLAND are less favorable for ES detection than in Borexino, harder analysis cuts are required to obtain a data sample of com- parable quality.

For instance, the considerably larger external background requires the definition of a tight fiducial volume (FV): A software cut limits the analysis to a cylindrical volume of 6\,m in height and diameter aligned to the center of the detector, corresponding to a target mass of 130\,t. Due to a substantial contamination of the liquid with  {$^{208}$Tl} , the analysis threshold was set to 5.5\,MeV. Finally, the lower rock shielding at Kamioka results in a substantially higher cosmic muon flux and thus enhanced production rates for cosmogenic radioisotopes {$^{11}$Be},  {$^{8}$Li}  and {$^8$B}. The veto based on time-coincidences between parent muon and daughter decay must such provide higher rejection efficiency, resulting in greater life time loss. The residual background and signal rates in KamLAND are listed in table \ref{tab:b8}, along with the corresponding values of the Borexino analysis.
\medskip\\
{\bf Result.} The {$^8$B} signal rate of $1.49 \pm 0.14_{(stat)} \pm 0.17_{(syst)}$ events per day and kiloton has been determined based on a likelihood fit to the signal and background energy spectra that is shown in figure \ref{fig:kamlandb8} \cite{Abe:2011em}. While the statistical uncertainty on rate and flux is smaller than in the case of the Borexino result (sect.~\ref{sec:b8leta}), the systematic uncertainty is in fact larger. As a consequence, the accuracy of both analyses is roughly on par in the energy range above 5.5\,MeV.

\subsection{$^7$Be-neutrinos}
\label{sec:be7}

In the 1990's, a measurement of the {$^7$Be} $\nu_e$ flux was considered as a possible key to reveal oscillations as the source of the solar neutrino problem. The relatively low value of $P_{ee} \approx (0.33 \pm 0.06)$ measured by the Chlorine Experiment (sect.~\ref{sec:homestake}) as compared to $P_{ee} \approx (0.60 \pm 0.06)$ provided by GALLEX and SAGE (sect.~\ref{sec:gallium}) hinted at a strong suppression of $P_{ee}$ at the energy of the {$^7$Be}-I line (862\,keV) \cite{Bahcall:1999cb}. An almost complete conversion of $\nu_e \to \nu_{\mu,\tau}$ by resonant matter oscillations was in fact predicted by the Small Mixing Angle (SMA) solution that preferred a somewhat lower $\Delta m^2_{21}$ value than the LMA solution but a $\theta_{12}$ value of merely $\sim5^\circ$ \cite{GonzalezGarcia:1999aj}. Therefore, a measurement of the {$^7$Be} line was set as a primary scientific objective of Borexino \cite{Ranucci:1992dx}.

After SNO, SK and KamLAND had provided clear evidence in favor of the LMA solution (sect.~\ref{sec:b8sno}), the expectation for the {$^7$Be} rate was drastically increased because the neutrino energy now corresponded to the regime of vacuum oscillations (sect.~\ref{sec:oscmat}). In the following, both Borexino and KamLAND performed a measurement of the {$^7$Be} flux to obtain a result for $P^{\rm vac}_{ee}$. Beyond oscillations, an accurate {$^7$Be} measurement also of interest for the determination of solar metal abundances (sect.~\ref{sec:statsolcomp}).

\subsubsection{Borexino}
\label{sec:be7bx}

\begin{figure}[ht]
\centering
\includegraphics[width=\textwidth]{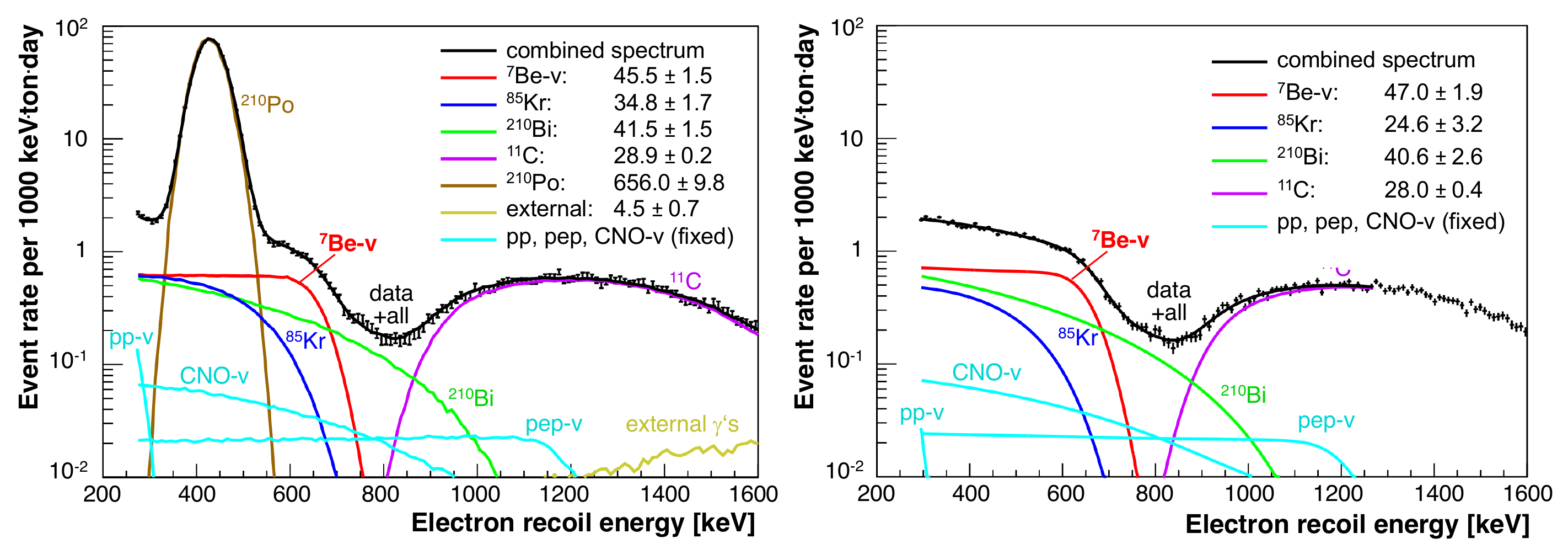}
\caption{Electron recoil spectra in the energy region of interest for {$^7$Be} neutrino detection in Borexino.The solid lines represent the fit spectra of neutrino signal and background components to the overall spectrum (data points). While the left panel shows the full data spectrum inside the fiducial volume, the right demonstrates the efficient removal of $\alpha$-decays from {$^{210}$Po} by pulse shape discrimination \cite{Bellini:2013lnn}.}
\label{fig:be7spectrum}
\end{figure}

In 2007, only three month after the start of data taking, the Borexino collaboration was able to report the positive detection of the {$^7$Be} neutrino line \cite{Arpesella:2007xf}. This early success was achieved because of the excellent radiopurity of the liquid scintillator (LS) that substantially exceeded the prior specifications. Thus, the distinct shoulder in the visible energy spectrum caused by the Compton-like electron recoils (fig.~\ref{fig:solarnuspectrum}) became immediately apparent.
\medskip\\
{\bf Selection cuts.} Figure \ref{fig:be7spectrum} displays the visible energy spectrum in the region of interest for the entire data set of Phase I, corresponding to 741 live days \cite{Bellini:2013lnn}. To obtain this spectrum, a number of selection cuts are applied to the raw data: The most relevant are the rejection of muon events via the anti-coincidence veto with the Outer Detector\footnote{and pulse-shape discrimination}, removal of fast coincident events originating from radioactivity, and a fiducial volume cut at 3\,m radius that rejects the external background events dominant in the outer layers of the LS target (plus stricter cuts on the upper and lower end caps). The analysis is thus restricted to the innermost $\sim$70 tons of the LS target. In addition, pulse shape discrimination can be applied that removes the residual background from  {$^{210}$Po}  $\alpha$-decays from the spectrum of otherwise $\beta$-like signal and background events.
\medskip\\
{\bf Spectral fit.} Figure \ref{fig:be7spectrum} displays the visible energy spectrum in the region of interest from 0.2 to 1.6\,MeV, once with and once without applying pulse shape discrimination: The residual background levels are sufficiently low to directly identify the spectral excess created by the {$^7$Be}-induced electron recoil shoulder. The exact ES rate from {$^7$Be} neutrinos has been determined by a spectral fit to the data, separating the contributions of signal and several classes of residual background events. The underlying spectra have been obtained from the nuclear data and theory, adding the energy resolution of the detector. In case of figure \ref{fig:be7spectrum}, the detector response function has been modelled by Monte Carlo, while other analyses use an analytic description \cite{Bellini:2011rx}.
\medskip\\
{\bf Rate result.} In figure \ref{fig:be7spectrum}, the fit results for all spectral contributions are indicated by the colored solid lines\footnote{Note that the neutrino fluxes from other spectral branches have been set to their standard SSM value, including the known oscillation parameters. However, there contribution is clearly subdominant in the region of interest.} \cite{Bellini:2013lnn}. While the dominant background contribution in the {$^7$Be} energy range arises from the $\alpha$-decay of  {$^{210}$Po} , its impact on the precision of the measurement is minor due to the distinctive difference in spectral shape. More problematic are the $\beta^-$-emitters  {$^{210}$Bi} and {$^{85}$Kr} that both contribute at visible energies around the {$^7$Be} shoulder. However, both backgrounds are well constrained by the specific spectral shapes.

From the fit, the {$^7$Be} ES rate has been determined to $46.0\pm1.5(stat)^{+1.5}_{-1.6}(syst)$ per day and 100 ton. The systematic uncertainties were derived from varying the definition of the FV, the fit ranges and the energy resolution within the boundary conditions obtained from calibration. This rate result corresponds to a {$^7$Be} neutrino flux of $(2.78\pm0.13)\cdot10^9$\,cm$^{-2}$s$^{-1}$ when assuming a pure $\nu_e$ flux. Interpreting this flux deficit in terms of flavor oscillations, the comparison to the SSM prediction yields a survival probability $P^{\rm vac}_{ee} = 0.51 \pm 0.07$ \cite{Bellini:2011rx}.
\medskip\\
{\bf Day-night variations.} Borexino performed several analyses searching for a possible day-night modulation in the {$^7$Be} ES rate \cite{Bellini:2011yj}. The most sensitive method does not rely on direct spectral fits to the day- and night-time energy data but relies on their spectral subtraction: The corresponding data sets are classified by the zenith position of the Sun above or below the horizon and scaled for the relative exposure times. 

\begin{figure}[ht]
\centering
\includegraphics[width=0.6\textwidth]{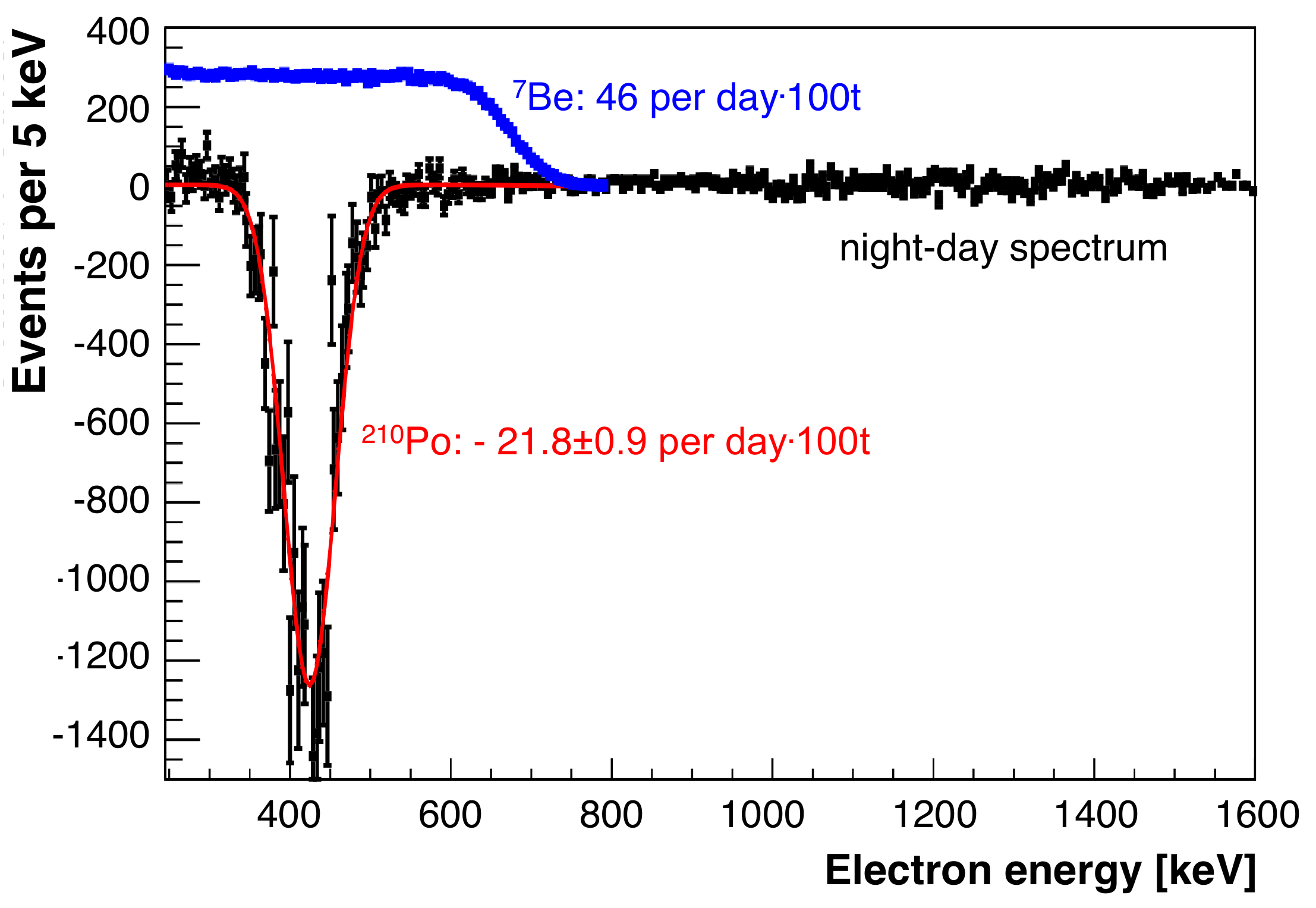}
\caption{(Absolute) spectral difference between day- and night-time data sets in the energy region of {$^7$Be} neutrinos measured in Borexino: Apart from an over-representation of  {$^{210}$Po}  background events during day-time. A fit of the  {$^{210}$Po}  ({\it red line}) and {$^7$Be} ({\it blue line, not scaled}) spectra to the residual difference returns a negligible {$^7$Be} residual rate, leading to a tight-limit on a possible day-night rate asymmetry \cite{Bellini:2011yj}.}
\label{fig:bxdnas}
\end{figure}

The resulting (absolute) night$-$day difference spectrum is displayed in figure \ref{fig:bxdnas}:  The observed energy distribution is flat, apart from a striking asymmetry at low energies that is introduced by the slowly decaying  {$^{210}$Po}  background that was highest during the long days of the summer months following the initial filling of Borexino. All other background sources are expected to be virtually stable over time. A possible {$^7$Be} rate asymmetry is extracted by a spectral fit to this difference spectrum in the energy range from 250 to 800\,keV, that includes both the Be7 shoulder and the  {$^{210}$Po}  peak.

The fit returns a {$^7$Be} rate difference of $R_{\rm diff} = 0.04 \pm 0.57_{(stat)}$ per day and 100\,ton, corresponding to a residual asymmetry value of $A = 0.001 \pm 0.012_{(stat)} \pm 0.007_{(syst)}$ [eq.~(\ref{eq:daynight})] that is well compatible with zero modulation. This is in agreement with the predictions of the MSW-LMA solution, for which the Earth-matter $\nu_e$ regeneration effect at {$^7$Be} energies is much less pronounced than for {$^8$B} neutrinos (sect.~\ref{sec:b8sk}). However, the findings are not compatible with the large asymmetry expected in case of the LOW oscillation scenario which can therefore be rejected \cite{Bellini:2011yj}.

\noindent{\bf Annual modulation.} In 2017, the Borexino collaboration published positive evidence for the presence of an annual modulation in the detected {$^7$Be} rate that is induced by the variation of Earth's distance from the Sun \cite{Agostini:2017iiq}. The analysis was based on a 4-year data set (2011$-$2015) from Borexino Phase-II in which the stability and signal-to-background ratio was significantly improved compared to Phase-I. The time bins of such a modulation analysis are of the order of 20 days and such too short to apply a spectral fit to extract the {$^7$Be} rate. Instead, the analysis was performed on a $\beta$-like event sample selected by a fiducial volume and tight energy (215$-$715\,keV) cuts, subtracting  {$^{210}$Po}  $\alpha$-background based on an improved pulse shape discrimination using a Multi-Layer Perceptron machine learning algorithm and correcting for the residual background rate of  {$^{210}$Bi}  $\beta$-decays included in the selected data.    

\begin{figure}[ht]
\centering
\includegraphics[width=0.55\textwidth]{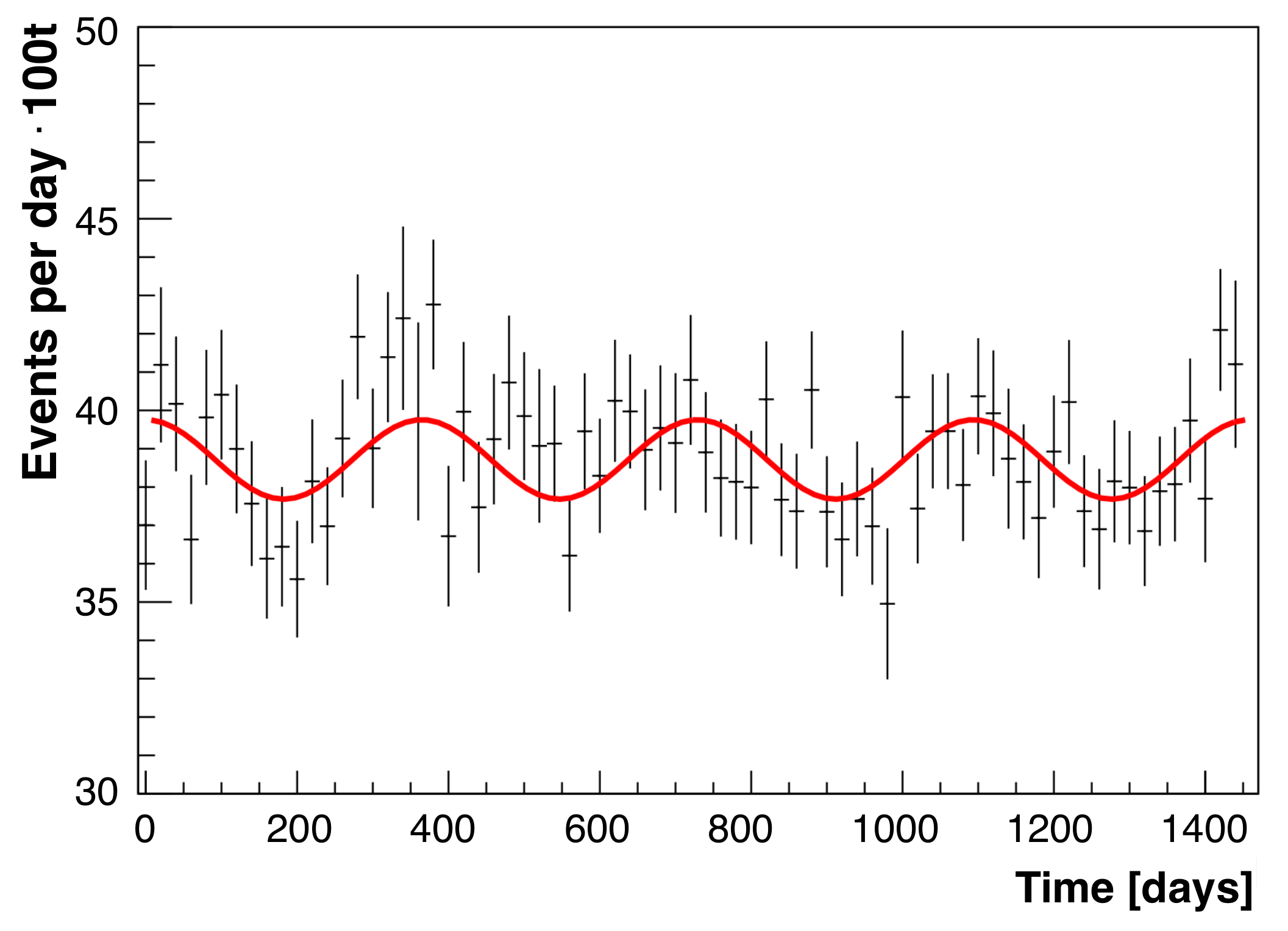}
\caption{The {$^7$Be} interaction rate as a function of time during Borexino Phase-2. The sinusoidal fit applied to the date ({\it red line}) is in perfect agreement with the annual modulation of the solar neutrino flux induced by the eccentricity of the terrestrial orbit \cite{Agostini:2017iiq}.}
\label{fig:bxam}
\end{figure}

The presence of an annual modulation in the resulting rate data was validated by a sinusoidal fit (fig.~\ref{fig:bxam}), based on the Lomb-Scargle method and the Empirical Mode Decomposition technique: The absence of an annual modulation is rejected with a confidence level of 99.99\,\%. The direct fit to the event rate returns a modulation amplitude corresponding to an eccentricity of $(1.66\pm0.45)$\,\% of the terrestrial orbit. Moreover, the Lomb-Scargle periodogram features a clear spectral maximum at 1-year modulation period \cite{Agostini:2017iiq}. This finding is a direct confirmation of the solar origin of the observed {$^7$Be} signal, comparable to the solar peak in the angular distribution of ES events in Water Cherenkov Detectors. 

\subsubsection{KamLAND}
\label{sec:be7kl}

\begin{figure}[ht]
\centering
\includegraphics[width=0.55\textwidth]{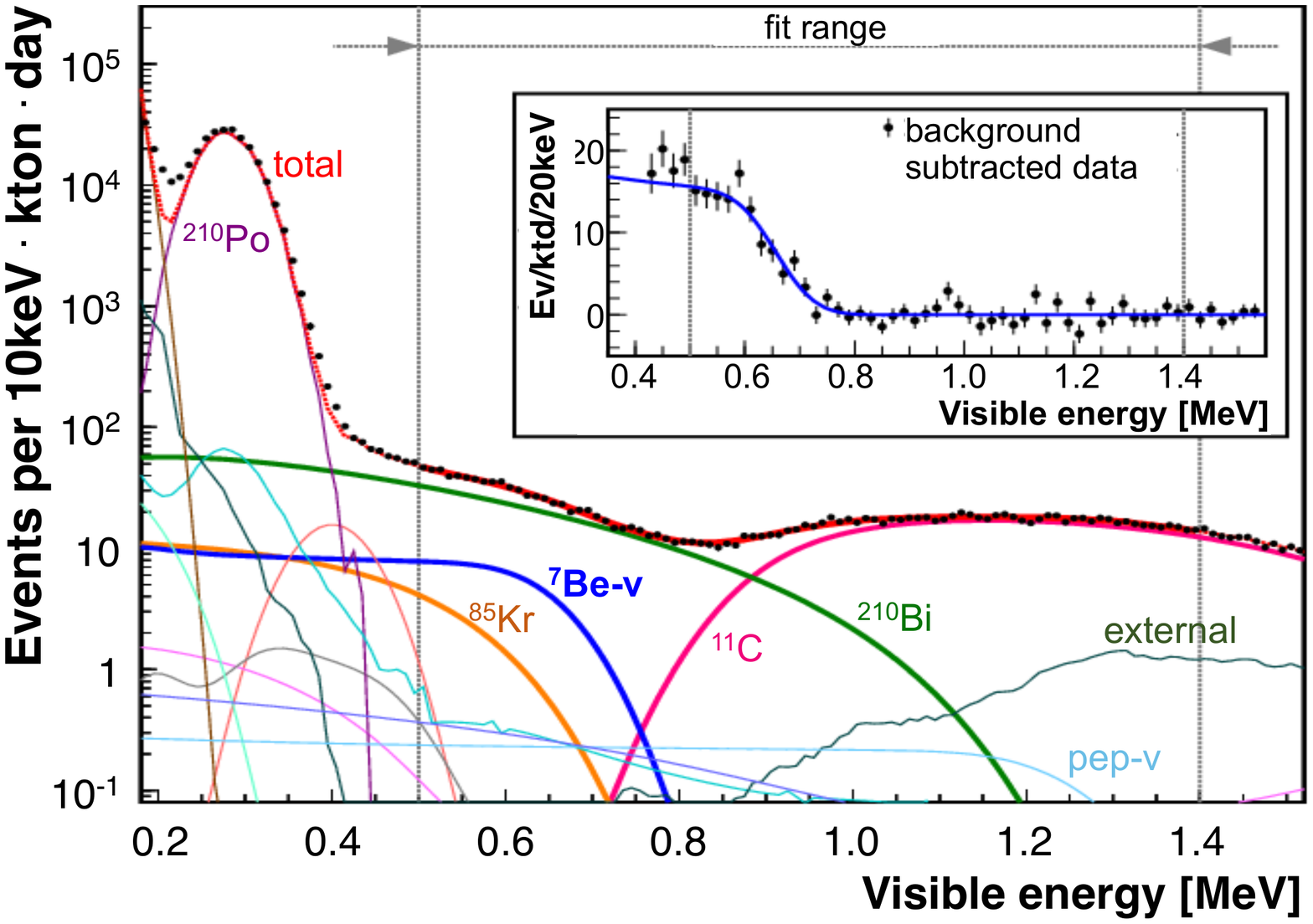}
\caption{Rank-1 data spectrum in the energy region of interest for {$^7$Be} neutrino detection in KamLAND. The data was selected for its low  {$^{210}$Bi}  background (see the text for details). The solid lines represent the fit spectra of neutrino signal and background components to the overall spectrum ({\it data points}). The inset depicts the residual {$^7$Be} recoil spectrum after statistical subtraction of all backgrounds \cite{Gando:2014wjd}.}
\label{fig:kamlandbe7}
\end{figure}

In the time period 2007-2009, the KamLAND collaboration performed two purification campaigns of its liquid scintillator to reduce the background levels from radioisotopes dissolved in the target. This effort was very successful, reducing the residual contamination levels of the three isotopes most relevant for {$^7$Be} detection,  {$^{85}$Kr},  {$^{210}$Bi}  and  {$^{210}$Po}, to $6\cdot10^{-6}$, $8\cdot10^{-4}$ and $5\cdot10^{-2}$ of their prior presence \cite{Gando:2014wjd}. To extract the {$^7$Be} ES rate from the data collected in the following period, an analysis based on a spectral fit was performed.
\medskip\\
{\bf External radioactivity.} As for the {$^8$B} neutrino analysis, the $\gamma$-rays emitted by the surrounding detector materials created a substantial background. A spherical fiducial volume of 9\,m diameter was chosen for self-shielding, thereby reducing the spectral contribution of the external background to a negligible level.
\medskip\\
{\bf Spallation products.} While cosmogenic isotopes play virtually no role for the Borexino analysis, the somewhat lower energy resolution of KamLAND leads to a smear-out of the  {$^{11}$C}  visible energy spectrum from its lower endpoint at 1.02\,MeV into the {$^7$Be} signal detection region. Therefore,  {$^{11}$C}  had to be included in the fit.
\medskip\\
{\bf Internal radioactivity.} Despite of the purification efforts, background rates in KamLAND remained substantially above the levels achieved in Borexino. While the beforehand prominent $\beta^-$-decay rate of  {$^{85}$Kr}  was reduced to a virtually negligible (4-19)\,Bq/m$^3$, and also {$^{238}$U} and {$^{232}$Th} were brought down to levels comparable to Borexino Phase-I, the purification was less successful in reducing  {$^{210}$Pb},  {$^{210}$Bi}  and  {$^{210}$Po}  concentrations: A relatively large $\alpha$-decay rate of (2.4-4.8)\,mBq/m$^3$ was measured for  {$^{210}$Po} , while the $\beta$-decay rate of  {$^{210}$Bi}  (fed by the decay of the long-lived  {$^{210}$Pb}) was determined to (0.04-0.68)\,mBq/m$^3$.

The wide activity ranges given reflect the strong spatial inhomogeneity of the  {$^{210}$Bi}  and  {$^{210}$Po}  contaminations over the detection volume. To extract the {$^7$Be} signal despite of these large backgrounds, a grid of volume parcels inside the FV was set up. An unbiased selection algorithm classified all parcels into seven ranks describing their relative radiopurity. The corresponding data sets were then subjected to a simultaneous spectral fit, the rank-1 data featuring the lowest  {$^{210}$Bi}  contamination level enacting the strongest constraint on the {$^7$Be}-rate result \cite{Gando:2014wjd}.
\medskip\\
{\bf Result.} Figure \ref{fig:kamlandbe7} shows the energy spectrum of the rank-1 data  as well as the  signal and background components of the simultaneous fit to the complete data set (all ranks). Even in the rank-1 set, the  {$^{210}$Bi}  background is much more prominent than in Borexino (fig.~\ref{fig:be7spectrum}) and almost covers the {$^7$Be} electron recoil shoulder. However, the simultaneous fit is indeed sensitive to the presence of the {$^7$Be} component, excluding an omission of the solar neutrino contribution at $8.2\sigma$ level. The corresponding ES rate amounts to $(582 \pm 94)$\,counts/(kt$\cdot$d)$^{-1}$, fully compatible with but less accurate than the latest Borexino result (sect.~\ref{sec:be7bx}). Compared to SSM predictions, the $\nu_e$ survival probability is $P_{ee} = 0.88\pm0.15$ \cite{Gando:2014wjd}.

\subsection{$pep$ and CNO neutrinos}
\label{sec:pepcno}

\begin{figure}[ht]
\centering
\includegraphics[width=0.8\textwidth]{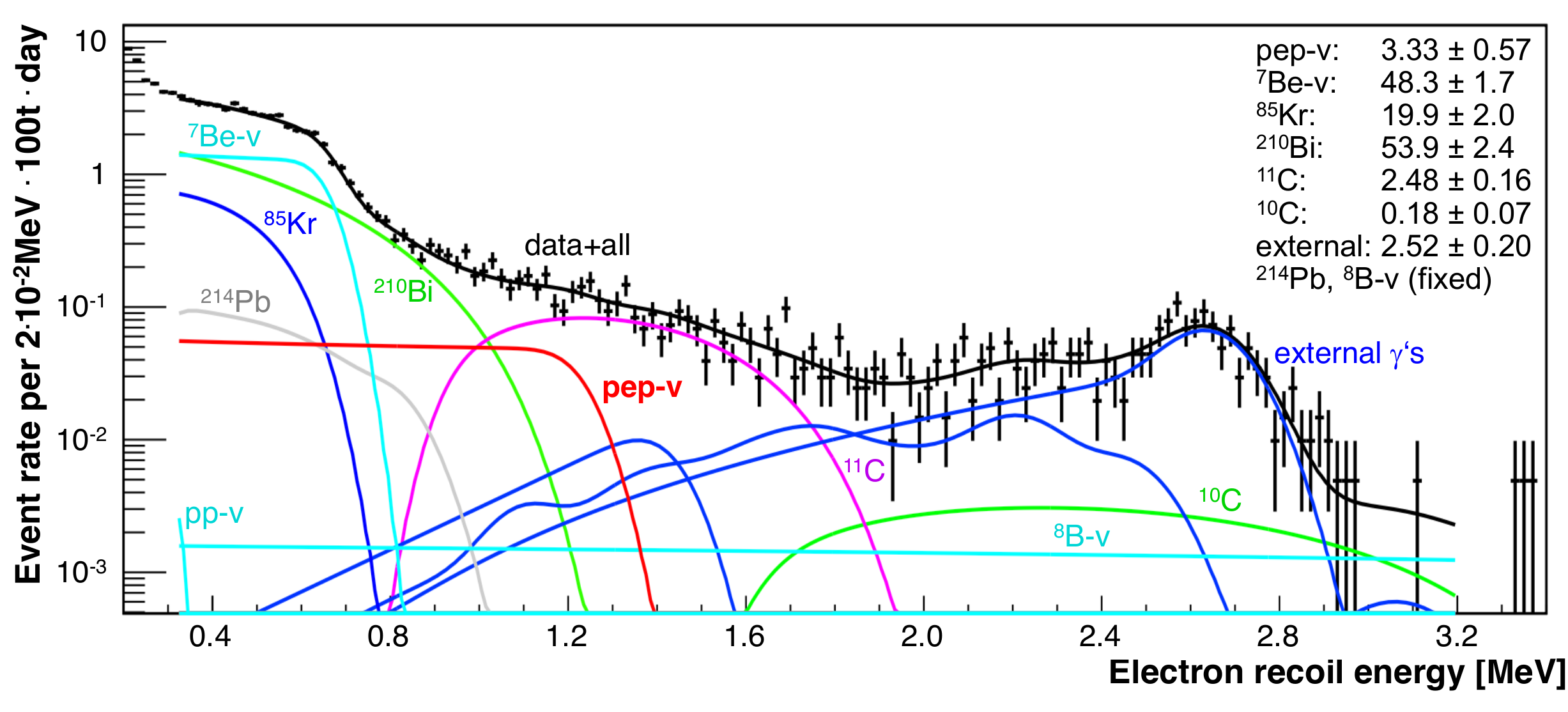}
\caption{Electron recoil spectrum after application of the threefold coincidence veto (see the text): The data points are reproduced by a fit containing several neutrino and background spectra. A non-zero contribution from $pep$ neutrinos is required at $2.7\sigma$ level \cite{Bellini:2013lnn}.}
\label{fig:borexinopep}
\end{figure}

While $pep$ and CNO-neutrinos originate from rather different nuclear processes, for purposes of solar neutrino detection they are interlinked by the almost complete overlap of their energy spectra \cite{Raffelt:1996wa}. As their origins, the basic motivations for their measurement are rather different: Because of the well-known cross-section for $p + e + p$ fusion to deuterons and its role as one of the two basic reactions in the start of the $pp$-chain, the SSM predicts the $pep$ neutrino flux very precisely \cite{Bahcall:2005va}. Due to the two-body final state, $pep$ neutrinos are emitted at a fixed energy of $E_{pep} = 1.44$\,MeV \cite{Raffelt:1996wa}, located in the low-energy onset region of the vacuum-matter-transition of the MSW-LMA solution\footnote{It should be noted that the $pep$ neutrinosphere is at larger solar radii than for the {$^7$Be} neutrinos, leading to a comparable value of $P_{ee}$ because of the lower matter densities (fig.~\ref{fig:raddists}).}: A precise $pep$ flux measurement thus bears the potential for a very accurate measurement of $P_{ee}(E_{pep})$ \cite{Friedland:2004pp, Minakata:2010be}.

Contrariwise, CNO neutrinos are mostly interesting from a perspective of solar and stellar astrophysics: The SSM prediction of the solar CNO fusion rate features large uncertainties induced both by uncertainties on the relevant nuclear cross-sections and the measured value of solar metallicity (sect.~\ref{sec:solnurates}). Therefore, a determination of the CNO rate will provide a very relevant addition to our understanding of solar fusion and $-$ given the much greater importance of the CNO cycle for older and heavier stars than the Sun $-$ main sequence stars in general \cite{Serenelli:2012zw}.
\medskip\\
{\bf Detection window.} What unites these two very different neutrino species is their common energy window for detection. The Compton shoulder induced by $pep$ neutrinos at 1.22MeV electron recoil energy is barely exceeded by the smooth neutrino spectra from the CNO cycle, their spectral endpoints ranging from about 1.20MeV to 1.74MeV \cite{Raffelt:1996wa}. What is more, their relative spectral contributions are on a comparable level, both ES rates about an order of magnitude lower than for {$^7$Be} neutrinos. Therefore, the lower halves of the $pep$ and CNO-induced electron recoil spectra are effectively blocked for detection, the signal becoming accessible only above the energy of the {$^7$Be} Compton shoulder at 662\,keV. Currently, Borexino is the only ex- periment that features sufficiently low threshold and radioactive background levels to provide information on these neutrinos in the energy window from $\sim$0.7 to 1.7\,MeV \cite{Collaboration:2011nga}.
\medskip\\
{\bf Fiducial volume.} The analysis in Borexino has to overcome several challenges in order to isolate the neutrino signal \cite{Collaboration:2011nga}: External background from $\gamma$-rays plays a larger role in the $pep$/CNO energy region than for {$^7$Be} neutrinos (sect.~\ref{sec:be7}). Therefore, the FV radius is defined to only 2.8\,m from the center, with tighter cuts in the end-cap region.
\medskip\\
{\bf Cosmogenic background.} Figure \ref{fig:borexinopep} illustrates the visible energy spectrum inside the detection window. Above $\sim$0.9\,MeV, the ES spectrum is dominated by the $\beta^+$-decay of the cosmogenic isotope  {$^{11}$C}  . While the maximum kinetic energy of the positron is only 961\,keV, the combined signal of the kinetic energy and the annihilation of the $e^+$ in the scintillator results in a visible energy of (1$-$2)\,MeV, broadened further by energy smearing. The long  {$^{11}$C}  live-time of 29.4\,min forbids a simple timing veto following each potential parent muon $-$ the average time difference between subsequent muons in Borexino is only 20\,s. Instead, a sophisticated three-fold coincidence (TFC) veto had to be developed. The TFC is based on the production mechanism of  {$^{11}$C}  from  {$^{12}$C}  spallation that is accompanied by the emission of a neutron in 95\,\% of all cases. The veto relies not only on timing but also on a set of finely tuned spatial cuts that regard the distances of a possible  {$^{11}$C}  event to both the parent muon track and the neutron capture vertex. Figure \ref{fig:borexinopep} depicts the data spectrum after applying the TFC veto. In this plot, the residual  {$^{11}$C}  rate is to $(2.5\pm0.3)$ per day and 100\,t, corresponding to a reduction to $(9\pm1)$\,\% of the original rate, while 48.5\,\% of the original exposure is preserved, bringing background and signal roughly on par \cite{Collaboration:2011nga}.
\medskip\\
{\bf Electron/positron discrimination.} In addition, the pulse shape of the reconstructed events is used in a multi-variate spectral fit to differentiate between the $pep$/CNO ES recoil electrons and the positrons emitted in  {$^{11}$C}  decay: The positrons form positronium with electrons of the scintillator. In 25\,\% of all cases, the relatively long-lived ortho-configuration is formed, introducing a slight delay between the otherwise virtually synchronous signals from $e^+$ ionization and annihilation. The corresponding substructure in the observed scintillation pulse shape can be exploited on a statistical basis to discriminate neutrino-induced electron recoils from positrons emitted in  {$^{11}$C}  decays \cite{Collaboration:2011nga}. 
\medskip\\
{\bf Results.} In the residual spectrum shown in figure \ref{fig:borexinopep}, a slight event excess is found between 1.0 and 1.2\,MeV that is induced by the $pep$-neutrino ES recoil shoulder \cite{Bellini:2013lnn}. In fact, the spectral contributions of both $pep$ and CNO neutrinos are extracted by a multivariate fit relying on event energy, radial position and $e^\pm$ pulse shape, simultaneously fitting the  {$^{11}$C}  depleted and enriched data samples. The best-fit value for $pep$ neutrinos corresponds to $3.1\pm0.6_{(stat)}\pm0.3_{(syst)}$ per day and 100\,t, representing the first direct evidence of the presence of the $pep$ neutrino signal at 98\,\% confidence level \cite{Collaboration:2011nga}.

The situation is less clear in case of the CNO neutrino signal: Due to the lack of a distinct spectral feature, the CNO ES signal is much harder to separate from the residual backgrounds, most noteworthy from the $\beta^-$ decay of  {$^{210}$Bi}  that features similar spectral shape and endpoint. Due to its comparatively large  {$^{210}$Bi}  decay rate of $55^{+3}_{-5}$ per day and 100\,t, the analysis provides only an upper limit of 7.9 CNO neutrino events per day and 100\,t (95\,\% C.L.).

Including electron scattering cross-sections and the oscillation probabilities predicted by the MSW-LMA solution, the corresponding neutrino fluxes can be calculated to $\Phi_{pep}=(1.6\pm0.3)\cdot 10^8$\,cm$^{-2}$s$^{-1}$, both compatible with the SSM prediction \cite{Collaboration:2011nga}. The limit obtained for $\Phi_{\rm CNO}$ exceeds the expected flux of the high-metallicity SSM only by about 50\,\% (sect.~\ref{sec:solnurates}).

\subsection{$pp$-neutrinos}
\label{sec:pp}

\begin{figure}[ht]
\centering
\includegraphics[width=0.6\textwidth]{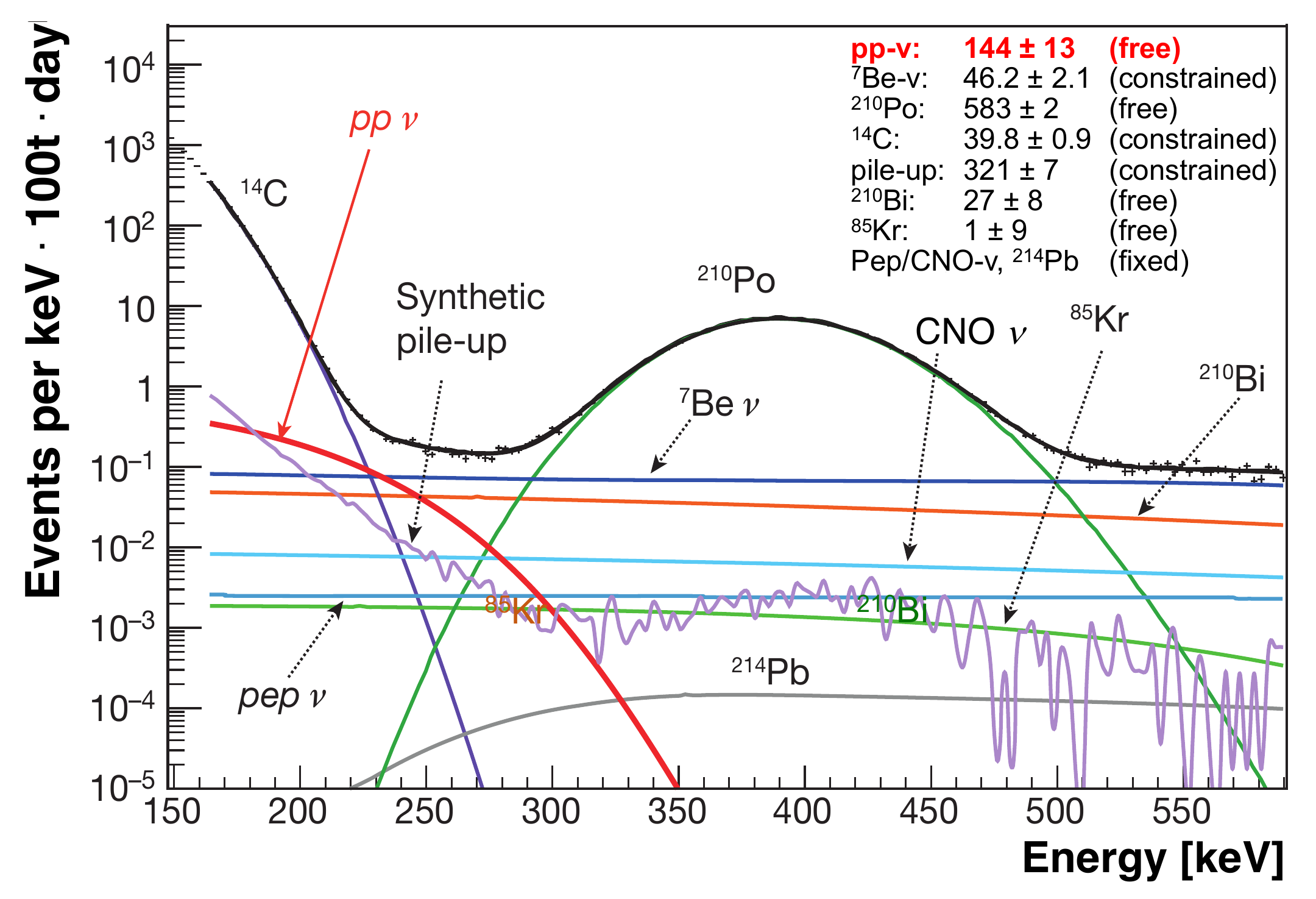}
\caption{The Borexino spectral fit to the electron recoil spectrum in the region of interest for $pp$-neutrinos. The high decay rate of  {$^{14}$C}  as well as the resulting pile-up events constitute the primary background \cite{Bellini:2014uqa}.}
\label{fig:ppspectralfit}
\end{figure}

The neutrinos emitted in the primary proton-proton fusion reaction are both the most ubiquitous and the hardest to measure component of the solar neutrino spectrum. From the early 90's, the $pp$-flux at Earth could be derived from the combination of data taken by the radiochemical and Water Cherenkov experiments: The value was determined by subtracting {$^7$Be} and {$^8$B} results from the Homestake and Kamiokande experiments from the integral rate measured in the gallium experiments \cite{Bahcall:2003ce}. Only in 2014, Borexino was able to provide a first direct measurement of the $pp$-flux, resolving the $pp$ recoil spectrum from the background formed by internal backgrounds and other neutrino signals \cite{Bellini:2014uqa}.
\medskip\\
{\bf Spectral signature.} The analysis was performed based on a spectral fit to the events in the innermost 75.7 tons of LS. Cosmic background and fast coincidence signals from the U/Th de- cay chains can be removed by standard cuts and play only a minor role. The primary challenge is set by the extremely low endpoint of the $pp$-neutrino spectrum of 423\,keV. The corresponding electron recoil energy spectrum terminates at 264\,keV and has no distinctive features. Including the lower ES cross-section, the expected event rate in Borexino of $\sim$150 counts/(day$\cdot$100 ton) is only about a factor 3 higher than for {$^7$Be} neutrinos. As illustrated by the visible energy spectrum of figure \ref{fig:ppspectralfit}, the $pp$-signal exceeds background levels only in a quite small region around 230\,keV.

Part of the background is formed by spectral contributions with relatively high end-points: {$^7$Be} neutrinos,  {$^{210}$Bi}  and  {$^{85}$Kr}  $\beta^-$-decays sum up to a flat background in the $pp$-region that
is of the same order of magnitude as the $pp$-signal \cite{Bahcall:1995mm}. These background fluxes can be well constrained by the spectral fit when including higher energy regions. Moreover, both  {$^{210}$Bi}  and especially  {$^{85}$Kr}  rates where substantially reduced in the 2010/11 purification campaign of Borexino, creating more favorable conditions for $pp$ detection.
\medskip\\
{\bf  {$^{14}$C}  decay.} The primary background is formed by the low-endpoint ($Q_\beta = 156$\,keV) but high-rate $\beta^-$-decay of the carbon isotope  {$^{14}$C}. Compared to the levels found in the Earth's atmosphere and living organisms, the crude oil from which the scintillator of Borexino was produced features a very low  {$^{14}$C}  content of $2.7\cdot10^{-18}$ relative to  {$^{12}$C}. However, given a target mass of $\sim$100\,t, this amounts to a rate of $\sim$30\,Bq above the detection threshold of $\sim$50keV, dominating the overall data spectrum and especially the $pp$ ES rate by orders of magnitude. Broadening of the spectrum by energy smearing results in an almost complete overlay of the $pp$ and  {$^{14}$C}  spectra.

Therefore, the first challenge to the analysis is to constrain the  {$^{14}$C}  decay rate by independent means to allow an extraction of the underlying $pp$-signal. Rate determination is complicated by the fact that a relevant part of the  {$^{14}$C}  decay spectrum is below trigger threshold. While an extrapolation of the complete  {$^{14}$C}  rate from the detectable range is in principle possible, it will introduce a number of systematic uncertainties that are caused by the energy-dependence of the trigger efficiency and the corresponding distortion of the  {$^{14}$C}  spectrum. To avoid these,  {$^{14}$C}  rate and spectrum are determined in a parasitic measurement mode: Once the DAQ is triggered by a fast coincidence of 25 PMT hits or more, all hits following in the next 16\,s will be acquired. Within this time gate, a sample of random events is provided that extends well below the trigger threshold. These secondary events are dominantly  {$^{14}$C}  decays and thus provide an undistorted $\beta$-spectrum as well as an overall rate estimate.

Secondly, the  {$^{14}$C}  rate is sufficiently large to create physical pile-up with other uncorrelated single events in the detector, of which in turn the majority originates from  {$^{14}$C}  decays. Especially this low-energy pile-up is very dangerous as its shape extends somewhat above the single  {$^{14}$C}  spectrum and so mimics the slightly harder $pp$ recoil spectrum. In order to obtain an effective spectral shape for these coincidences, the undistorted event spectrum from secondary triggers is used to create a high-statistics sample of 'synthetic' pile-up events. For those, the hit-time patterns of two random events are artificially combined, varying the degree of signal overlap by varying the relative start times. The rate and energy spectrum of this synthetic double events are both used as input for the spectral fit.
\medskip\\
{\bf Result.} Fig.~\ref{fig:ppspectralfit} displays the data spectrum as well as the result of the spectral fit \cite{Bellini:2014uqa}. The parameters representing the rate normalizations of  {$^{14}$C},  {$^{14}$C}  pile-up and {$^7$Be} neutrinos were constrained by independent measurements, while  {$^{210}$Bi}  and  {$^{85}$Kr}  are left floating and thus provide flexibility for the "constant" background to adjust over the whole energy range regarded in figure \ref{fig:ppspectralfit}. The resulting $pp$ rate is $144 \pm 13_{(stat)} \pm 10_{(syst)}$ per day and 100\,t, corresponding to a survival probability of $P^{\rm vac} = 0.64 \pm 0.12$ when compared to the SSM prediction of $\Phi_{pp} = (5.98 \rm 0.04) \cdot 10^{10}$\,cm$^{-2}$s$^{-1}$ \cite{Bellini:2014uqa}. Within errors, it is thus in good agreement with both the SSM and the MSW-LMA oscillation solution. 
\medskip\\
{\bf Gallium results.} Borexino's result is as well consistent with the effective $pp$ rate that can be inferred from the radiochemical gallium experiments (sect.~\ref{sec:gallium}): Starting from the total rate of the integral measurement, equivalent rate contributions from {$^8$B}, {$^7$Be}, $pep$ and CNO can be subtracted based on the results reported in sections \ref{sec:b8}$-$\ref{sec:pepcno} \cite{Abdurashitov:2009tn}. The resulting value for the terrestrial $pp$ neutrino flux features an accuracy similar to the Borexino result, $\sim$13\,\% \cite{Abdurashitov:2009tn}.

%
%

\section{Current status and open issues}
\label{sec:status}

After decades in which the seeming discrepancy between SSM predictions and measured neutrino rates in terrestrial detectors had persisted, the SNO result demonstrated beyond doubt that the rate deficit was neither a flaw in solar modelling nor in experimental techniques but a harbinger of exiting new neutrino properties. By requiring non-zero neutrino masses, flavor oscillations necessitated the first extension of the Standard Model of particle physics since its founding days in the 1970's (e.g.~\cite{Raby:1997bn}).

In the conjuncture with a broad variety of results from atmospheric, reactor and accelerator neutrino experiments, the basic picture of flavor oscillations is by now well established. However, the exploration of the connected parameters and particle properties is far from complete: The first section \ref{sec:statsolnuosc} will compile our current picture of solar neutrino oscillations and discuss the issues still open for experimental scrutiny. On the other hand, section \ref{sec:statsolprop} will highlight the possibilities for using solar neutrinos as probes to access physics of the SSM. Open aspects from both fields are motivating the currently on-going measurement campaigns of SK and Borexino. What is more, they have triggered a development towards new detection concepts that will be described in section \ref{sec:newexp}.

\subsection{Solar neutrino oscillations}
\label{sec:statsolnuosc}

\begin{figure}[ht!]
\centering
\includegraphics[width=0.7\textwidth]{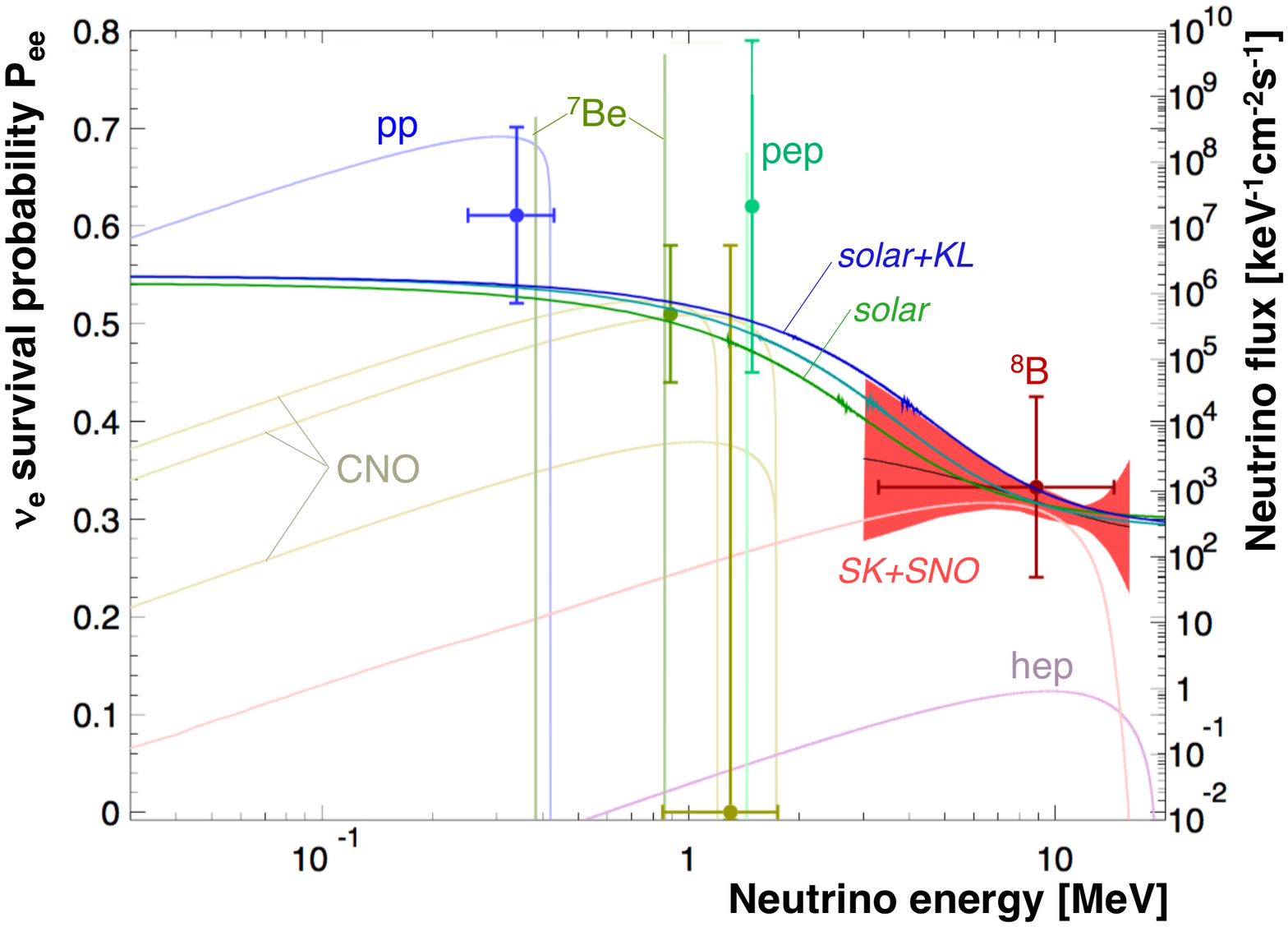}
\caption{Combined results of all solar neutrino experiments (data points) on the energy- dependence of the electron-neutrino survival probability $P_{ee}(E)$ . The solid lines depict the predicted course of $P_{ee}$ for SK-IV best-fit oscillation parameters ({\it green}) and for solar+KamLAND parameters ({\it blue}). The $1\sigma$-band from the combined SK+SNO analysis is shown in red \cite{Abe:2016nxk}.}
\label{fig:skpeefit}
\end{figure}

\begin{figure}[ht!]
\centering
\includegraphics[width=0.49\textwidth]{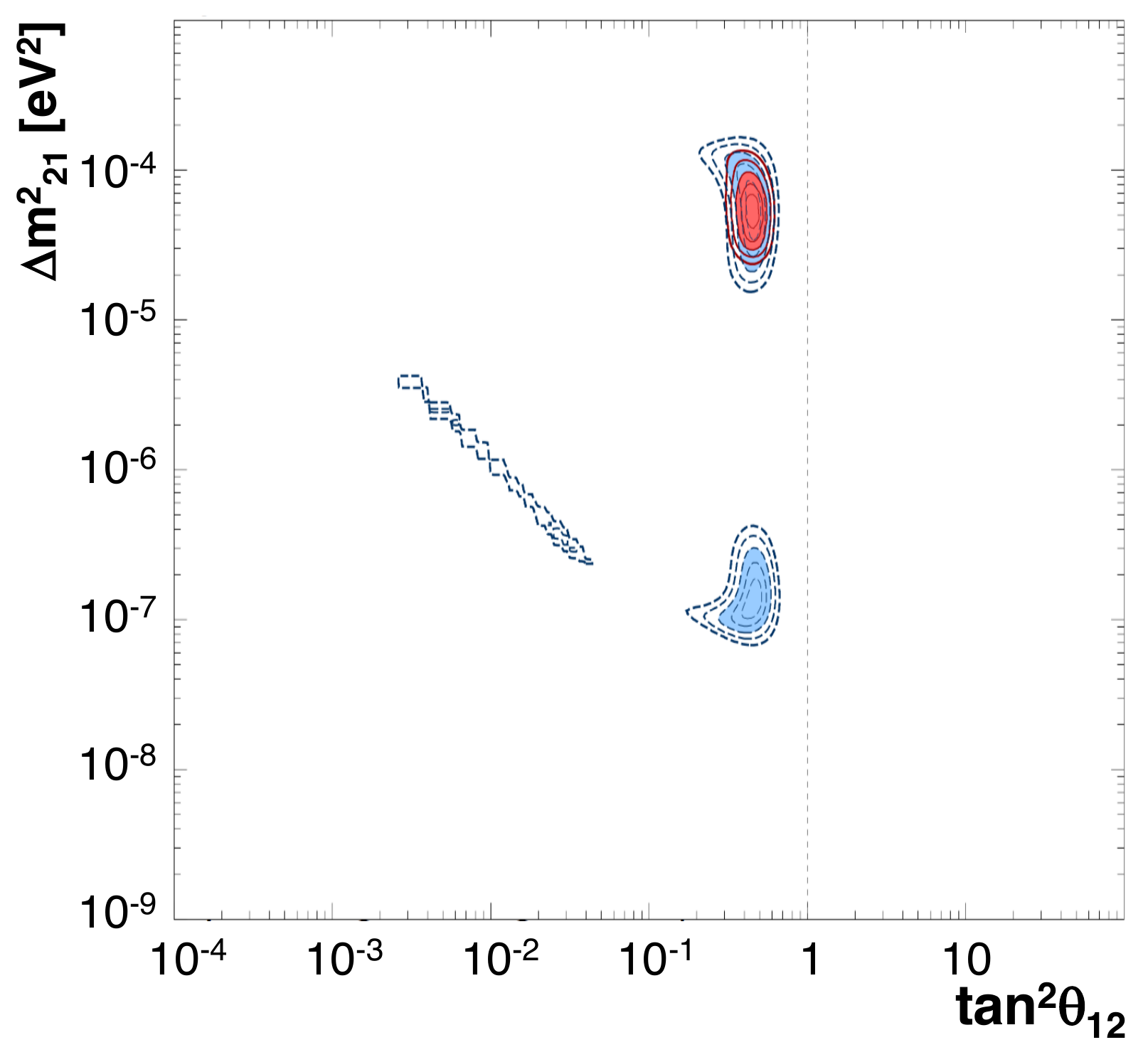}
\includegraphics[width=0.49\textwidth]{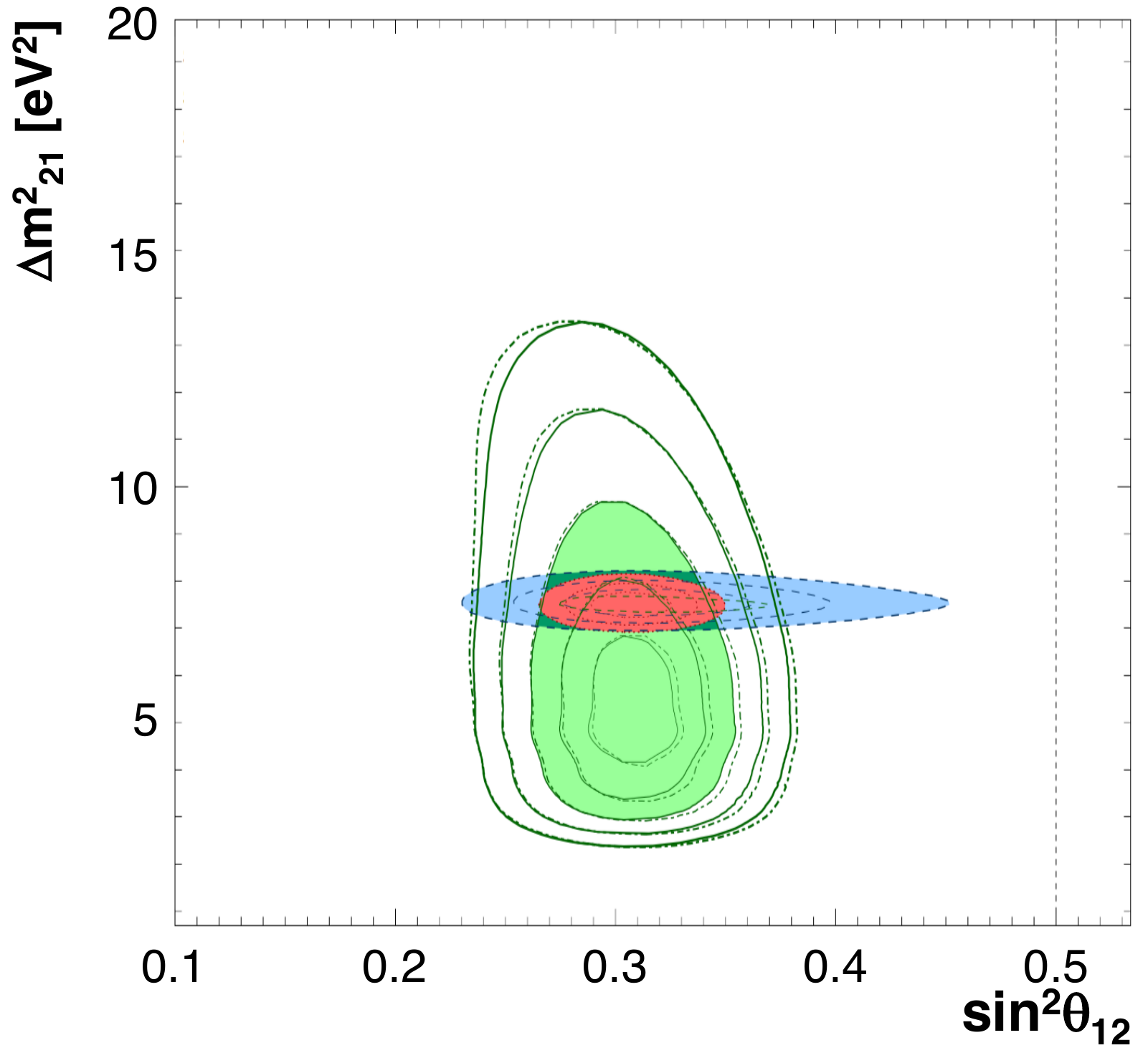}
\caption{Current contours for the solar oscillation parameters $\theta_{12}$ and $\Delta m_{21}^2$: The left panel depicts allowed regions based on SNO ({\it blue}) and the combination of solar experiments ({\it red}). The right panel shows a zoom of the solar best fit region ({\it green}), the KamLAND reactor antineutrino constraints ({\it blue}) as well as the combined result ({\it red}). The shaded regions delineate the $3\sigma$ contour levels \cite{Abe:2016nxk}.}
\label{fig:oscpar}
\end{figure}

As a result of the spectral measurements that have determined the solar $\nu_e$ survival probability $P_{ee}(E)$ as a function of energy $E$, the basic paradigm of a large mixing angle (LMA) combined with solar matter effects (MSW) driving the conversion $\nu_e \to \nu_{\mu,\tau}$ has been established beyond doubt. Figure \ref{fig:skpeefit} exemplifies this by overlaying the course of $P_{ee}(E)$ predicted by the MSW-LMA scenario and the current best-fit values for $\theta_{12}$, $\Delta m_{21}^2$ with the data points provided by SNO, SK and Borexino \cite{Abe:2016nxk}. The experiments verify the MSW-LMA solution by measuring two distinctively different values of the survival probability at both ends of the spectrum, i.e. a low value of $P^m_{ee} \approx 0.33$ in the high-energy {$^8$B} region (SK, SNO) contrasting a high $P^{\rm vac}_{ee} \approx 0.55$ for $pp$ and {$^7$Be} neutrinos (Borexino, gallium experiments).
\medskip\\
{\bf Mixing angle $\theta_{12}$.} The determination of the solar mixing angle $\theta_{12}$ is dominated by the precise measurements of the survival probability for high-energy {$^8$B} neutrinos. As laid out in sect.~\ref{sec:oscmat}, $P^m_{ee} = \sin^2\theta_{12}$ features little energy-dependence above 10\,MeV. By comparing the ES and CC rate measurements of SNO as well as the ES result by SK to the NC result of SNO, a value of $\sin^2\theta_{12} = 0.310 \pm 0.014$ is obtained \cite{Abe:2016nxk}, corresponding to $\theta_{12} = (33.8 \pm 0.9)^\circ$.
\medskip\\
{\bf Mass-squared difference $\Delta m_{21}^2$.} As the distance from the Sun to terrestrial detectors exceeds the neutrino oscillation length by far, a determination of $\Delta m_{21}^2$ via measuring the oscillation phases is impractical. Instead, the access to $\Delta m_{21}^2$ is somewhat more indirect: The positive sign, i.e.~$(m_2)^2 > (m_1)^2$, can be deduced directly from the mere presence of the matter resonance: The suppression of $P^m_{ee}$ at {$^8$B} energies compared to the vacuum value $P^{\rm vac}_{ee}$ would otherwise not occur.\\
The numerical value of $\Delta m_{21}^2$ derived for solar neutrinos relies on two sources: It can be extracted from the energy dependence observed for $P_{ee}(E)$, the dominant information arising from the precise measurement of the {$^8$B} neutrino spectrum. As described in section \ref{sec:b8leta}, all solar neutrino experiments find no deviation from the SSM-predicted shape down to the lowest accessible {$^8$B} energies ($\sim$3\,MeV). This allows to set an upper limit on $\Delta m_{21}^2$. On the other hand, a lower limit arises from the observation of vacuum oscillations for $pp$ and {$^7$Be} neutrino energies.\\ 
The second constraint arises from the day-night asymmetry in the {$^8$B} neutrino rate observed by Super-Kamiokande: The value of the day-night amplitude is strongly related to the realized value of $\Delta m_{21}^2$.
The recent result provided by SK, $A(^8{\rm B})_{\rm SK} = -0.033 \pm 0.010_{(stat)} \pm 0.005_{(syst)}$, in fact drives the current best-fit value of $\Delta m_{21}^2 = (4.8 \pm 0.6) \cdot 10^{-5}\,{\rm eV}^2$ for solar data only \cite{Abe:2016nxk}.\\ 
On the other hand, the MSW-LMA predicts a vanishing day-night asymmetry for the {$^7$Be} neutrino rates. In accordance, Borexino sets a tight limit of $A(^7{\rm Be}) = 0.001 \pm 0.012_{(stat)} \pm 0.007_{(syst)}$ and is able to rule out at high significance the LOW oscillation scenario (sect.~\ref{sec:be7bx}) \cite{Bellini:2011yj}.\\
Globally, the most accurate value of $\Delta m_{21}^2 = (7.49^{+0.19}_{-0.18}) \cdot 10^{-5}\,{\rm eV}^2$ is obtained \cite{Abe:2016nxk} when combining all solar data with the reactor $\bar\nu_e$ results from KamLAND that provides $P_{\bar e\bar e}(E,L)$ at long ($\sim$200\,km) oscillation baselines \cite{Gando:2013nba}. Here, greater accuracy can be gained by a direct measurement of the oscillation frequency in eq.~(\ref{eq:oscformula}). Contrariwise, solar neutrinos provide better constraints on the mixing angle $\theta_{12}$. Best-fit contours of both parameters, individually and combined for solar and KamLAND data, are displayed in figure \ref{fig:oscpar} \cite{Abe:2016nxk}.

\subsubsection{Future oscillation program}

{\bf Improving solar $\Delta m_{21}^2$ result.} The low accuracy on the solar measurement of $\Delta m_{21}^2$ when compared to the KamLAND reactor data \cite{Gando:2013nba} is mainly caused by the still relatively large uncertainty on the amplitude of the day-night asymmetry that is observed in the {$^8$B} neutrino rate  (sect.~\ref{sec:pee}). The corresponding measurement accuracy in SK is still statistics-limited, so that a slow but steady improvement is expected by a continuation of the data taking. A corresponding measurement in the much larger Hyper-Kamiokande experiment would likely saturate the accuracy at the 0.5\,\% level implied by the systematics of the measurement (sect.~\ref{sec:futurewcd}) \cite{Abe:2011ts}.\\
A further handle on $\Delta m_{21}^2$ is offered by a precise measurement of the energy-dependence of $P_{ee}(E)$ in the vacuum-matter transition region from $\sim$1 to 5\,MeV (neutrino energy): The best conceivable access points are the low-energy region of the {$^8$B} neutrino spectrum below 5\,MeV as well as the $pep$-neutrino line at 1.44\,MeV \cite{Friedland:2004pp, Robertson:2012ib, Gann:2015yta}. While information on both is available (sects.~\ref{sec:b8leta}+\ref{sec:pepcno}), current experimental results still lack the precision necessary to challenge the KamLAND result on $\Delta m_{21}^2$.
\medskip\\
{\bf Exploring new physics in the transition region.} An improved measurement of $\Delta m_{21}^2$ by solar experiments is considered especially appealing because of the slight strain that exists between solar and KamLAND best-fit values \cite{Abe:2016nxk}. Applying the somewhat lower $\Delta m_{21}^2$ value obtained from reactor $\bar\nu_e$'s to predict the course of $P_{ee}(E)$, the spectrum of low-energy {$^8$B} neutrinos is expected to feature an upturn towards larger survival probabilities in the 3$-$5\,MeV range. Instead, the low-energy {$^8$B} data acquired by SK, SNO and Borexino is more compatible with a constant value of $P_{ee}(E) = P^m_{ee}$. Solar data prefers such either a lower value for $\Delta m_{21}^2$ or requires the presence of additional novel physics processes suppressing the upturn.\\
For instance, the weak matter potential acting on $\nu_e$'s when traversing solar matter might be larger than expected. As a consequence, neutrinos at lower energies would be affected and the upturn of $P_{ee}(E)$ shifted accordingly. Such an effect could for instance be caused by non-standard interactions of neutrinos with quarks or electrons. The magnitude of the effect required would have to increase the potential by about 60\,\% compared to the standard weak potential \cite{Friedland:2004pp, Minakata:2010be}.\\
On the other hand, there might be an additional light sterile (!) neutrino eigenstate resulting in a new neutrino mass eigenstate $m_4$ located right in between the $m_1$ and $m_2$ masses. This allows for an additional resonance occurring for medium-energy $\nu_e$, resulting in a suppression of $P_{ee}$ in the transition region \cite{deHolanda:2010am}.
\medskip\\
Nevertheless, the statistical significance of the discrepancy between solar and global best-fit values is for now only on the $2\sigma$-level (fig.~\ref{fig:oscpar}). Likely, a future generation of experiments will be required to clarify the situation, either via considerably larger ES event numbers for {$^8$B} and $pep$ neutrinos or by employing a suitable charged-current reaction for neutrino detection that provides far better energy resolution at low neutrino energies (sect.~\ref{sec:newexp}) \cite{Gann:2015yta}.

\subsection{Solar properties}
\label{sec:statsolprop}

\subsubsection{Solar composition}
\label{sec:statsolcomp}

\begin{figure}[ht!]
\centering
\includegraphics[width=0.49\textwidth]{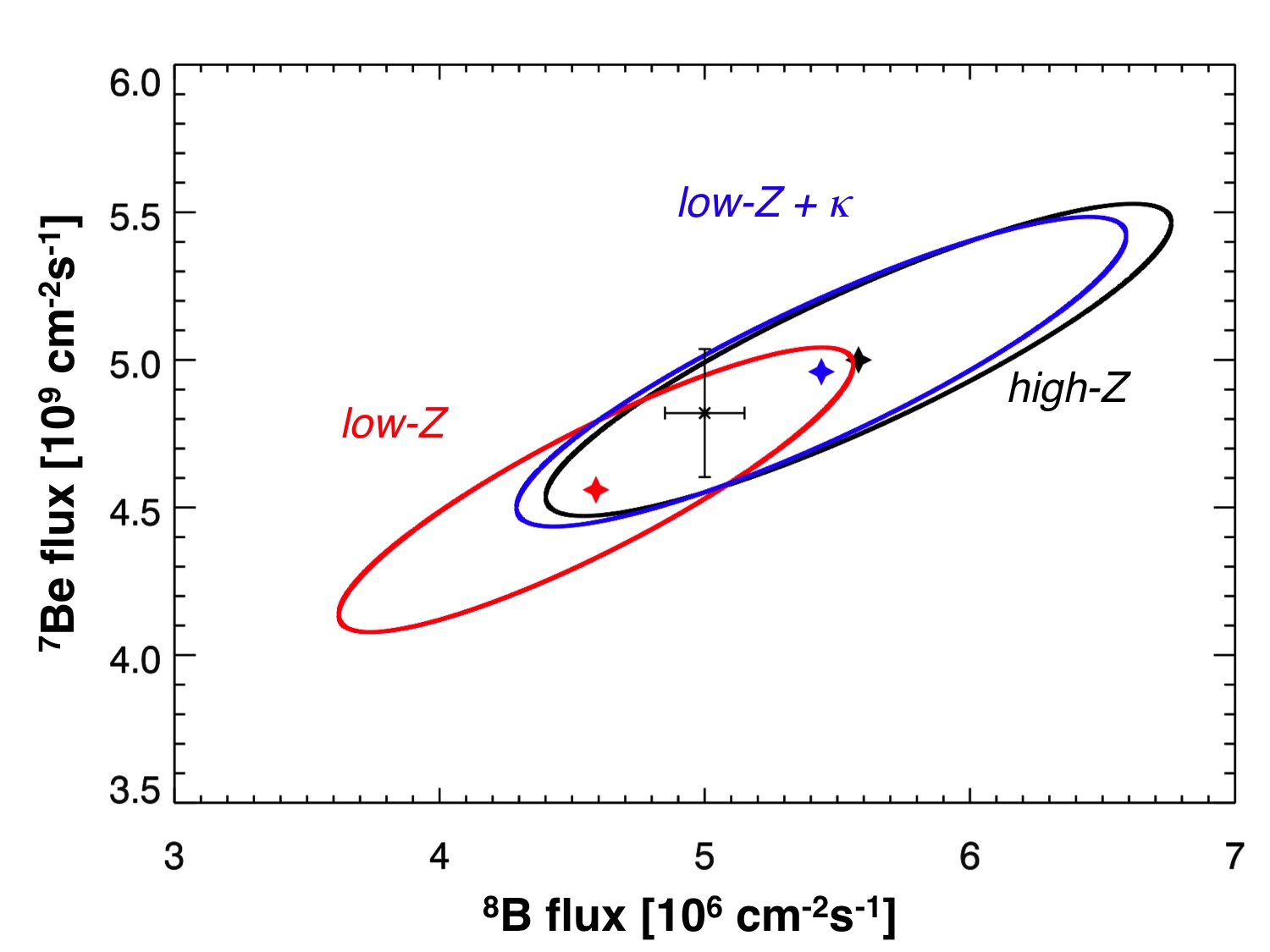}
\includegraphics[width=0.49\textwidth]{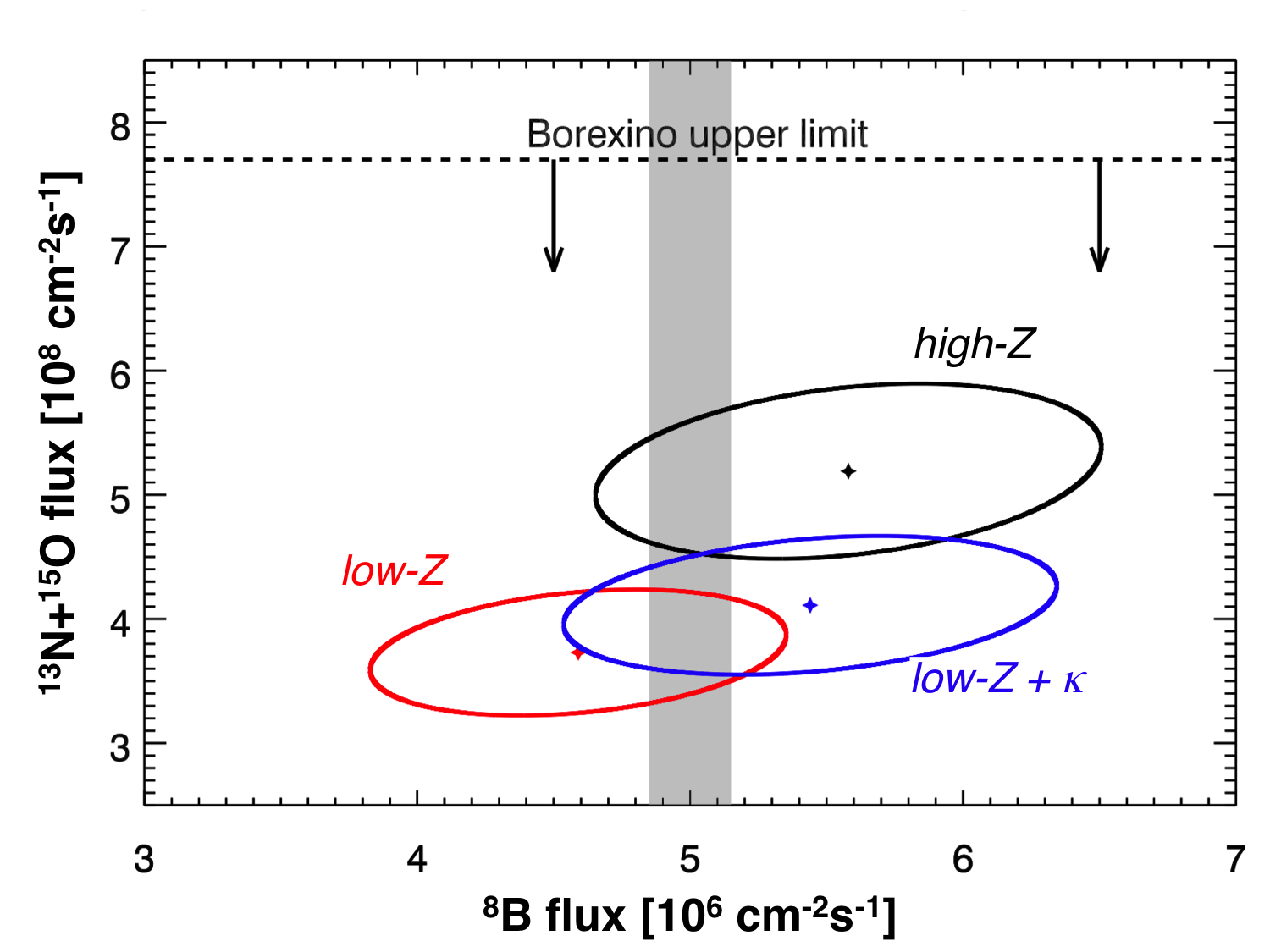}
\caption{The potential of solar neutrino spectroscopy to resolve the Solar Composition Problem. The left panel shows {$^8$B}/{$^7$Be} neutrino flux predictions ($1\sigma$ allowed regions) from the SSM, based on differing assumptions on solar metallicity ({\it low-Z} and {\it high-Z}) and modified opacities $\kappa$. Current measurement results (data point) are indecisive. The right panel illustrates the potential of a future combined measurement of CNO and {$^8$B} neutrinos. The current experimental status is represented by the upper Borexino bound and the shaded area for {$^8$B} neutrinos \cite{Serenelli:2009ww, An:2015jdp}.}
\label{fig:metallicity}
\end{figure}

\noindent During the days of the Solar Neutrino Problem, a source of great confidence in the neutrino flux predictions of the SSM was the excellent agreement of solar parameters derived from spectroscopic measurements of the photosphere on the one side and helioseismic observations on the other \cite{Bahcall:1996qw}. As laid out in section \ref{sec:solnurates}, this formerly high concordance has been significantly reduced by a re-evaluation of the abundances of volatile elements (C, N, O, Ne) based on photospheric absorption lines \cite{Asplund:2009fu}, resulting in reduced abundance values. Often cited, the corresponding radial sound speed profile predicted by the SSM is showing much larger deviations from helioseismic measurements for this new {\it low-Z} model than for the former {\it high-Z} SSM (fig.~\ref{fig:solarcsound}). This discrepancy has given rise to the so-called Solar Composition Problem \cite{PenaGaray:2008qe, Serenelli:2009yc}.
\medskip\\
{\bf Solar opacities.} The elemental composition of the solar plasma affects the speed of sound mostly via its influence on the radial temperature profile: Larger metal abundances increase the opacity of solar matter for electromagnetic radiation and thus steepen the gradients of both temperature and sound speed. On the other hand, the SSM extrapolates the measured temperature of the solar photosphere to obtain a value for the solar core temperature $T_c$. As the fusion rates of both $pp$ and CNO chain are connected to $T_c$ by high power laws, the emitted neutrino fluxes basically act as precise thermometers (fig.~\ref{fig:nufluxvstc}). As a consequence, the expected neutrino signals in terrestrial detectors vary substantially between the scenarios illustrated by table \ref{tab:ssm}, that only differ in the input metallicity in otherwise identical solar models \cite{Asplund:2009fu, Serenelli:2009ww, PenaGaray:2008qe}.
\medskip\\
{\bf Experimental results on {$^7$Be} and {$^8$B} neutrinos.} According to table \ref{tab:ssm}, the magnitude of the flux difference depends on the specific fusion reaction. In fact, sufficient experimental accuracy is today merely reached in case of {$^7$Be} and {$^8$B} neutrinos. In order to perform a comparison between experimental rates and SSM predictions, the effect of neutrino flavor oscillations has to be taken into account. This is rather straight-forward in the case of {$^8$B} neutrinos where the SNO NC measurement provides a flavor-independent result for the total flux. Contrariwise, the rate measurement of {$^7$Be} neutrinos by Borexino is scaled for $P_{ee}$ using the fit value of $\theta_{12}$, that is mostly determined by the SNO and SK measurements \cite{Serenelli:2009ww}.\\
Unfortunately, the interpretation of the combined results shown in figure \ref{fig:metallicity} is rather ambiguous: While experimental accuracy is roughly on par with the predicted flux difference, both {$^7$Be} and {$^8$B} flux results fall right into the middle ground between the predicted absolute values for {\it low-Z} and {\it high-Z} scenarios. What is more, the discrepancies in $T_c$ and thus neutrino flux predictions between {\it low-Z} and {\it high-Z} can be somewhat alleviated when changing the input values for elemental opacities $\kappa$ to the SSM \cite{Serenelli:2009ww}. So, while confirming the general consistency of the neutrino results with the SSM, current results cannot contribute towards resolving the composition problem.\\
To improve the discrimination power of the {$^7$Be} and {$^8$B} flux measurements, two requirements will have to be fulfilled: Not only experimental accuracy on the measured fluxes has to be improved, but also the uncertainties inherent to the SSM have to be reduced because they will otherwise dominate the comparison. The SSM prediction depends here on a number of input parameters, not only concerning elemental opacities both also uncertainties in the cross-sections of the involved fusion reactions. These would have to be addressed in dedicated laboratory experiments to reduce the uncertainties \cite{Serenelli:2009ww}.
\medskip\\
{\bf Potential of CNO neutrinos.} A promising alternative test is in principle provided by a measurement of the CNO neutrino flux \cite{Serenelli:2009ww}: Due to the direct dependence of reaction rates on the abundance of carbon and nitrogen isotopes in the solar core, the predicted flux difference of $\geq$30\,\% is much more prominent than for the reactions of the $pp$-chain. The sensitivity to solar composition can be further enhanced by a combination of {$^8$B} and CNO flux measurements that reduces underlying SSM uncertainties. As illustrated by figure \ref{fig:metallicity}, accurate experimental results might be used to discern not only {\it low-Z} and {\it high-Z} scenarios but provide also some sensitivity on the opacities $\kappa$ \cite{Serenelli:2009ww}.\\
However, an accurate measurement of the CNO neutrino flux is experimentally challenging. In the range of currently running experiments, only Borexino provides sensitivity in the relevant spectral region. Unfortunately, the present best upper limit on the CNO flux by Borexino still fails to provide a meaningful constraint on the SSM metallicity (fig.~\ref{fig:metallicity}).\\
Borexino's measurement accuracy is essentially limited by the internal radioactive background levels and event statistics (sec.~\ref{sec:pepcno}): While the cosmogenic  {$^{11}$C}  can be reduced to a sufficiently low level for a meaningful application of a spectral fit to extract the $pep$ neutrino rate, the $\beta^-$ decay spectrum of  {$^{210}$Bi}  dissolved in the scintillator is quasi-degenerate with the CNO recoil spectrum (fig.~\ref{fig:be7spectrum}). While the purification campaigns performed before the beginning of Phase-II have brought down the  {$^{210}$Bi}  rate, it is still considerably larger than the CNO signal. It is worthwhile to note that the real culprit is not the relatively short-lived  {$^{210}$Bi}  ($T_{1/2}=120$\,h) but the long-lived mother isotope  {$^{210}$Pb}  ($T_{1/2}=22.2$\,yrs) dissolved in the scintillator or attached to the inner surface of the nylon vessel.\\
Due to the spectral degeneracy, a further reduction of the  {$^{210}$Bi}  background (e.g.~by a further purification campaign) will most likely only improve the upper limit on the CNO neutrino flux. A positive measurement of the signal requires to reduce the uncertainty on the  {$^{210}$Bi}  decay rate present to a level substantially smaller than the expected CNO signal rate. Such a feat might be achieved by linking the  {$^{210}$Bi}  rate to that of neighbouring isotopes in the uranium decay chain. Unfortunately, scintillator chemistry favors  {$^{210}$Pb}  (the mother isotope) and  {$^{210}$Po}  (the daughter isotope) over other isotopes in the decay chain so that secular equilibrium was broken during filling and re-purification. Moreover, the endpoint of the  {$^{210}$Pb}  $\beta^-$ decay spectrum, $Q_\beta=63.5$\,keV, is below detection threshold.\\ 
A better chance is offered by the measurement of the  {$^{210}$Po}  $\alpha$-decays that are easily identified by their prominent peak around 425\,keV in the visible energy spectrum (fig.~\ref{fig:be7spectrum}). However, the initial  {$^{210}$Po}  decay rate in the scintillator after filling Borexino exceeded the  {$^{210}$Bi}  rate by more than a factor 20, and was replenished during the 2011/12 purification campaign. Based on the  {$^{210}$Po}  half-life of 138\,days, several years of waiting time are required before secular equilibrium of  {$^{210}$Po}  and  {$^{210}$Bi}  is achieved. Moreover, it has become clear that convective motion within the scintillator may introduce new contaminants from the vessel surface into the detection volume, disturbing the equilibrium. To prevent convection, the Borexino collaboration has insulated the detector tank in early 2015, establishing a steep temperature gradient over the height of the detector featuring the lowest temperature at the bottom.\\
It remains to be seen if the Borexino Phase-II measurement will be able to improve further on the upper limit of the CNO neutrino flux established during Phase-I or in the future might even be able to perform a positive detection. The potential of other experimental concepts proposed to detect CNO neutrinos is laid out in section \ref{sec:newexp}.

\subsubsection{Solar luminosity}
\label{sec:sollum}

A basic assumption made in many predictions for solar neutrino fluxes and also often in the interpretation of the data is the solar luminosity constraint: It links the fusion rates of $pp$-chain and CNO-cycle to the total output of fusion energy in the solar core and thus to the electromagnetic luminosity irradiated from the solar surface \cite{Bahcall:2001pf}. Testing the validity of this assumption might be interesting for two reasons:
\medskip\\
{\bf Long-term uniformity.} While solar neutrinos leave the Sun virtually instantaneously following their production in the solar core, the released binding energy needs several ten thousand years of radiative and convective transport to dissipate to the solar photosphere \cite{Raffelt:1996wa}. Sufficiently accurate measurements of photon and neutrino luminosities provide one past and one present snap-shot of solar fusion activity, thus allowing to check the for a variation of the energy output on this historically long (but astronomically vanishing) time scale.
\medskip\\
{\bf Novel physics.} On the other hand, imposition of the luminosity constraint can be employed to test solar energy release for non-standard physics: If neutrino luminosity exceeds the electromagnetic, this will provide a hint for an additional unknown cooling mechanism that could for instance be realized by irradiation of novel types of particles not detected in terrestrial detectors. On the other hand, a surplus of electromagnetic luminosity would hint at an additional process of energy generation in the Sun \cite{Raffelt:1996wa}.
\medskip\\
{The most sensitive experimental tests of neutrino luminosity are provided by measurements of the $pp$ and $pep$ neutrino fluxes. As the primary steps of the $pp$ chain, both reactions are closely linked to the total energy output by fusion and consequently feature only small uncertainties ($\leq1\,\%$) on their SSM flux predictions (tab.~\ref{tab:ssm}). As in the case of solar composition tests, the measured experimental rates have to be corrected for oscillation probabilities before comparing to the SSM value. Current measurements on both $pp$ and $pep$ fluxes by Borexino (sects.~\ref{sec:pepcno}+\ref{sec:pp}) feature relative uncertainties on the 11\,\% resp.~20\,\% level and thus provide at best a coarse consistency test \cite{Collaboration:2011nga, Bellini:2014uqa}. Future experiments will have to reach accuracies in the 1\,\% regime to perform meaningful comparisons of measurements and predictions.

\section{Detection concepts for future experiments}
\label{sec:newexp}

Although the basic mystery of the vanishing neutrinos has been solved, solar neutrinos remain a worthwhile subject to study for both particle and astrophysics (sect.~\ref{sec:status}) \cite{Robertson:2012ib, Gann:2015yta}. As a consequence, both SK and Borexino continue their solar neutrino programs under improved measurement conditions: While SK further pursues the measurement of low-energy {$^8$B} neutrinos and the consolidation of the day-night asymmetry in the {$^8$B} ES rate, Borexino explores the possibility to detect CNO neutrinos under the improved background conditions of its second data taking phase.

On the other hand, beyond these two veteran experiments a range of future projects is being devised that aim at a methodical improvement of the experimental sensitivity. This final section aims to review the experimental concepts discussed for upcoming and future detectors and their primary features. However, the persisting interest in the field results results in a large variety of experiments proposed, too many to hope that the overview given here is truly comprehensive. The following sections will thus address only the main experimental techniques along with a selection of representative projects.

\subsection{Organic liquid scintillators}

After Borexino, the main improvements to be considered for novel organic liquid scintillator experiments are significant increases in target mass, lower cosmogenic background levels or improved background discrimination capabilities.
\medskip\\
{\bf SNO+.} The SNO collaboration ended data taking in 2006, after which the heavy water was removed from the acrylic sphere and the light water drained. In the follow-up, the collaboration decided to embark on the successor experiment SNO+, replacing the D$_2$O by a liquid-scintillator target \cite{Kraus:2010zzb}. The refurbishment of the detector as well as the installation of additional infrastructure is far progressed, so that the SNO+ detector is scheduled to come online during 2017. Provided its large target mass (900\,kg) and extremely low-background levels (especially what regards shielding from cosmic muons), the experiment potentially has excellent prospects for a refined measurement of the CNO and $pep$ neutrino fluxes. As in Borexino, the possibility for a positive CNO measurement will largely depend on the radioactive background levels achieved in the scintillator, especially what concerns  {$^{210}$Bi}  (sect.~\ref{sec:statsolcomp}).\\
Nevertheless, the current priority of the physics program is on the loading of the scintillator with the $\beta\beta$-unstable tellurium isotope {$^{130}$Te}, aiming at a timely experiment searching for the neutrino-less double-beta decay \cite{Biller:2014eha}. Unfortunately, the $2\nu-\beta\beta$-decay of {$^{130}$Te} will dominate the solar neutrino ES signal in the energy range below $\sim$2.5\,MeV, covering most spectral components with the notable exception of {$^8$B} neutrinos. Apart from a short initial solar neutrino phase scheduled before the {$^{130}$Te} loading, a full solar neutrino phase of SNO+ seems conceivable only after the extraction of the $\beta\beta$-isotope.
\medskip\\
{\bf JUNO.} The recent discovery of a relatively large value of $\theta_{13}$ \cite{An:2012eh} and the still unresolved question of the neutrino mass hierarchy has triggered the realization of at least one medium-baseline reactor neutrino experiment: The Jiangmen Underground Neutrino Observatory (JUNO) is currently under construction in southern China, at identical baselines of 53\,km from to nuclear power complexes on the coast of the Chinese Sea \cite{Djurcic:2015vqa, An:2015jdp}. Excavation of the underground lab was begun in early 2015, the start of data taking is foreseen for 2020. A similar experimental setup under the name of RENO-50 has been considered for South Korea \cite{Kim:2014rfa}.\\
As the determination of the mass hierarchy relies on a highly resolved measurement of subdominant modes in the oscillation pattern, the antineutrino detector will have to provide both excellent energy resolution and high statistics \cite{An:2015jdp, Ghoshal:2010wt}. As a consequence, the target volume of JUNO consists of 20\,kt of LAB-based liquid scintillator, contained within an acrylic sphere of 35\,m diameter. The vessel is surrounded by 17\,000 20" PMTs (and a comparable number of 3" PMTs) to reach a photoelectron yield of more than 1\,100 per MeV and an energy resolution of 3\,\% at 1\,MeV. The high p.e.~yield is expected to enhance as well the capability for pulse shape discrimination. Both factors favor solar neutrino detection. On the other hand, at a depth of 700\,m the rock shielding provided is much lower than e.g.~in Borexino, and also the required specifications for scintillator radiopurity are only on the level of $10^{-15}$ to $10^{-17}$ for U/Th.\\
As a consequence, it is hard to tax the full potential of JUNO with regards to solar neutrinos: A high-precision measurement of the low-energy {$^8$B} neutrino spectrum seems well within reach, utilizing both elastic scattering and a CC interaction on  {$^{13}$C}  (sect.~\ref{sec:dopedls}) \cite{Mollenberg:2014mfa}. Depending on intrinsic background levels, also a measurement of {$^7$Be} or even $pp$ neutrinos might be possible. On the other hand, preconditions for the detection of $pep$ and CNO neutrinos will be less than ideal given the comparatively shallow depth \cite{An:2015jdp}.
\medskip\\
{\bf LENA.} A liquid-scintillator based observatory on the scale of SK and dedicated to Low-Energy Neutrino Astronomy (LENA) was first proposed in 2004 but not realized. The focus of the potential measurement program was primarily on the exploration of astrophysical neutrino sources, encompassing Supernova neutrinos (burst and diffuse), geoneutrinos, solar neutrinos and many more; a detailed description can be found in \cite{Wurm:2011zn}.\\
Far from nuclear reactors and at great depth (e.g.~in the 1400\,m deep Pyh\"asalmi mine in central Finland), the experiment was designed to optimize the conditions for rare event searches: The centerpiece were 50\,kt of ultrapure organic liquid scintillator housed in a detector 100\,m high and 38\,m in diameter. Other than JUNO, scintillator and light detection properties were optimized not primarily to increase energy resolution but to enhance pulse shape discrimination of background events. Moreover, the great target volume would enable full exploitation of the self-shielding capabilities in LS: Restricting the analysis to the innermost 30\,kt reduces external background to solar neutrino detection to a negligible level. Corresponding event rates range from $\sim10^2\,{\rm d}^{-1}$ for {$^8$B} to $10^4\,{\rm d}^{-1}$ for {$^7$Be} neutrinos \cite{Mollenberg:2014mfa}.\\
LENA would have the potential to address most of the open physics items laid out in section \ref{sec:status}: The {$^8$B} detection threshold was as low as 2\,MeV, enabling a very sensitive test of $P_{ee}(E)$ in the transition region by both elastic scattering and  {$^{13}$C}  CC reactions (sec.~\ref{sec:dopedls}). The time stability of the {$^7$Be} rate against periodic modulations could be tested on the 0.1\,\% level \cite{Wurm:2010mq}. Finally, CNO and $pep$ neutrino signals could be extracted at high accuracy even if internal background conditions were significantly inferior to Borexino. First studies indicate that  {$^{210}$Bi}  levels two orders of magnitude higher than in Borexino would still allow for a few-\% measurements of the CNO neutrino fluxes.

\subsection{Water Cherenkov detectors}
\label{sec:futurewcd}

The two main possibilities for a future detector to improve over the current sensitivity of SK is either to lower the detection threshold or two increase event statistics. A substantial decrease in detection threshold seems difficult to achieve both due to the intrinsically low light yield of the Cherenkov technique and the technological challenge to purify water to the level of e.g.~organic liquids. However, a substantial increase in target mass is much more straight-forward to achieve. 
\medskip\\
{\bf Hyper-Kamiokande (HK).} As direct successor to SK, the next-generation detector aims to improve the experimental sensitivity to a variety of neutrino sources by a substantial increase of the detection volume. The currently proposed configuration of two underground tanks holding 260\,kt of light water (187\,kt fiducial volume) each, the target mass available for solar neutrino detection would amount to roughly 17 times that of SK. Moreover, the current design foresees a slight increase in light yield, improving both the detection threshold and background discrimination capabilities for low-energy neutrinos \cite{Abe:2011ts,Abe:2016ero}.\\
HK features a broad physics program that encompasses atmospheric neutrino measurements, the potential long-baseline oscillation experiment T{$^{2}$H}K to determine the CP-violating phase in the neutrino mixing matrix as well as an intensified search for proton decay. While solar neutrino detection will still be restricted to the {$^8$B} neutrinos, an important contribution to be expected is an even more precise measurement of the day-night asymmetry in the ES rate. Given the considerably larger event statistics, the sensitivity will be eventually determined by the systematic uncertainty of the measurement. Optimistically, this might reach the level of SK at $\sim$0.5\,\% \cite{Abe:2011ts}. It should be noted, though, that background conditions might be less favorable in HK due to the shallower depth. A meaningful constraint on the value of the solar $\Delta m^2_{21}$ will be especially interesting if a non-standard course of $P_{ee}(E)$ is observed by other experiments in the vacuum-matter transition region (sect.~\ref{sec:statsolnuosc}). HK features as well a realistic chance for the first detection of $hep$ neutrinos \cite{Abe:2011ts}.

\subsection{Water-based or Slow Liquid Scintillators}
\label{sec:wbls}

A further promising path forward is opened by the relatively recent detector concept aiming at the simultaneous detection of scintillation and Cherenkov photons. Regular organic scintillators are foremost optimized for high light yield. The equally produced Cherenkov light  is either absorbed by the organic scintillator solvent or absorbed and re-emitted by the fluor (and possible secondary wavelength-shifter) that feature absorption bands in the near-UV. However, a larger fraction of the Cherenkov light can be preserved when somewhat compromising on the scintillation properties.
\medskip\\
{\bf Slow liquid scintillators} feature only a reduced concentration of a single fluor like PPO, greatly enhancing the transparency in the near-UV while decreasing the scintillation output. The reduced scintillation efficiency results as well in a relative delay in the scintillation signal, widening the gap to the quasi-instantaneous Cherenkov signals to several nanoseconds. This is sufficient for a time-based discrimination of the two signals based on currently available fast PMTs \cite{Bignell:2015oqa}.
\medskip\\
{\bf Water-based Liquid Scintillators (WbLS)} vary this approach by diluting the organic solvent to reduce the absorption of Cherenkov photons in the bulk material. The composite material is in fact a colloidal solution of organic scintillator droplets with an added surfactant in water \cite{JinpingNeutrinoExperimentgroup:2016nol}. Both light yield and self-absorption scale roughly linearly with the percentage of organic scintillator. As a low-cost material, WbLS has the potential to provide very large detection volumes. The reduction in carbon content will reduce the backgrounds of  {$^{14}$C}  and muon spallation products from  {$^{12}$C}  that beset the $pp$ and $pep$/CNO neutrino detection in organic scintillators. In practice, Rayleigh scattering on the scintillator droplets and radio-purification of the water phase are challenges that still need to be addressed, as well as production of the material on kiloton scale.
\medskip\\
A common feature of both variants of the technique is the added benefit of Cherenkov directionality in a scintillator detector, adding the possibility to discriminate solar neutrino events by their peaked recoil distribution. Moreover, the relative ratio of Cherenkov and scintillation light output can be used as an effective mean of particle identification: at comparable visible energies in the MeV range, $\beta$-like signals will feature a sizable Cherenkov component while $\alpha$-decay backgrounds won't, providing a clear signature for discrimination. However, this considerable advantages have to be carefully balanced with the drawbacks of reduced energy resolution and potentially inferior pulse shape discrimination capabilities. Currently, this novel detection approach is followed up in two projects:
\medskip\\
{\bf Jinping Neutrino Experiment (JNE).} The construction of a low-background neutrino detector based on 4\,kt of water-based or slow organic liquid scintillator (2\,kt fiducial mass) is foreseen for the currently excavated second phase of the China JinPing Laboratory (CJPL) \cite{JinpingNeutrinoExperimentgroup:2016nol}. Based in the foothills of the Himalayas, the CJPL features a formidable rock overburden of up to 2,400\,m. Compared to both Borexino and SNO, solar neutrino detection in the JNE will profit from both larger target size and lower cosmogenic background levels, as well as the added benefits of Cherenkov directionality. The expected reduction in light yield is to be compensated by a particularly dense instrumentation with PMTs. Assuming an effective photoelectron yield of 500\,p.e.\,per MeV and 1,500 days of exposure, and sufficient intrinsic radiopurity, the JPE measurement has the potential to greatly reduce the uncertainty on $P_{ee}(E)$ (fig.~\ref{fig:jinping}). First sensitivity studies point to relative flux uncertainties as low as 1\,\% for $pp$, better than 1\,\% for {$^7$Be}, $\sim$3\,\% for {$^8$B} and as low as 3\,\% for $pep$ neutrinos. Moreover, a positive detection is expected for the {$^{15}$O} component of the CNO neutrino spectrum \cite{JinpingNeutrinoExperimentgroup:2016nol}.
\medskip\\
\begin{figure}[ht!]
\centering
\includegraphics[width=0.6\textwidth]{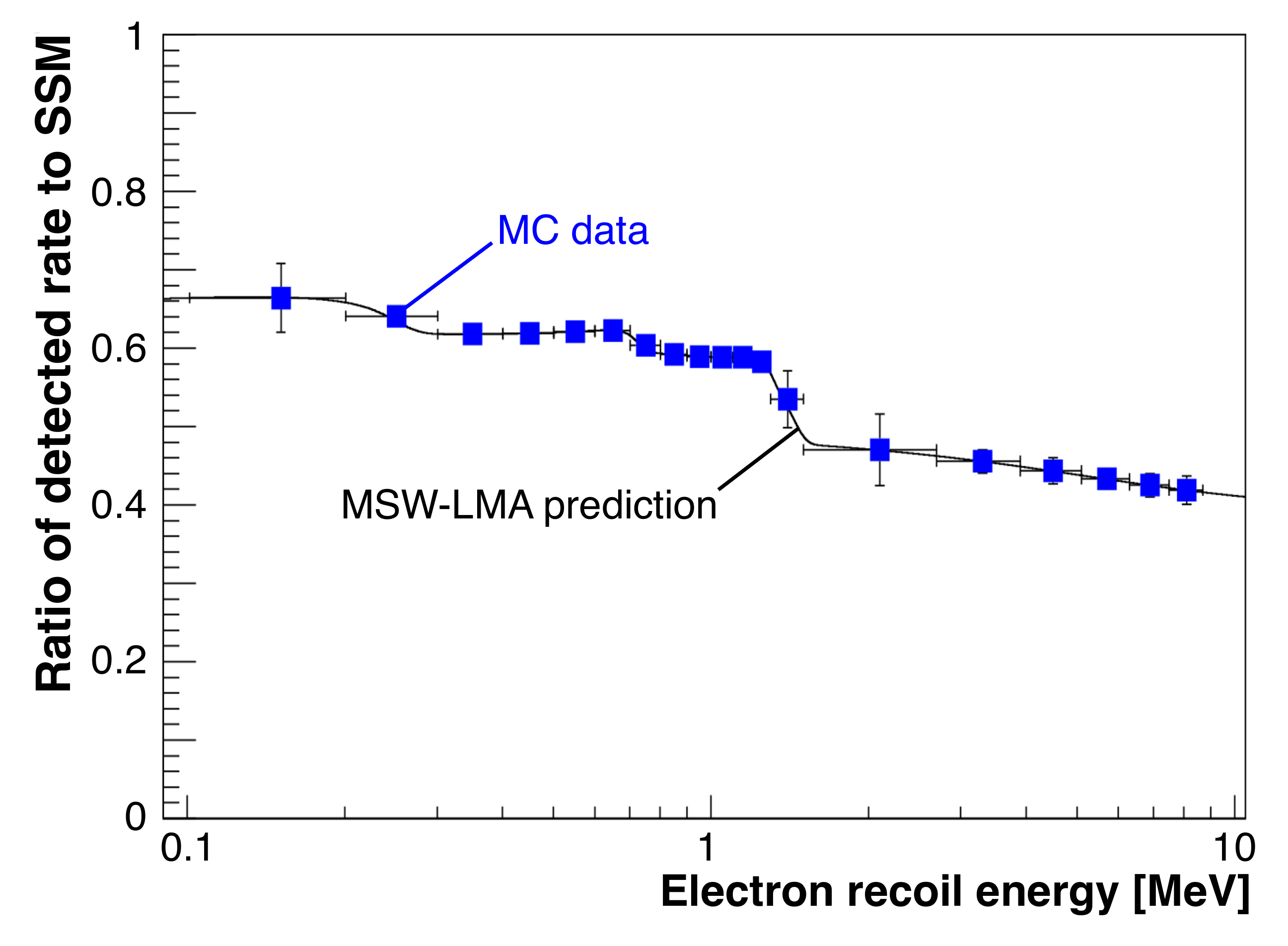}
\caption{A measurement of the solar neutrino spectrum in the Jinping Neutrino Experiment has the potential to greatly reduce the uncertainty on the energy-dependence of $P_{ee}$: The solid line represents the SSM({\it low-Z})+MSW-LMA prediction for $P_{ee}$ as a function of the electron recoil energy, while the data points represent the measurement results expected for 2\,kt fiducial mass, 1,500-day measuring time and 500\,p.e./MeV. The statistical uncertainties shown correspond to the results of a spectral fit to signal and backgrounds but do not include systematic uncertainties on the overall normalization \cite{JinpingNeutrinoExperimentgroup:2016nol}.}
\label{fig:jinping}
\end{figure}
\noindent{\bf THEIA.} The 50-100kt WbLS experiment is proposed as a low-energy neutrino observatory located at the Sanford Underground Research Facility (SURF) in the Homestake mine \cite{Gann:2015yta, Alonso:2014fwf, Gann:2015fba}. The envisioned experimental program comprises three stages of changing detector configuration: In the initial phase, THEIA would serve as a far detector for the future LBNF beam from Fermilab to SURF. For this purpose, a thin WbLS mixture of relatively low scintillation light yield would be used, limiting solar neutrino detection to {$^8$B} energies. In a second stage, the organic fraction of the WbLS would be increased (maybe even to 100\,\%), either in the total volume or in an inserted balloon holding several kilotons of target material. This second phase would allow for a full-fledged low-energy neutrino program, the possibilities for solar neutrino detection closely resembling JNE: The slightly lower rock-overburden would be balanced by significantly enhanced statistics due to the larger fiducial mass (cf.~LENA). Moreover, a 1\,\% loading of the WbLS with the isotope  {$^{7}$Li}  could be foreseen, leading to a substantial improvement in solar neutrino energy resolution (sect.~\ref{sec:dopedls}). In a final third stage, the scintillator inside the balloon could be loaded with an isotope suitable for $0\nu\beta\beta$-decay search.\\   
Utilizing recent technical developments in light detection, THEIA might be (partially) equipped with Large-Area Picosecond PhotoDetectors (LAPPDs). These novel MCP-based photosensors provide a time resolution of 50 picoseconds and thus offer a considerable potential for the discrimination of Cherenkov and scintillation components of the dual light signal even for fast scintillators with nanosecond-scale fluorescence constants \cite{Adams:2013nva}. Commercial production of LAPPDs has started in 2016, albeit in low quantity.

\subsection{Doped scintillators}
\label{sec:dopedls}

While 'plain' water and liquid-scintillator have proven immensely successful in the detection of solar neutrinos, they still face a significant limitation in resolving the shape of the neutrino energy spectra: ES does not allow for event-by-event energy reconstruction. To overcome this restriction, a nuclear CC reaction is required in which the neutrino is captured and its energy is transferred to the final-state charged lepton (reduced by the $Q$-value of the reaction), thus allowing for an unambiguous energy measurement.
\medskip\\
{\bf JUNO.} In principle, large scale organic liquid scintillator detectors like JUNO feature already now the possibility for CC interactions on  {$^{13}$C} . The reaction ${^{13}{\rm C}} + \nu_e \to {^{13}{\rm N}}+e^-$ features an energy threshold of 2.2\,MeV and provides direct access to the interacting $\nu_e$'s energy. Background can be reduced by the delayed coincidence provided by the re-decay of {$^{13}$N}. However, the natural isotopic abundance of  {$^{13}$C}  is only $\sim$1\,\%, and thus the CC channel provides only reduced statistics compared to ES \cite{Mollenberg:2014mfa}. 
\medskip\\
{\bf THEIA.} Water-based liquid scintillator (sect.~\ref{sec:wbls}) features an increased chemical solubility for metals, providing the possibility for a substantial loading with foreign isotopes. For THEIA,  {$^{7}$Li}  loading has been proposed to provide a target for charged-current interactions of $\nu_e$ with a threshold of 862\,keV. While this excludes $pp$ and {$^7$Be} neutrinos from detection, a large fiducial mass combined with moderate loading levels ($\sim$1\,\%) has the potential for a high-rate measurement of CNO neutrinos. Due to the inherent energy resolution of the reaction, the  {$^{7}$Li} -doped detector might even allow to discriminate the partial spectra of {$^{13}$N} and {$^{15}$O} neutrino captures (fig.~\ref{fig:theia}) \cite{Gann:2015yta, Alonso:2014fwf, Gann:2015fba}.
\medskip\\

\begin{figure}[ht!]
\centering
\includegraphics[width=0.6\textwidth]{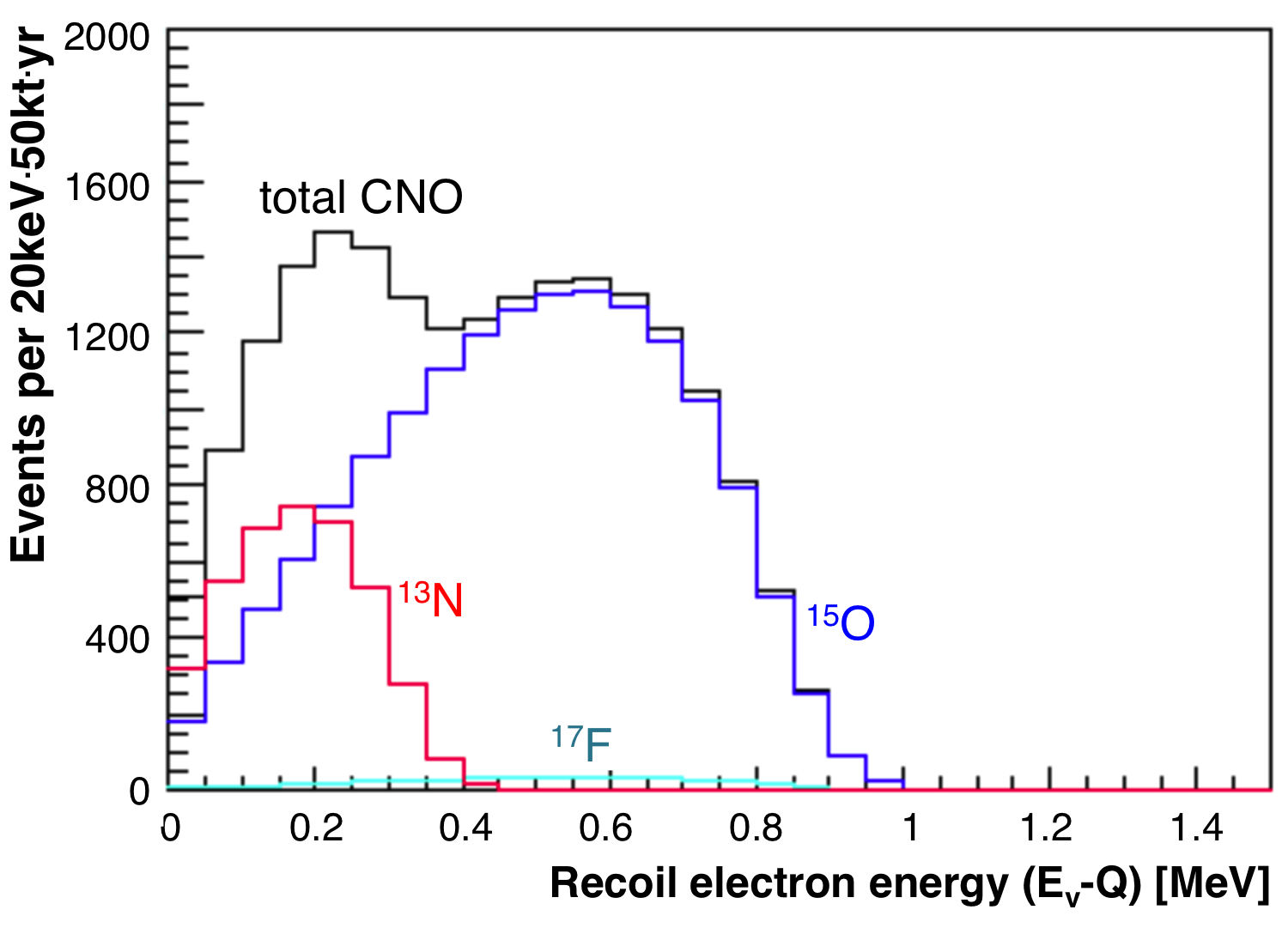}
\caption{Expected electron energy spectrum for the  {$^{7}$Li}  CC reaction channel in THEIA. Due to the intrinsic energy resolution of the reaction, the {$^{13}$N} and {$^{15}$O} contributions to the overall CNO neutrino spectrum could be resolved \cite{Alonso:2014fwf}.}
\label{fig:theia}
\end{figure}

\noindent{\bf LENS.} The LENS (Low Energy Neutrino Spectroscopy) project followed the idea of an organic scintillator loaded with the indium isotope {$^{115}$In} \cite{Grieb:2006mp}. The detection reaction for solar neutrinos, 
$\nu_e + {^{115}{\rm In}} \to {^{115}{\rm Sn}^*} + e^-$, features a very low threshold of 115\,keV, potentially allowing for a direct detection of $pp$-neutrinos. What is more, the reaction is followed by the decay of the excited tin nucleus to the ground state, featuring two low-energy gamma rays (116\,keV, 497\,keV) that provide a fast coincidence tag for the identification of neutrino-induced signals.\\ 
Adversely for detection, {$^{115}$In} in itself is $\beta$-unstable and decays with a half life of $6.4\cdot10^{14}$\,yrs to the ground state of {$^{115}$Sn} ($Q = 499$\,keV). As large quantities of indium are needed to allow for solar neutrino detection, this natural decay rate constitutes a sizable intrinsic background to detection that can only be overcome by optimizing the detector for identification of the neutrino-induced triple coincidence. The latest LENS proposal would foresee a highly segmented detector dividing the scintillator volumes into rectangular compartments, the separating walls acting as light guides that enhance the spatial resolution for event vertices (fig.~\ref{fig:lens}). This concept would provide the necessary discrimination power to discern three nearby $e^-/\gamma$ signals from the random coincidences of three independent {$^{115}$In} decays distributed over the detection volume.\\
A detector based on 10 tons of natural indium would result in an intrinsic background from 2.5\,MBq caused by {$^{115}$In} decays. Provided a sufficiently sophisticated background discrimination, CC reactions of solar neutrinos on {$^{115}$In} would result in the prompt event energy spectrum depicted in figure \ref{fig:lens}: Clear signals are expected for the $pp$ neutrinos, for which a 2.5\,\% measurement of the rate would require 5 years of exposure, and for {$^7$Be} neutrinos. In principle, CNO neutrinos would as well induce an easy-to-separate signal. However, the detection rate would not be sufficient for an actual CNO rate measurement \cite{Grieb:2006mp}.

\begin{figure}[ht!]
\centering
\includegraphics[width=0.9\textwidth]{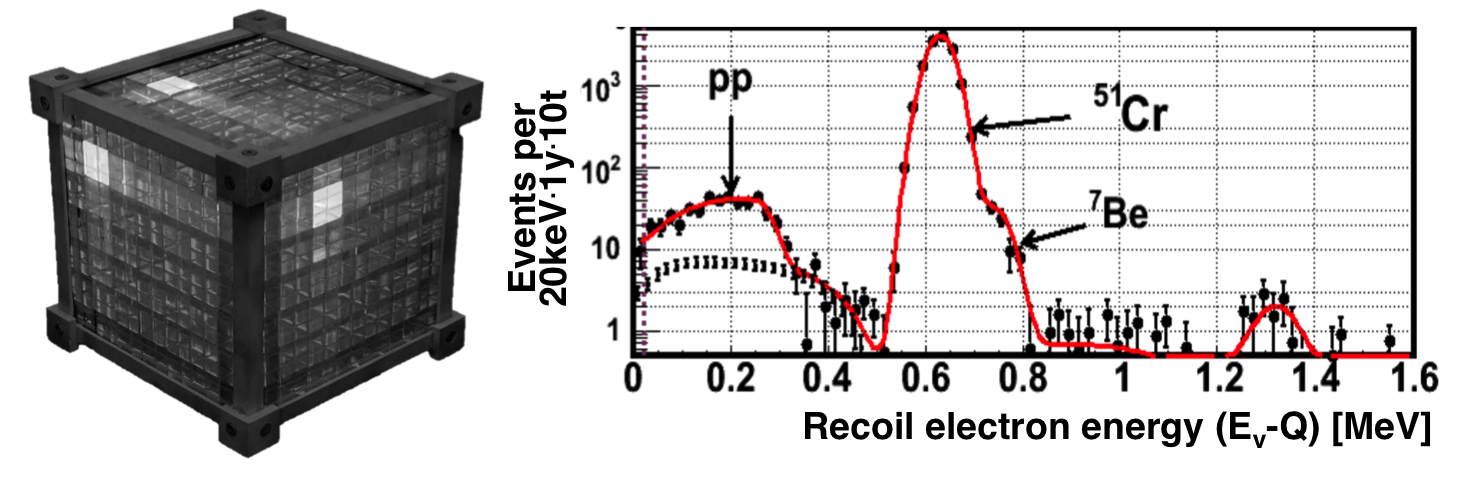}
\caption{Layout of the segmented LENS detector and expected energy spectrum of signal electrons. The CC interaction on {$^{115}$In} provides a direct relation to neutrino energy \cite{Grieb:2006mp}.}
\label{fig:lens}
\end{figure}

\subsection{Liquid noble gases}

Over the last decade, liquid noble gas detectors have made their entrance as large low-threshold and low-background detectors. With target masses on the scale of tens to hundreds of kilograms, these experiments have contributed both to dark matter and neutrino-less double-beta decay searches. Based on their considerable success and progress in experimental technique, liquid-noble gas detectors on the ton-scale are now in their start-up phase, and target masses of 10 tons or more seem technically feasible. This opens up a new pathway to the detection of solar neutrinos via ES in noble liquids.

While technically more demanding than water or organic scintillator detectors, the (still) smaller volumes and higher cost could potentially be offset by a number of advantageous detector properties for solar neutrino detection: substantially higher light yield, better background discrimination and maybe most importantly target materials like neon that feature no long-lived radioactive isotopes. In the following, several experimental approaches are discussed that have been proposed for future detectors.

\subsubsection{Single-phase scintillation detectors}

The most direct way towards a liquid noble gas neutrino detector is a monolithic single-phase target volume surrounded by light sensors to read out the scintillation light. Compared to organic scintillators, the expected light yield is considerably larger: for instance, liquid argon (LAr) releases about 40\,000 photons per MeV. The gases are transparent to their own scintillation light, so that the photons will propagate to the verge of the detection volume. However, the emission spectrum is in the far or extreme UV and thus at too short wavelengths for conventional photosensors. Depending on the noble liquid, either PMTs with UV-transparent windows (liquid xenon, LXe) or coated with a thin layer of fluorescence molecules (LAr) are employed for light detection. In present-day small detectors, e.g. DarkSide-50 and MicroClean, photoelectron yields of 6\,000$-$8\,500 per MeV have been achieved for liquid argon, exceeding the value realized for Borexino by more than an order of magnitude \cite{Franco:2015pha}. This translates directly to a better energy resolution, but has as well beneficial consequences for other detector properties like pulse shape discrimination that depends critically on the available amount of photons.
\medskip\\
{\bf DEAP/CLEAN.} With the DEAP-3600 experiment, a single-phase LAr detectors of 3.6\,t mass is currently going into operation at SNOLab \cite{Amaudruz:2014nsa}. However, the DEAP measurement program and thus the experimental layout is focussed on direct dark matter detection.\\
Contrariwise, the closely related CLEAN project is directed towards solar neutrino detection \cite{McKinsey:2004rk}. The detector concept pursued is based on a spheric target volume containing 300\,t of liquid neon (LNe). With a density of 1.2\,g/l, LNe offers better self-shielding properties than water or organic scintillator. Neon can be effectively cleaned from radioactive contaminants by purification in cryogenic traps. Most noteworthy, neon features $-$ other than argon, xenon and notably carbon $-$ no long-lived radioactive isotope. Therefore, the combination of self-shielding and intrinsic radiopurity allows for virtually background-free measurements of neutrinos in the sub-MeV range. If CLEAN design goals on background levels can be fulfilled, the elastic scattering rate from $pp$-neutrinos can be measured at $\sim$1\,\% accuracy, thereby permitting a meaningful comparison to the SSM prediction for the $pp$-neutrino flux and the solar luminosity constraint \cite{McKinsey:2004rk}.

\subsubsection{Liquid-argon TPCs}

\noindent In current or upcoming detectors, liquid argon (LAr) is mostly used as a target medium for large-scale Time Projection Chambers (TPCs) in which a homogenous electric drift field is applied across a kilogram to kiloton-scale detection volume. Detectors are either single-phase or dual-phase, changing the readout scheme for the drift signal: In the first case, readout wires are directly inserted into the liquid phase, while in the second option the wires or GEMs are mounted in an argon gas layer above the LAr bulk volume, adding a gas amplification stage for the drift signal. In both cases, the direct scintillation signal from LAr is mainly used as start signal for the measurement of the drift time, while the ionization drift signal offers by comparison better energy resolution \cite{Agnes:2014bvk}.
\medskip\\
{\bf Low-threshold TPCs.} Relatively small-scale detectors optimized for dark matter search, e.g.~the dual-phase DarkSide experiment at LNGS \cite{Agnes:2014bvk}, feature detection thresholds sufficiently low for the detection of recoil electrons from ES of solar neutrinos. Unlike neon, atmospheric argon contains a non-negligible amount of the radioactive $\beta$-emitter  {$^{39}$Ar} . Its decay rate defines a lower detection threshold of 0.57\,MeV. This intrinsic background is thus likely to prevent any attempts on the detection of $pp$-neutrinos, while the feasibility to detect the Compton edge of {$^7$Be} neutrinos at 662\,keV depends on the energy resolution achieved. A further potential background, cosmogenic  {$^{42}$Ar}, can be avoided by using argon extracted from underground reservoirs \cite{Franco:2015pha}.

\begin{figure}[ht!]
\centering
\includegraphics[width=0.6\textwidth]{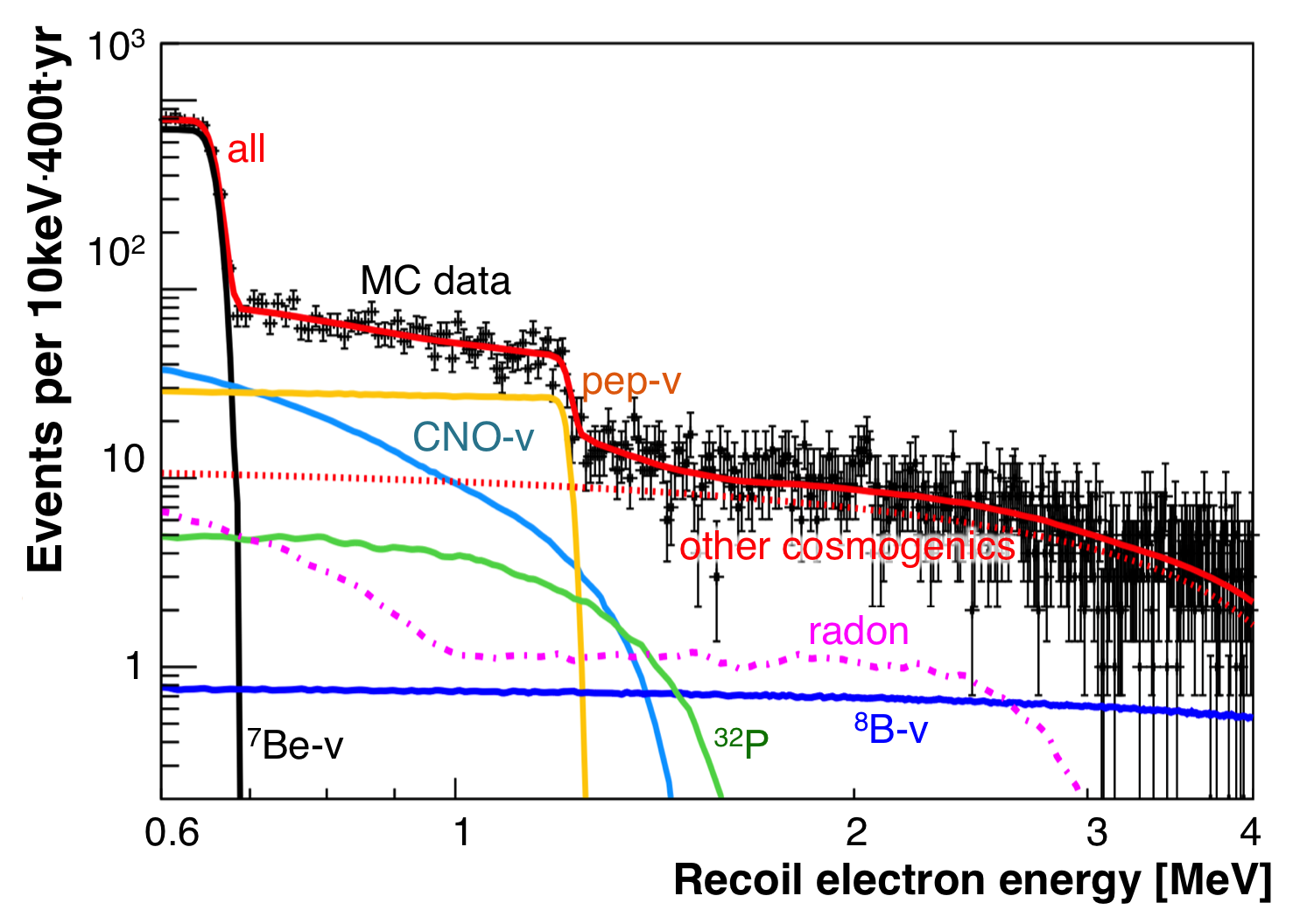}
\caption{Electron recoil spectrum expected in a 400-ton liquid-argon TPC. Background levels correspond to the rock shielding at LNGS and a radon background level of 10\,\textmu Bq per 100\,t \cite{Franco:2015pha}.}
\label{fig:lar}
\end{figure}

\noindent Figure \ref{fig:lar} depicts the electron recoil spectrum above the  {$^{39}$Ar}  endpoint: The most important contributions of background arise from cosmogenic activation of argon, other radioactive noble gases ({$^{85}$Kr}  and  {$^{222}$Rn}) dissolved in the LAr and external $\gamma$-rays from materials of the detector periphery. Compared to LS detectors, a two-phase LAr TPC features an important advantage for the latter contribution: The spatial resolution of the drift signal is sufficient to discriminate a multi-site $\gamma$-quantum that scatters several times inside the target volume from the single vertex of a neutrino interaction. Most likely, the sensitivity to solar neutrinos will depend primarily on the level of  {$^{222}$Rn}  and its decay daughters in the detection volume, which in turn depends on the effectivity of argon purification by charcoal filters or similar techniques \cite{Franco:2015pha}.\\
Unfortunately, none of the currently running LAr TPCs features a sufficiently large target volume for solar neutrino detection: Even the upcoming DarkSide-20k experiment based on 20 tons of LAr is likely too small.\\
In \cite{Franco:2015pha}, a dual-phase LAr TPC dedicated to solar neutrino detection has been proposed, featuring a target mass of 400\,t and operated at the LNGS: The expected background levels are discussed in detail, the corresponding visible energy spectrum is shown in figure \ref{fig:lar}. Depending on the internal radioactivity levels achieved, the authors predict a measurement of the {$^7$Be} elastic scattering rate at the $\sim$2\,\% level, $\sim$10\,\% in the case of $pep$ and $\sim$15\,\% for CNO neutrinos. Therefore, the sensitivity compared to an organic LS detector of the same target mass (Borexino) would be substantially improved.
\medskip\\
{\bf High-threshold TPCs.} A second detector category aims on the instrumentation of 0.1$-$10\,kt of LAr for the detection of GeV neutrinos in long-baseline neutrino oscillation experiments. Prominent examples of running experiments are ICARUS at LNGS (soon Fermilab) and MicroBooNE at Fermilab \cite{Amerio:2004ze,Acciarri:2016smi}. Given the considerably larger volume and reduced instrumentation density, the detection threshold is in fact above solar neutrino energies. This concerns especially the light collection for the prompt scintillation signal that is not needed when the detector is operated in coincidence with a pulsed neutrino beam but a necessity for the observation of solar neutrinos.\\
Similar to the aforementioned experiments, the future DUNE detectors at SURF will primarily aim at the detection of neutrinos from the LBNF beam. However, the DUNE physics scope envisages as well a low-energy neutrino program, including the detection of Supernova neutrinos at 10-MeV energies \cite{Acciarri:2015uup}.  Up to four 10-kt LAr TPCs will be placed at a depths of 1450\,m, the large rock overburden and hence low cosmogenic background levels offering favorable conditions for solar neutrino detection. Solar neutrino sensitivity will lastly depend on the instrumental detection threshold and the intrinsic radioactivity levels realized, as well as the capabilities for the reconstruction of the light and drift signals at low energies.\\
Provided the energy threshold is in the range of a few MeV, a high-statistics measurement of the high-energy part of the {$^8$B} neutrino spectrum seems well feasible. What is more, the most abundant isotope  {$^{40}$Ar}  features a nuclear cross-section for $\nu_e$ CC detection, albeit with a relatively high energy threshold of 4.4\,MeV. Provided the very large target mass envisioned, this CC channel bears the potential for a clean measurement of the {$^8$B} spectrum and day-night asymmetry \cite{Acciarri:2015uup}.

\subsubsection{Liquid-xenon TPCs}

\begin{figure}[ht!]
\centering
\includegraphics[width=0.6\textwidth]{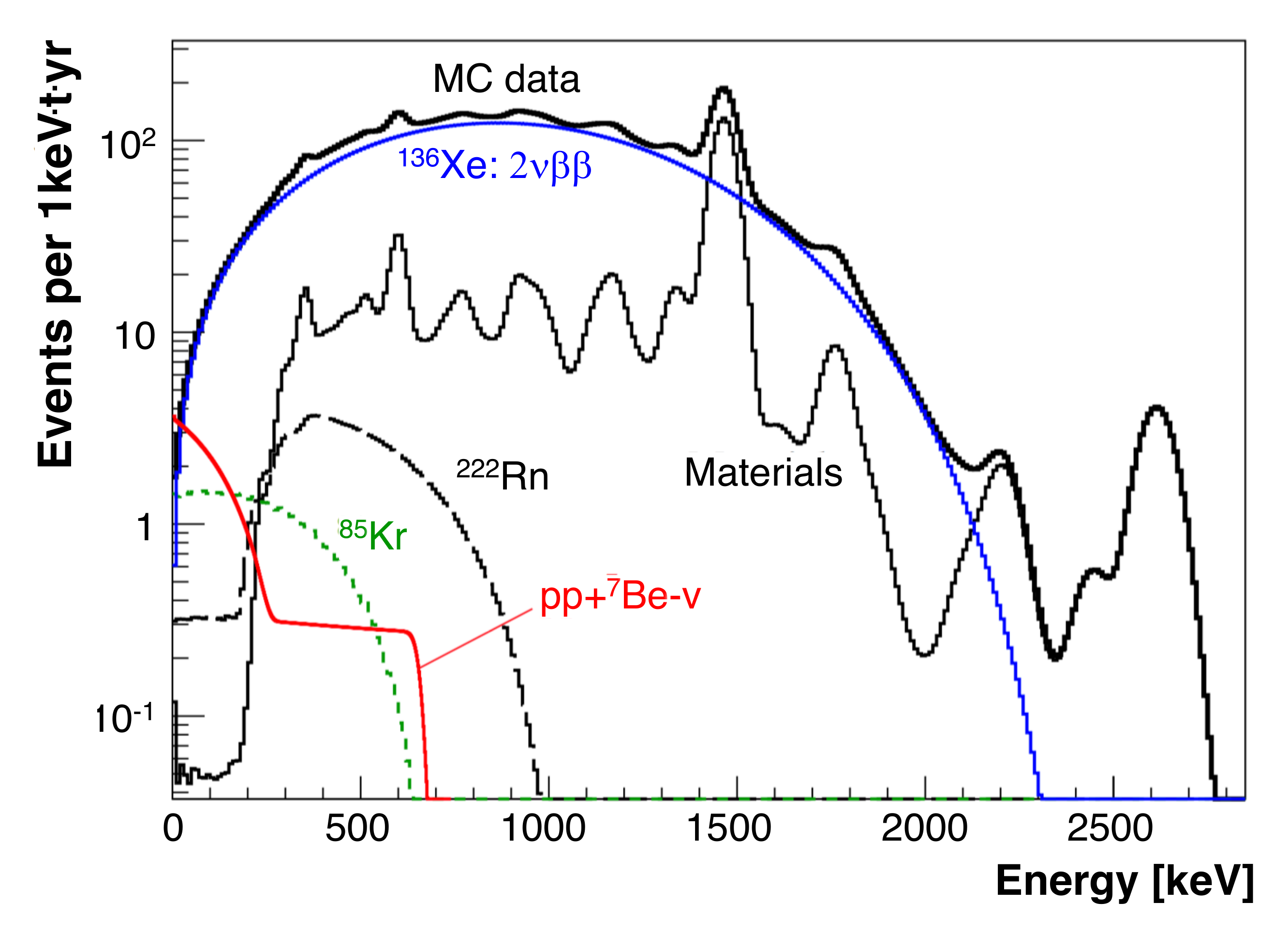}
\caption{Electron energy spectrum in a large liquid-xenon TPC. The solar neutrino signal is covered by the $2\nu2\beta$-decay of {$^{136}$Xe} for all but the lowest energies. Nevertheless, a detection of $pp$-neutrino scattering events will be possible in the energy window from 2 to 30\,keV \cite{Baudis:2013qla}.}
\label{fig:lxe}
\end{figure}

\noindent The experimental focus of current LXe single-phase or double-phase TPCs is mostly on the search for the neutrino-less double beta decay of {$^{136}$Xe} or for nuclear recoils from dark matter (DM). In both respects, LXe detectors have proven very successful and are currently at the forefront of their respective fields \cite{Auger:2012ar, Akerib:2015rjg}. While target masses of today's experiments are still on the scale of several hundred kilograms, the XENON1T experiment has recently started to operate at the LNGS, featuring a detection/fiducial volume of 1 ton for DM interactions \cite{Aprile:2015uzo}. Preparations for the next generation of such experiments (LZ and XENONnT) featuring target volumes of $\sim$5 tons have already begun and will most likely be in operation by 2020 \cite{Akerib:2015cja}. With the DARWIN detector, an even larger project on the scale of 20 tons is being discussed \cite{Baudis:2012bc}.

While the primary motivation of these projects is the direct search for DM, they offer as well a low-background environment for solar neutrino detection. LXe features excellent self-shielding capabilities due to the high density of 3.1\,g/l. Moreover, DM search requires excellent internal purity with regards to  {$^{85}$Kr}  and  {$^{222}$Rn} , setting a standard sufficient for solar neutrino detection. As illustrated by figure \ref{fig:lxe}, the main difficulty arises from the presence of the $\beta\beta$-isotope {$^{136}$Xe} intrinsic to the target material that will dominate the solar neutrino signal at all but the lowest electron recoil energies \cite{Baudis:2013qla}.

Therefore, solar neutrino detection in natural LXe will be restricted to the low-energy regime of $pp$-neutrinos. Given the ultra-low threshold of $\sim$2\,keV that is achieved in current-day detectors, a large and up to now unobserved regime of the $pp$ ES spectrum can be detected, the rate amounting to $\sim$1 count per day and ton \cite{Baudis:2013qla}. Considering as well an upper energy limit of 30\,keV from which backgrounds begin to dominate, an exposure of 70\,t$\cdot$yrs is necessary to achieve a measurement of the $pp$ flux on the 1\,\% level. While clearly outside the reach of current day experiments, this roughly corresponds to 5\,yrs of measurement in the future DARWIN detector \cite{Baudis:2013qla}.

It should be noted that LXe depleted in {$^{136}$Xe} might become available in large quantities as a 'waste product' of the isotopic enrichment performed for $0\nu\beta\beta$-experiments. In this case, an extension of the detection range to higher recoil energies might become possible.

\section{Conclusions and Outlook}

In 2015, Arthur MacDonald and Takaaki Kajita were co-awarded the Noble Price of Physics 'for the discovery of neutrino oscillations, which shows that neutrinos have mass'. Almost two decades have passed since the defining measurements of Super-Kamiokande and SNO that finally lifted the secret. Their remarkable success has set the starting point for the present period of solar neutrino observations: Increasingly accurate measurements of the solar neutrino fluxes and spectral shapes provide not only information on the energy-dependence of oscillation probabilities but also on the parameters of the underlying solar model.

In all these efforts, detectors capable of spectroscopic measurements have played an ever-increasing role. Today, there are two primary players in the field: Huge Super-Kamiokande provides a high-statistics measurement of the rate, the time modulation and the spectral shape of the {$^8$B} neutrino flux. On the low-energy end, the ultra-low background Borexino detector has by now resolved the $pp$, {$^7$Be} and $pep$ components of the spectrum. Between the two of them, they have firmly anchored the basic MSW-LMA oscillation scenario in its vacuum and matter-dominated energy regimes.

Despite this tremendous advances, the story of solar neutrino observation is far from over: Quite fortunately, the transition from vacuum to matter-dominated oscillations falls within the energy range of the solar neutrino spectrum. However, a precise measurement of the $\nu_e \to \nu_e$ survival probability $P_{ee}$ in the thinly populated intermediate region of the neutrino spectrum still poses an experimental challenge. The comparison of future, more precise results on $P_{ee}(E)$ from low-energy {$^8$B} and $pep$ neutrinos and the amplitude of the {$^8$B} day-night asymmetry with the $\bar\nu_e \to \bar\nu_e$ data collected by KamLAND will provide a precision test of the standard MSW-LMA scenario and excellent sensitivity to new neutrino or weak-interaction related phenomena.

On the other hand, further measurements of {$^8$B}, {$^7$Be} and $pp$-neutrino fluxes as well as a first positive detection of CNO neutrinos will increase our knowledge on the rates and ratios of the basic fusion processes and as a consequence a better understanding of the conditions and parameters of the solar interior. In particular, neutrinos might resolve the contradictory findings on solar metallicity from (photon) spectroscopic and helioseismic observations.

Even more than before, solar neutrino detectors able to provide energy-resolved information will be vital to make further progress on these topics. High-statistic measurements in very large conventional water or organic scintillator detectors, direct energy measurements via CC interactions by loaded detectors or new approaches exploiting the advantages of noble liquids offer an exciting range of technologies that can be put to use in solar neutrino spectroscopy. In time, a next generation of detectors will (hopefully) provide a wealth of new results, offering either further confirmation of or entirely new insights into the mechanisms underlying neutrino flavor oscillations and the workings of our central star.


\bibliographystyle{h-physrev}
\bibliography{solarnuspec}

\end{document}